\documentclass[final,11pt]{article}
\usepackage{multirow}
\usepackage[nottoc]{tocbibind}
\usepackage{geometry}
\usepackage{subcaption}
\usepackage{placeins}
\usepackage[affil-it, auth-sc]{authblk}
\usepackage[english]{babel}
\usepackage[utf8]{inputenc}
\usepackage{graphicx}
\usepackage{amssymb}
\usepackage{amsthm}
\usepackage{amsmath}
\usepackage{amsfonts}
\usepackage{verbatim}
\usepackage{lineno}
\usepackage{todonotes}
\presetkeys{todonotes}{inline, color=blue!30, size=\small}{}
\usepackage{slashed}
\usepackage{color, soul}
\definecolor{darkred}{rgb}{0.9, 0.0, 0.0}
\definecolor{darkgreen}{rgb}{0.0, 0.5, 0.0}
\usepackage[colorlinks,bookmarks,linkcolor=darkred, citecolor=darkgreen]{hyperref}
\usepackage{eso-pic}
\usepackage[sort&compress,numbers]{natbib}
\usepackage{doi}
\usepackage{paralist,epsfig}
\usepackage{relsize}
\usepackage{nicefrac,esint}
\usepackage{float}
\usepackage{cleveref}
\usepackage{marginnote}

\newcommand{\slashpi}{\protect{\slash\hspace{-0.5em}\pi}}

\begin{document}

\AddToShipoutPictureFG*{\AtPageUpperLeft{\put(-60,-60){\makebox[\paperwidth][r]{LA-UR-23-28357, FERMILAB-PUB-24-0532-T, INT-PUB-24-054}}}}

\title{\bf Effective field theory for radiative corrections \\
to charged-current processes II: Axial-vector coupling}
\author[1]{Vincenzo~Cirigliano \thanks{cirigv@uw.edu}}
\affil[1]{Institute for Nuclear Theory, University of Washington, Seattle WA 98195, USA \vspace{1.2mm}}
\author[1]{Wouter~Dekens \thanks{wdekens@uw.edu}}
\author[2]{Emanuele~Mereghetti \thanks{emereghetti@lanl.gov}}
\affil[2]{Theoretical Division, Los Alamos National Laboratory, Los Alamos, NM 87545, USA \vspace{1.2mm}}
\author[2]{Oleksandr~Tomalak \thanks{sashatomalak@icloud.com}}

\date{\today}

\maketitle

We discuss the hadronic structure-dependent radiative corrections to the axial-vector coupling that controls single-nucleon weak charged-current processes -- commonly denoted by $g_A$. We match the Standard Model at the GeV scale onto chiral perturbation theory at next-to-leading order in the one-nucleon sector, in the presence of electromagnetic and weak interactions. As a result, we provide a representation for the corrections to $g_A$ in terms of infrared finite convolutions of simple kernels with the single-nucleon matrix elements of time-ordered products of two and three quark bilinears (vector, axial-vector, and pseudoscalar). We discuss strategies to determine the required nonperturbative input from data, lattice QCD (+QED), and possibly hadronic models. This work paves the way for a precise comparison of the values of the ratio $g_A/g_V$ extracted from experiment and from lattice-QCD, which constrain physics beyond the Standard Model.

\newpage

\tableofcontents

\section{Introduction}
\label{sec1}

Precision studies of charged-current processes involving baryons at low energies are well motivated and timely. On the one hand, precision measurements of neutron decay observables~\cite{Markisch:2018ndu,Dubbers:2018kgh,UCNt:2021pcg}  provide information on the fundamental parameters of the Standard Model (SM) and nucleon nonperturbative structure. On the other hand, antineutrino-proton scattering, so-called inverse beta decay, allows us to precisely constrain the antineutrino fluxes from nuclear reactors~\cite{Reines:1959nc,DayaBay:2012fng,RENO:2012mkc,DoubleChooz:2011ymz,JUNO:2015zny,KamLAND:2002uet,Hayes:2016qnu,STEREO:2018rfh,NEUTRINO-4:2018huq,PROSPECT:2018dtt}. Both processes achieve subpercent level of precision and call for precise predictions of the corresponding decay rates and cross sections.

At low energy, up to recoil corrections and an overall factor proportional to $G_F V_{ud}$, with the Fermi coupling constant $G_F$~\cite{Fermi:1934hr,Feynman:1958ty,vanRitbergen:1999fi,MuLan:2012sih} and the matrix element of the Cabibbo-Kobayashi-Maskawa quark mixing matrix $V_{ud}$~\cite{Cabibbo:1963yz,Kobayashi:1973fv,Hardy:2020qwl,ParticleDataGroup:2020ssz}, nucleon charged-current semileptonic reactions are described by two coupling constants: vector $g_V$ and axial-vector $g_A$. The ratio of these coupling constants can be extracted from measurements of the beta asymmetry in polarized neutron decay~\cite{Markisch:2018ndu,Dubbers:2018kgh}, while their absolute values determine the neutron lifetime~\cite{UCNt:2021pcg}. One of the main interests of the theoretical community is the determination of both vector and axial-vector coupling constants starting from the Standard Model Lagrangian. In the traditional approach~\cite{Sirlin:1967zza,Abers:1968zz,Sirlin:1977sv,Sirlin:1981ie,Czarnecki:2004cw}, the coupling constants are evaluated in the full Standard Model using current algebra techniques, with nonperturbative input from dispersion theory~\cite{Seng:2018yzq,Seng:2018qru,Shiells:2020fqp,Hayen:2020cxh}, hadronic models~\cite{Czarnecki:2019mwq}, or lattice QCD~\cite{Chang:2018uxx,Ma:2023kfr}.

More recently, Refs.~\cite{Cirigliano:2022hob,Cirigliano:2023fnz} presented a complementary approach to this problem,  based on effective field theory (EFT) methods.  Working at the level of an EFT involving nucleons, pions, photons, and light leptons,  Ref.~\cite{Cirigliano:2022hob} uncovered potentially large structure-dependent radiative corrections to the axial-vector coupling constant. Ref.~\cite{Cirigliano:2023fnz} presented a systematically-improvable top-down EFT approach for low-energy charged-current semileptonic interactions, which led to a precise determination of the vector coupling constant $g_V$. Our approach uses a tower of EFTs, connecting physics from the electroweak scale, $M_W\sim 100$ GeV, all the way to the scale of a few MeV. As the first step, we integrate out the electroweak-scale physics and match the SM to the low-energy effective field theory (LEFT), which describes weak processes through effective four-fermion interactions. Subsequently, we determine the low-energy couplings (LECs) that determine the strengths of the interactions in heavy-baryon chiral perturbation theory (HBChPT) by matching it to the LEFT. This matching is nonperturbative, and is performed by specifying the quantum chromodynamics (QCD) correlation functions  that describe the short-distance hadronic contributions to HBChPT LECs.  This approach was pioneered in Ref. \cite{Descotes-Genon:2005wrq} in the mesonic sector, and extended to $g_V$ in Ref. \cite{Cirigliano:2023fnz}. Next, we integrate out the pion fields and arrive at the effective Lagrangian~\cite{Ando:2004rk,Falkowski:2021vdg,Cirigliano:2022hob}
\begin{equation}
\mathcal L_{\slashpi} = - \sqrt{2} G_F V_{ud} \, \overline{e} \gamma_\rho P_L \nu_e  \, \overline{N}_v \left(  g_V v^\rho - 2 g_A S^\rho \right) \tau^+  N_v   + \mathcal{O} \left(  \alpha, \epsilon_{\rm recoil} , \epsilon_{\slashpi}, \epsilon_\chi \right)  + \mathrm{h.c.},\label{eq:Lagrangian_at_leading_order}
\end{equation}
where $N_v = \left( p,~n \right)^T$ denotes the heavy-nucleon field doublet, $v^\rho = (1,0,0,0)$ is the nucleon velocity, and $S^\rho = \left( 0, \sigma^\rho/2 \right)$ denotes the nucleon spin. $\sigma$ and $\tau$ are the Pauli matrices in spin and isospin space, respectively. The scale-dependent effective couplings $g_V$ and $g_A$ have an expansion in powers of the electromagnetic coupling constant $\alpha = e^2/\left(4 \pi \right)$ with the electric charge $e$, $\epsilon_{\rm recoil} = q_{\rm ext}/m_N$, which describes small kinematic corrections, $\epsilon_\slashpi = q_{\rm ext}/m_\pi$, which captures the radiative pion contributions, and the HBChPT expansion parameter $\epsilon_\chi = m_\pi/\Lambda_\chi$ with the scale $\Lambda_\chi = 4 \pi F_\pi \approx 1~\mathrm{GeV}$, where $F_\pi$ is the pion decay constant. $m_N$ and $m_\pi$ denote the nucleon and pion masses, respectively, and $q_{\rm ext}$ represents the typical kinematic scale in the neutron decay or low-energy (anti)neutrino-nucleon scattering. In Ref.~\cite{Cirigliano:2023fnz}, we derived a representation for the vector coupling constant $g_V$ in terms of QCD objects and predicted the neutron lifetime in terms of the theoretically determined $g_V$ and the ratio $g_A/g_V$ extracted from experiment. Improving on the traditional approach, we provided a seamless interface between nonperturbative input (from dispersive or lattice calculations) and the resummation of large logarithms in the next-to-leading logarithmic approximation. The goal of this paper is to complete the matching of LEFT to HBChPT and pionless EFT by providing representations for the structure-dependent radiative corrections to $g_A$ in terms of QCD correlation functions.
 
The effective coupling $g_A (\mu_\chi)$ appearing in the pionless EFT Lagrangian can be expressed as a series in small parameters~\cite{Cirigliano:2022hob}:
\begin{align}
g_A (\mu_\chi) &= g_A^{(0)} C_\beta^r(\mu)\left[1 + \sum_{n = 2}^\infty \Delta^{(n)}_{A, \chi} + \frac{\alpha}{2 \pi} \sum_{n = 0}^\infty \Delta^{(n)}_{A, \rm em}  (\mu, \mu_\chi) \right],\label{eq:axial_coupling_expansion}
\end{align}
where $g_A^{(0)}$ denotes the axial-vector coupling in the chiral limit and in absence of electromagnetic corrections. $C_\beta^r (\mu) = 1 + \left( \alpha/\pi \right) \ln \left(M_Z/\mu\right) + ... $, with the mass of $Z$ boson $M_Z$, denotes the matching coefficient between the Standard Model and the four-fermion theory below the electroweak scale, with $\mu$ the renormalization scale in LEFT, which should of course cancel in observables. $\Delta^{(n)}_{A, \chi}$ denote nonperturbative QCD corrections proportional to the quark mass. They have been calculated up to the fourth order, i.e., $n=4$, in HBChPT~\cite{Bernard:2006te}, but, for our purposes, they can be absorbed into a definition of $g_A$ in the isospin limit, which we denote by $g_A^{\rm QCD} \equiv g_A^{(0)} \left( 1 + \sum_n \Delta^{(n)}_{A,\chi} \right)$. Finally, $\Delta^{(n)}_{A, \rm em} (\mu, \mu_\chi)$ denote electromagnetic corrections and depend on both the LEFT and HBChPT renormalization scales, denoted by $\mu$ and $\mu_\chi$, respectively. These corrections were computed in Ref.~\cite{Cirigliano:2022hob} up to next-to-leading order, i.e., $n=1$, and  include loops proportional to the pion-mass splitting $m_{\pi^\pm}^2 - m^2_{\pi^0} = 2 e^2 F_\pi^2 Z_\pi$ (with the low-energy coupling (LEC) $Z_\pi$)  and the short-distance contributions parameterized by the  LECs $c_{3,4}$ and $\hat{C}_A(\mu, \mu_\chi)$:
\begin{align}
\Delta^{(0)}_{A, \rm em}(\mu, \mu_\chi) &= Z_\pi \left[\frac{1+ 3 \left( g_A^{(0)} \right)^2}{2} \left(\ln \frac{\mu^2_\chi}{m_\pi^2} -1 \right) - \left( g_A^{(0)} \right)^2 \right] + \hat{C}_A(\mu, \mu_\chi), \label{eq:leading_order_correction_gA} \\
\Delta^{(1)}_{A, \rm em} &=   4  Z_\pi \pi m_\pi  \left[c_4 - c_3  + \frac{3}{8 m_N} + \frac{9}{16} \frac{\left( g_A^{(0)} \right)^2}{m_N} \right]. 
\end{align}
The combination of LECs $\hat{C}_A$ is given by
\begin{align}    
    \hat{C}_{A} &= 8 \pi^2 \left[ -\frac{X_6}{2} + 2 \left( A_1 + A_2 + A_3 + A_4 \right) + \frac{1}{g_A^{(0)}} \left( g_1 + g_2 + g_{13} + \frac{g_{11}}{2} \right) \right],\label{eq:axial_contribution_LECs_v1}
\end{align}
cf. Section~\ref{sec:subsec31} for the definitions of the coupling constants $X_6, A_{1,2,3,4}, g_{1,2,11,13}$. The goal of this work is to express the combinations of LECs in $\hat C_A$ in terms of two- and three-point QCD correlation functions of quark bilinears.\footnote{In this paper, we refer to the matrix element of $n$ quark bilinears between initial and final nucleon states as the $n$-point function.} The correlation functions can be calculated via nonperturbative techniques, such as lattice QCD. Recent evaluations within the traditional approach~\cite{Gorchtein:2021fce,Hayen:2021iga} consider only two-point correlation functions, while we demonstrate that there are nonvanishing contributions from the three-point correlation functions as well~\cite{Cirigliano:2022hob,Tomalak:2023xgm}. Our results have a straightforward interpretation in terms of the perturbative insertion of QED currents in hadronic matrix elements (see Ref.~\cite{Carrasco:2015xwa} for the lattice-QCD formulation of a similar strategy), but can also be interpreted as the leading term in a nonperturbative lattice-QCD+QED approach~\cite{Endres:2015gda}. Therefore our approach is complementary to the one proposed in Ref.~\cite{Seng:2024ker}, which relies on input from future lattice QED+QCD calculations. We will discuss in detail the connections between our results and those of Ref.~\cite{Seng:2024ker} in Subsection~\ref{sec:subsec52}.

The paper is organized as follows. In Section~\ref{sec2}, we review the matching of the Standard Model to the charged-current four-fermion interactions in the LEFT and introduce the LEFT spurion fields and external sources that will be used to perform the matching onto HBChPT. Section~\ref{sec3} describes the connection of LEFT to HBChPT. First, we provide the relevant terms of the HBChPT Lagrangian at leading and next-to-leading order and introduce the spurions, external sources, as well as a complete basis of electro(weak) HBChPT operators. In  Subsection~\ref{sec:subsec31}, we define the contribution from the two-point and three-point correlation functions $\left[\hat{C}_A \right]_{\mathrm{2pt}}$ and $\left[\hat{C}_A \right]_{\mathrm{3pt}}$, respectively, that enter the axial-vector coupling constant. First, we evaluate the electromagnetic coupling constants that are responsible for the renormalization of the isoscalar nucleon mass in Subsection~\ref{sec:subsec315}. We determine the electroweak coupling constants and the combination of electromagnetic coupling constants $g_{11} + g_{12}$ in  Subsections~\ref{sec:subsec32} and~\ref{sec:subsec33}, respectively. Combining these results in Subsection~\ref{sec:subsec34}, we present the two-point contribution $\left[\hat{C}_A \right]_{\mathrm{2pt}}$. In Subsection~\ref{sec:subsec35}, we determine the remaining combination of HBChPT electromagnetic coupling constants in terms of the LEFT three-point correlation functions. Here, we adopt two strategies, involving insertions of the pseudoscalar density  (Subsection ~\ref{sec:subsec351}) or insertions of the axial-vector current (Subsection ~\ref{sec:subsec352}). For completeness, in Section~\ref{sec4}, we describe the QED contributions to the evolution of the axial-vector coupling constant between hadronic and pion-mass scales. We summarize our results in Subsection~\ref{sec:subsec51}, compare to the recent framework of Ref.~\cite{Seng:2024ker} in Subsection~\ref{sec:subsec52}, and present the connection with future lattice-QCD and dispersive approaches in Subsection~\ref{sec:subsec53}. Conclusions and outlook are presented in Section~\ref{sec6}.

We relegate some more technical aspects of our work to a number of Appendices. In Appendix~\ref{app:relations}, we present the invariant amplitude decomposition for the two-point correlation functions of the axial-vector and vector currents, derive novel relations between two-point invariant amplitudes, present a new sum rule for these amplitudes, and determine the correlation functions of pseudoscalar and vector currents as well as axial-vector and vector currents in terms of invariant amplitudes. In Appendix~\ref{app:IR}, we derive the infrared behavior for all nonperturbative correlation functions relevant to our analysis. In Appendix~\ref{app:Ward}, we present the Ward identities in LEFT and HBChPT and derive some relevant relations between one-, two-, and three-point correlation functions in Apendices~\ref{app:Ward_one_point},~\ref{app:Ward_two_point}, and~\ref{app:Ward_three_point}, respectively. In Appendix~\ref{app:NLOChPT}, we determine all next-to-leading order HBChPT coupling constants and present the expression for the radiative corrections to the nucleon axial-vector charge without any LECs from HBChPT.

\section{Step I: matching the Standard Model to LEFT}
\label{sec2}

The matching between the SM and the LEFT follows the same steps as for the vector coupling constant in Ref.~\cite{Cirigliano:2023fnz}. In this Section, we introduce notations and provide the most important details of the matching. After integrating out the electroweak gauge bosons, the Higgs field, and the top quark, the LEFT Lagrangian that describes muon and $\beta$ decays is
\begin{equation}
    {\cal L}_{\rm LEFT}  =   - 2 \sqrt{2} G_F \ \overline{e}_L \gamma_\rho \mu_{L}  \,  \overline{\nu}_{\mu L} \gamma^\rho \nu_{eL} - 2 \sqrt{2} G_F  V_{ud} \ C_\beta (a,\mu)  \ \overline{e}_L \gamma_\rho \nu_{eL}  \,  \overline{u}_{L} \gamma^\rho d_L + \ {\rm h.c.} + ...,\label{eq:LLEFT}
\end{equation}
where $G_F$ is the scale-independent Fermi constant~\cite{Fermi:1934hr,Hill:2019xqk} that is extracted from precise measurements of the muon lifetime~\cite{vanRitbergen:1999fi,FAST:2007rsc,Casella:2013bla,MuLan:2012sih}, and $V_{ud}$ is the matrix element of the Cabibbo-Kobayashi-Maskawa quark mixing matrix~\cite{Cabibbo:1963yz,Kobayashi:1973fv,Hardy:2020qwl,ParticleDataGroup:2020ssz}. The semileptonic matching coefficient $C^r_\beta \left(a, \mu \right)$ depends on the scale and on the renormalization scheme. We use the Naive Dimensional Regularization (NDR) scheme for $\gamma_5$ in $d \neq 4$ dimensions and use an arbitrary parameter $a$ to define the evanescent operator $\mathrm{E}(a)$ that appears in the tensor product of $3$ gamma matrices in $d$ dimensions~\cite{Buras:1989xd,Dugan:1990df,Herrlich:1994kh}:
\begin{equation}
    \gamma^\alpha \gamma^\rho \gamma^\beta P_L \otimes \gamma_\beta \gamma_\rho \gamma_\alpha P_L = 4 \left[ 1 + a \left( 4-d\right) \right] \gamma^\rho P_L \otimes \gamma_\rho P_L + \mathrm{E} \left( a \right),
\end{equation}
where $\mathrm{E}(a)$ has a vanishing matrix element in $d=4$. Within this scheme, near $\mu\simeq M_W$, the renormalized semileptonic coefficient $C_\beta^r$ can be conveniently expressed as~\cite{Hill:2019xqk,Dekens:2019ept,Sirlin:1981ie}
\begin{align}
    C^r_\beta \left( a, \mu \right) = 1 + \frac{\alpha}{\pi} \, \ln \frac{M_Z}{\mu} + \frac{\alpha}{\pi} B \left( a \right)  - \frac{\alpha \alpha_s}{4 \pi^2} \ln \frac{M_W}{\mu}+ \mathcal{O} \left( \alpha \alpha_s \right) + \mathcal{O} \left( \alpha^2 \right), \qquad B \left( a \right) = \frac{a}{6} - \frac{3}{4},
\end{align}
where we only keep logarithmic contributions at $\mathcal{O}(\alpha \alpha_s)$, as the finite matching coefficients and the corresponding three-loop anomalous dimensions are not known, see Ref. \cite{Moretti:2024cbe} for progress in this area, $M_Z$ and $M_W$ are the masses of $Z$ and $W$ bosons, respectively, and $\alpha_s$ is the strong coupling constant at the scale $\mu$. The evolution of the Wilson coefficient $C_\beta^r$ between electroweak and GeV scales is described in Refs.~\cite{Cirigliano:2023fnz,Aebischer:2025hsx}. In this paper, we start from the $C_\beta^r$ value at low energies.

\subsection{External sources and spurions}
\label{sec:subsec21}

The matching of LEFT to HBChPT is conveniently performed by introducing classical source fields $\overline{l}^\mu(x)$ and $\overline{r}^\mu (x)$ for the left- and right-handed chiral currents of quarks as well as electromagnetic left ${\bf q}_L$ and right ${\bf q}_R$ spurions, and the weak spurion ${\bf q}_W$~\cite{Urech:1994hd,Moussallam:1997xx,Knecht:1999ag,Descotes-Genon:2005wrq}. For completeness, we introduce scalar and pseudoscalar classical sources, $\overline{s}(x)$ and $\overline{p}(x)$, respectively, which play an important role in the derivation of Ward identities and in the matching calculation of Section~\ref{sec:subsec351}. The source term for currents, the QED and weak charged-current interactions of the light quarks $q^T = (u,d)$ can be written as
\begin{align}
    {\cal L}_{\rm LEFT} &= \overline{q}_L \slashed{\overline{l}} q_L + \overline{q}_R \slashed{\overline{r}} q_R - e \left( \overline{q}_L {\bf q}_L \slashed{A} q_L + \overline{q}_R {\bf q}_R \slashed{A} q_R \right)  +   \left( \overline{e}_L \gamma_\rho \nu_{eL} \, \overline{q}_L {\bf q}_W \gamma^\rho  q_L + \mathrm{h.c.}\right) \nonumber \\
    & - \overline{q}_R (\mathcal M + \overline{s} + i \overline{p}) q_L - \overline{q}_L (\mathcal M + \overline{s} - i \overline{p}) q_R, \label{eq:LEFT2}
\end{align}
where $A^\mu$ denotes the photon field, and $\mathcal M$ is the quark-mass matrix.

The Lagrangian in Eq.~\eqref{eq:LEFT2} is invariant under local $G=SU(2)_L \times SU(2)_R \times U(1)_V$ transformations
\begin{equation}
    q_L \to L (x)e^{i\alpha_V(x)}  q_L, \quad q_R \to R(x) e^{i\alpha_V(x)}  q_R,
\end{equation}
with $L,R \in SU(2)_{L,R}$, provided ${\bf q}_{L,R}$,  ${\bf q}_W$, $\overline{s}$, and $\overline{p}$ transform as ``spurions" under the chiral group, namely ${\bf q}_{L,W} \to L {\bf q}_{L,W} L^\dagger$ and ${\bf q}_R \to R  {\bf q}_R R^\dagger$, $\mathcal M + \overline{s} + i \overline{p} \rightarrow R \left( \mathcal M +  \overline{s} + i \overline{p}\right) L^\dagger$. In addition, $\overline{l}_\mu$ and $\overline{r}_\mu$ transform as gauge fields under $G$. At the physical point, the spurions take the values,
\begin{equation}
    {\bf q}_L = {\bf q}_R = {\rm diag} (Q_u, Q_d),\qquad  {\bf q}_W = -2 \sqrt{2}  \mathrm{G}_\mathrm{F} V_{ud} C^r_\beta\, \tau^+, \qquad \mathcal M = \frac{m_u + m_d}{2} + \frac{m_u - m_d}{2} \tau^3,
\end{equation}
with $\tau^+ = \left(\tau^1 + i \tau^2 \right)/2$ in terms of the Pauli matrices $\tau^{a}$, while the remaining sources vanish. Note that we include the LEFT Wilson coefficient $C_\beta^r$ in the definition of the spurion ${\bf q}_{W}$. In our analysis we will neglect strong isospin breaking and use $m_d = m_u = \overline{m}$, as the corrections to $g_A$ are expected to arise at $\mathcal{O} \left( \left(m_u - m_d \right) m_\pi^2/\Lambda_\chi^3 \right)$~\cite{Cirigliano:2022hob}, or at $\mathcal{O} \left(\left(m_u - m_d \right)^2 /\Lambda_\chi^2 \right)$ according to Ref.~\cite{Chang:2018uxx}.

The $\mathcal{O}(e^2 )$ counterterms in the LEFT Lagrangian can be written in terms of spurion fields as~\cite{Descotes-Genon:2005wrq}
\begin{align}
{\cal L}^{\rm CT}_{\rm LEFT} &=  -2 e^2 Q^2_e g_{00} \, \overline{e} \left( i \slashed{\partial} - e Q_e \slashed{A} -m_e\right) e - i g_{23} e^2 \left( \overline{q}_L \left[ {\bf q}_L, D^\rho {\bf q}_L \right] \gamma_\rho q_L + \overline{q}_R \left[ {\bf q}_R, D^\rho {\bf q}_R \right] \gamma_\rho q_R \right) \nonumber \\
& +   e^2 Q_e \Big( \overline{e}_L \gamma_\rho  \nu_L \left( g_{02} \,  \overline{q}_L {\bf q}_W  {\bf q}_L \gamma^\rho q_L + g_{03}  \, \overline{q}_L {\bf q}_L  {\bf q}_W \gamma^\rho  q_L \right) + \mathrm{h.c.} \Big), \label{eq:CT}
\end{align}
where $g_{00}$ is the counterterm related to the electron wavefunction renormalization, $g_{02}$ and $g_{03}$ come from the counterterm of $C_\beta$, while $g_{23}$ includes contributions from both the counterterm of $C_\beta$ as well as divergences related to the quark wavefunction renormalization. Furthermore,
\begin{align} 
D^\rho {\bf q}_L  &\equiv  \partial^\rho {\bf q}_L - i \left[ l^\rho, {\bf q}_L \right], \label{eq:charge_covariant_derivative_left} \\
D^\rho {\bf q}_R &  \equiv  \partial^\rho {\bf q}_R - i \left[ r^\rho, {\bf q}_R \right], \label{eq:charge_covariant_derivative_right}
\end{align}
are chiral covariant derivatives, expressed in terms of the fields $l^\mu (x)$ and $r^\mu (x)$ that combine the classical sources, spurions, the electromagnetic field, and leptonic currents:
\begin{align}
    l_\mu &= \overline{l}_\mu - e {\bf q}_L A_\mu  + {\bf q}_W \, \overline{e}_L \gamma_\mu \nu_{eL} + {\bf q}_W^\dagger \overline{\nu}_{eL} \gamma_\mu e_L, \\
    r_\mu &= \overline{r}_\mu - e {\bf q}_R A_\mu.\label{eq:sources}
\end{align}
In the $\overline{\rm MS}$ scheme, the $g_{ij}$ couplings in Eq.~\eqref{eq:CT} are determined by the $1/\varepsilon$ divergences and can be written as
\begin{equation}
    g_{ij} = \frac{h_{ij}}{\left(4 \pi \right)^2} \left( \frac{1}{\varepsilon} - \gamma_E + \ln \left( 4 \pi \right) \right), \label{eq:gij}
\end{equation}
with $h_{00} = 1/2$, $h_{23}= (1/2) (1 - \alpha_s/\pi) $, $h_{02} = -1 -  \alpha_s/\pi$, and $h_{03} = 4 -2 \alpha_s/\pi$.

\section{Step II: matching LEFT to HBChPT}
\label{sec3}

In this Section, we provide details of the matching between LEFT and HBChPT. We start by discussing the relevant terms of the HBChPT Lagrangian at leading and next-to-leading order and introduce spurions, external sources, and the complete operator basis in Subsection~\ref{sec:subsec31}. In this Subsection, we also define the two-point $\left[\hat{C}_A \right]_{\mathrm{2pt}}$ and three-point $\left[\hat{C}_A \right]_{\mathrm{3pt}}$ components of the total correction to $g_A$ given in Eq.~\eqref{eq:axial_contribution_LECs_v1}. In Subsection~\ref{sec:subsec315}, we determine the combinations of the electromagnetic coupling constants that are responsible for the renormalization of the isoscalar nucleon mass. In Subsections~\ref{sec:subsec32} and~\ref{sec:subsec33}, we derive the representation for the semileptonic electroweak LECs and the combination $g_{11} + g_{12}$ of electromagnetic LECs, respectively. We combine these results in Subsection~\ref{sec:subsec34}, where we derive a representation for the two-point contribution $\left[\hat{C}_A \right]_{\mathrm{2pt}}$. In Subsection~\ref{sec:subsec35}, we determine the remaining combination of HBChPT electromagnetic LECs in terms of the LEFT three-point correlation functions. We express these correlators in terms of derivatives w.r.t. the pseudoscalar external source in Subsection~\ref{sec:subsec351} and with the help of derivatives w.r.t. the axial-vector external source in Subsection~\ref{sec:subsec352}, respectively.

\subsection{The chiral Lagrangian}
\label{sec:subsec31}

The chiral representation of the LEFT Lagrangian in Eqs.~\eqref{eq:LEFT2} and~\eqref{eq:CT} can be built by using standard spurion techniques. As in Eqs.~\eqref{eq:LEFT2} and~\eqref{eq:CT}, the chiral Lagrangian, originally constructed in Refs.~\cite{Meissner:1997ii,Muller:1999ww,Knecht:1999ag,Gasser:2002am,Cirigliano:2022hob}, contains purely leptonic operators, hadronic operators induced by purely electromagnetic corrections, and semileptonic operators involving hadrons, charged leptons, and neutrinos. As in Ref.~\cite{Cirigliano:2023fnz}, we impose the following minimal constraints on the spurions $\mathbf{q}_{W,L,R}$: $\textrm{Tr}(\mathbf{q}_{W})=0$ (the weak current has no singlet component under isospin) and $\textrm{Tr}(\mathbf{q}_{L} - \mathbf{q}_{R})=0$ (there exists a singlet vector current coupled to the photon). We do not impose additional conditions such as $\mathbf{q}_L \mathbf{q}_W = \frac{Y+1}{2} \mathbf{q}_W$ or $\mathbf{q}_W \mathbf{q}_L = \frac{Y-1}{2} \mathbf{q}_W$, which require specifying the isoscalar photon coupling (or equivalently the hadron weak hypercharge $Y$ via $Q = T_3 + \frac{Y}{2}$). Moreover, to avoid hadronic contributions to purely leptonic LECs, following Ref.~\cite{Descotes-Genon:2005wrq}, we do not use the equations of motion to reduce the set of operators. This weaker set of constraints requires an extension of the operator basis from Refs.~\cite{Gasser:2002am,Cirigliano:2022hob}.

We define vector ${\bf{q}}_V$ and axial-vector ${\bf{q}}_A$ charge spurions and external sources $\mathrm{v}_\rho$ and $\mathrm{a}_\rho$, respectively, as
\begin{equation}
    {\bf{q}}_V = {\bf{q}}_L + {\bf{q}}_R, \qquad {\bf{q}}_A = {\bf{q}}_L - {\bf{q}}_R, \qquad \mathrm{v}_\rho = l_\rho + r_\rho, \qquad \mathrm{a}_\rho = l_\rho - r_\rho,
\end{equation}
and split the electromagnetic charge spurions into isovector and isoscalar components
\begin{equation}
    {\bf{q}}^\mathrm{baryon}_J = {\bf q}^0_J + {\bf q}^a_J \tau^a, \qquad {\bf{q}}_J^\mathrm{quark} =  \frac{ {\bf q}_J^0}{3} + {\bf q}_J^a \tau^a,\label{eq:nucleon_and_quark_spurions}
\end{equation}
with $J \in \{ L, R, V, A\}$. The physical values are ${\bf q}_{L,R}^0 = {\bf q}_{L, R}^3 = \frac{1}{2}$ for the left and right spurions, ${\bf q}_V^0 = {\bf q}_V^3 = 1$ for the vector spurion, and ${\bf q}_A^0 = {\bf q}_A^3 = 0$ for the axial-vector case.\footnote{In what follows, we will omit the superscripts in the charge spurions: whenever ${\bf q}_{L,R,V,A}$ appears in the HBChPT Lagrangian, it is understood to be ${\bf q}_{L,R,V,A}^\mathrm{baryon}$.}

The building blocks for the construction of the chiral Lagrangian are the chiral covariant functions of the charges and of the corresponding covariant derivatives in Eqs.~\eqref{eq:charge_covariant_derivative_left} and~\eqref{eq:charge_covariant_derivative_right}
\begin{equation}
    {{\cal{Q}}}^W_L = u {\bf q}_W u^\dagger, ~~~ {{\cal{Q}}}_L = u {\bf{q}}_L u^\dagger, ~~  {{\cal{Q}}}_R = u^\dagger {\bf{q}}_R u, ~~ {{\cal{Q}}}_\pm = \frac{ {{\cal{Q}}}_L \pm {{\cal{Q}}}_R}{2}, ~~ c_\rho^\pm =  \frac{1}{2} \left( u \left(D_\rho  {\bf{q}}_L\right)  u^\dagger \pm u^\dagger \left(D_\rho {\bf{q}}_R\right)  u  \right),
\end{equation}
with $u^2 = U = \exp (i \boldsymbol{\pi} \cdot \boldsymbol\tau/F_\pi)$ and $F_\pi \approx 92$ MeV.

At the lowest order, the ChPT Lagrangian including mesons and nucleons in the heavy baryon formalism is given by
\begin{equation}
    \mathcal L^{p^2}_\pi + \mathcal L^{e^2}_\pi + \mathcal L^{p}_{\pi N} = \frac{F_\pi^2}{4} \langle u_\mu u^\mu + \chi_+  \rangle  + e^2 Z_\pi F_\pi^4 \langle {\cal Q}_L {\cal Q}_R  \rangle +  \overline{N}_v i v \cdot \nabla N_v + g^{(0)}_A \overline{N}_v S \cdot u N_v,\label{eq:LO}
\end{equation}
where $F_\pi$ denotes the pion decay constant, and $g_A^{(0)}$ is the nucleon axial-vector coupling constant in the chiral limit. $\langle \ldots \rangle$ denotes the trace operation in isospin space. The standard ChPT building blocks are given by
\begin{equation}
    u_\mu =  i \left[ u^\dagger (\partial_\mu - i r_\mu) u  - u (\partial_\mu - i l_\mu) u^\dagger\right], \qquad \chi_\pm = u^\dagger \chi u^\dagger \pm u \chi^\dagger u, \qquad \chi  =  2 B_0  (\overline{m}  + \overline{s} + i \overline{p}),
\end{equation}
with the isospin-averaged light-quark mass $\overline{m} = \frac{m_u + m_d}{2}$ and a LEC that has the dimension of mass $B_0$,\footnote{The LEC $B_0$ in Refs.~\cite{Cirigliano:2022hob,Cirigliano:2023fnz} is twice as large as the more conventional definition adopted here.} which is related to the quark condensate. We further introduce the nucleon chiral covariant derivative
\begin{eqnarray}
    \nabla_\mu N_v &\equiv & \left(\partial_\mu + \Gamma_\mu  \right) N_v,  \qquad \Gamma_\mu = \frac{1}{2} \left[ u (\partial_\mu - i l^{\rm baryon}_\mu) u^\dagger + u^\dagger (\partial_\mu - i r^{\rm baryon}_\mu) u \right],
\end{eqnarray}
where the superscript $\textit{baryon}$ indicates that the photon couples to the nucleon via the charge $\textbf{q}_V^{\rm baryon}$ from Eqs.~\eqref{eq:nucleon_and_quark_spurions}. In addition to the weak and electromagnetic interactions that arise from chiral covariant derivatives, Eq.~\eqref{eq:LO} contains electromagnetic effects mediated by high-momentum virtual photons via the coupling $Z_\pi$, which is related to the pion-mass splitting.

To obtain the axial-vector coupling constant in Eq.~(\ref{eq:axial_coupling_expansion}) at $\mathcal O(\alpha \epsilon_\chi)$ in HBChPT, we have to consider also the $\mathcal O(p^2)$ and $\mathcal O(e^2)$ pion-nucleon interaction Lagrangians~\cite{Gasser:1987rb,Krause:1990xc,Ecker:1995rk,Bernard:1995dp,Muller:1999ww}
\begin{align}
&  \mathcal L^{p^2}_{\pi N} + \mathcal L^{e^2}_{\pi N} =\overline{N}_v \Bigg[ \frac{(v \cdot \mathcal \nabla)^2 - \mathcal \nabla^2}{2 m_N}  - \frac{i g^{(0)}_A}{2 m_N} \left\{ S \cdot {\nabla}, v \cdot u\right\} + c_1 \langle \chi_+\rangle +  \left(c_2 - \frac{\left( g^{\left(0 \right)}_A\right)^2 }{8 m_N}\right) (v\cdot u)^2 + c_3 u \cdot u \nonumber \\
&  + \left(c_4 + \frac{1}{4 m_N} \right) \left[S^\mu, S^\nu\right] u_\mu u_\nu  + c_5 \tilde\chi_+ - \frac{i \left[S^\mu,S^\nu\right]}{4 m_N}   \left( (1+\kappa_1) f^+_{\mu\nu} + \frac{1}{2} (\kappa_0 - \kappa_1) \langle  f^+_{\mu\nu}\rangle \right) \Bigg] N_v \nonumber \\
& + e^2 F^2_\pi \, \overline{N}_v \left( f_1   \langle \tilde{{\cal  Q}}_+^2 - {\cal  Q}_-^2 \rangle + f_2 \tilde{{\cal  Q}}_+  \langle {\cal  Q}_+ \rangle +  f_3  \langle \tilde{{\cal  Q}}_+^2 + {\cal  Q}_-^2 \rangle + f_4 \langle {\cal  Q}_+ \rangle^2 + f_5 {\cal  Q}_-  \langle {\cal  Q}_+ \rangle  \right) N_v, \label{Eq:LagPiN_nlo}
\end{align}
with $\tilde{{\cal  O}} = {\cal  O} - \frac{1}{2} \langle {\cal  O} \rangle$ and $f^+_{\mu\nu} =  u^\dagger \left(  \partial_\mu r_\nu -  \partial_\nu r_\mu - i \left[ r_\mu, r_\nu \right] \right) u + u \left(  \partial_\mu l_\nu -  \partial_\nu l_\mu - i \left[ l_\mu, l_\nu \right] \right) u^\dagger$. The LECs $c_i$ have dimension GeV$^{-1}$ and scale as $\Lambda_\chi^{-1}$, while the anomalous magnetic moments $\kappa_{0,1}$ are dimensionless. $c_1$, $c_2$, $c_3$, and $c_4$ have been extracted from  pion-nucleon scattering data~\cite{Siemens:2016jwj}. $c_5$ is determined by the quark-mass contribution to the nucleon mass splitting, or, equivalently, by the isovector scalar charge $g^{u-d}_S$. $\kappa_{0,1}$ represent the leading contributions to the nucleon isoscalar and isovector anomalous magnetic moments. $f_2$ determines the electromagnetic contribution to the nucleon mass splitting, while $f_1$, $f_3,$ and $f_4$ shift the isospin-averaged nucleon mass, and are not well determined \cite{Gasser:2002am}.

The chiral counterterm Lagrangian at order $\mathcal O(G_F \alpha)$ is given by
\begin{equation}
    \mathcal L = \mathcal L^{CT}_{\rm lept} + \mathcal L^{e^2 p}_{\pi N} + \mathcal L^{e^2 p}_{\pi N \ell}.
\end{equation}
$\mathcal L^{CT}_{\rm lept}$ is a purely leptonic counterterm Lagrangian
\begin{equation}
    \mathcal L^{CT}_{\rm lept} = e^2 X_6 \overline{e} \left( i \slashed{\partial} + e \slashed{A}\right) e.\label{eq:LX6}
\end{equation}
The coupling $X_6$ is determined by computing the electron propagator in LEFT and chiral perturbation theory~\cite{Cirigliano:2023fnz} as
\begin{equation}
    X^r_6 \left( \mu_\chi ,\mu, \xi \right)=  \frac{\xi}{\left( 4 \pi \right)^2} \left( 1 - \ln \frac{\mu_\chi^2}{\mu^2} \right), \label{eq:X6_coupling_constant}
\end{equation}
in arbitrary $R_\xi$ gauge with the gauge parameter $\xi$, where $\mu$ and $\mu_\chi$ are the LEFT and HBChPT renormalization scales, respectively. $X_6^r(\mu_\chi,\mu)$ denotes the renormalized coupling, after subtraction of the $1/\varepsilon$ pole in the $\overline{\rm MS}_\chi$ scheme. Note that, following standard practice~\cite{Gasser:1983yg}, in the $\overline{\rm MS}_\chi$ scheme, we subtract
\begin{equation}
    \frac{1}{\varepsilon} - \gamma_E + \ln \left( 4 \pi \right) + 1,
\end{equation}
instead of the conventional $\overline{\rm MS}$ subtraction in LEFT:
\begin{equation} \label{eq:MSbar}
    \frac{1}{\varepsilon} - \gamma_E + \ln \left( 4 \pi \right).
\end{equation}
The electromagnetic Lagrangian with arbitrary spurion values and the constraint $\textrm{Tr}(\mathbf{q}_{L} - \mathbf{q}_{R})=0$ can be expressed in terms of $14$ operators:
\begin{equation}\label{eq:Lage2p}
{\cal L}_{\pi N}^{e^2 p}  =  e^2    \sum_{i=1}^{14} \  g_i \ \overline{N}_v \, O_i^{e^2 p} \, N_v.
\end{equation}
Eight of these operators contribute to the Gamow-Teller matrix elements at tree level:
\begin{align}
O_1^{e^2 p} &= \langle {\cal  Q}_+^2 - {\cal  Q}_-^2 \rangle \, S \cdot u , \\
O_2^{e^2 p} &= \langle {\cal  Q}_+ \rangle^2  \, S \cdot u  , \\
O_3^{e^2 p} &= \langle {\cal  Q}_+ \rangle  \, \langle {\cal Q}_+ S \cdot  u \rangle , \\
O_4^{e^2 p} &= {\cal  Q}_+  \, \langle {\cal Q}_+ S \cdot  u \rangle , \\
O_5^{e^2 p} &= {\cal  Q}_-  \, \langle {\cal Q}_- S \cdot  u \rangle , \\
O_{11}^{e^2 p} &= i  [ {\cal  Q}_+,  S \cdot c^- ] , \\
O_{12}^{e^2 p} &= i  [ {\cal  Q}_-,  S \cdot c^+ ] , \\
O_{13}^{e^2 p} &= \langle {\cal  Q}_+^2 + {\cal  Q}_-^2 \rangle \, S \cdot u.
\end{align}
With respect to the basis from Refs.~\cite{Gasser:2002am,Hoferichter:2009gn,Hoferichter:2016duk,Cirigliano:2022hob}, we add two more operators $O^{e^2 p}_{13}$ and $O^{e^2 p}_{14}$. Only the operator $O^{e^2 p}_{13}$ contributes to the axial-vector coupling constant, while the operator $O^{e^2 p}_{14}$ takes the form $O^{e^2 p}_{14}  = \langle {\cal{Q}}_+ {\cal{Q}}_- \rangle v \cdot u$. The LECs in Ref.~\cite{Gasser:2002am} are related to LECs in Eq.~\eqref{eq:Lage2p} by
\begin{equation}
    g_2 \left( [60] \right) = g_2 + g_{13}. \label{eq:relation_to_Meissner_basis}
\end{equation}

The most general weak-interaction Lagrangian in the heavy-baryon sector, only making use of the assumption $\langle{\bf{q}}_W\rangle = 0$, was presented in Refs.~\cite{Tomalak:2023xgm,Cirigliano:2023fnz}:
\begin{equation}
    \mathcal L^{e^2 p}_{\pi N \ell} = e^2 \sum \limits_{i=1}^{6} \overline{e}_L \gamma_\rho  \nu_{e L} \overline{N}_v \, \left(  {V}_i v^\rho - 2 {A}_i g^{(0)}_A S^\rho \right) {{\mathrm{O}_i }}  N_v  + \mathrm{h.c.}, \label{eq:Lagrangian_electroweak}
\end{equation}
where $g_A^{(0)}$ denotes the nucleon axial-vector coupling constant in the isospin limit, and the operator basis is chosen as
\begin{align}
    { \mathrm{O}_1 } &= [{{\cal{Q}}}_{L}, {\cal{Q}}^W_L], \qquad ~~ { \mathrm{O}_2 } = [{{\cal{Q}}}_{R}, {\cal{Q}}^W_L],  \\
    { \mathrm{O}_3 } &= \{ {{\cal{Q}}}_{L}, {\cal{Q}}^W_L \}, \qquad ~{ \mathrm{O}_4 } = \{ {{\cal{Q}}}_{R}, {\cal{Q}}^W_L \},  \\
    { \mathrm{O}_5 } &= \langle {{\cal{Q}}}_{L} {\cal{Q}}^W_L \rangle, \qquad \quad { \mathrm{O}_6 } = \langle {{\cal{Q}}}_{R} {\cal{Q}}^W_L \rangle.\label{eq:operators_electroweak}
\end{align}

Following the strategy from Refs.~\cite{Moussallam:1997xx,Descotes-Genon:2005wrq} in the mesonic sector, and from Ref.~\cite{Cirigliano:2023fnz} for the vector coupling that contributes to Fermi transitions, we can interpret amplitudes in LEFT and in HBChPT as functionals of the same charges ${\bf q}^{0,a} (x)$, promoted to be spacetime-dependent external fields. The matching between the LEFT and HBChPT is then obtained by equating functional derivatives of the effective action with respect to ${\bf q}^{0,a} (x)$ in both theories.

We reorganize the LEC contributions from Eq.~(\ref{eq:leading_order_correction_gA}) in terms of two combinations of coupling constants
\begin{align}\label{eq:split}
\hat C_A &= \left[\hat C_A\right]_{\rm 2pt} +\left[\hat C_A\right]_{\rm 3pt} \nonumber \\
&=   8 \pi^2 \left[ -\frac{X_6}{2} + 2 (A_1 + A_2 + A_3 + A_4) + \frac{g_{11}+ g_{12}}{4 g_A^{(0)}}  \right] + \frac{8 \pi^2}{g_A^{(0)}} \left( g_1 + g_2 + g_{13} + \frac{g_{11} - g_{12}}{4} \right),
\end{align}
and illustrate in the following Subsections~\ref{sec:subsec32}-\ref{sec:subsec34} that $[\hat C_A]_{\rm 2pt}$ can be represented in terms of one-nucleon matrix elements of the time-ordered product of only {\it two} quark bilinears. The derivation of the representation for $[\hat C_A]_{\rm 2pt}$ closely parallels the derivation of the corrections to the vector coupling $g_V$~\cite{Cirigliano:2023fnz}, and we identify the two-point contribution in terms of the same nonperturbative input as in the current-algebra evaluation of the axial-vector $\gamma W$ box diagram~\cite{Gorchtein:2021fce,Hayen:2021iga} as part of $[\hat C_A]_{\rm 2pt}$. In Subsections~\ref{sec:subsec351} and~\ref{sec:subsec352}, we express the combination of coupling constants $[\hat C_A]_{\rm 3pt}$ in terms of a single-nucleon matrix element of the time-ordered product of {\it three} quark bilinears.

\subsection{Electromagnetic coupling constants $f_1+f_3$ and $f_4$}
\label{sec:subsec315}

As we will illustrate in Subsection~\ref{sec:subsec351}, the matching conditions to determine $[\hat C_A]_{\rm 3pt}$ involve also the $\mathcal O(e^2)$ LECs $f_{1}$, $f_3$, and $f_4$, for which we provide a representation. To determine the combination of coupling constants $f_1 + f_3$ and $f_4$\footnote{In Appendix~\ref{app:NLOChPT}, we determine all LECs of the $\mathcal O(p^2)$ and $\mathcal O(e^2)$ pion-nucleon interaction Lagrangian.} that enter evaluation of HBChPT diagrams in Subsection~\ref{sec:subsec351}, we introduce the correlation functions $G_{VV}$ and $G_{VV,0}$ as integrals of the derivatives from the generating functional $W$:
\begin{align} \label{eq:GVV}
    G_{VV} &= \frac{\delta^{a b} \delta^{i j}\delta^{\sigma \sigma^\prime}}{12} \int \mathrm{d}^d x \langle N(k, \sigma^\prime, j ) |  \frac{\delta^2 W  \left( {\bf{q}}_V, {\bf{q}}_A, {\bf{q}}_W \right)}{ \delta {\bf{q}}^b_{V} \left( x \right) \delta {\bf{q}}^a_{V} \left( 0 \right)}   \Bigg|_{{\bf{q}}=0} | N(k, \sigma, i) \rangle , \\
    G_{VV,0} &= \frac{\delta^{i j}\delta^{\sigma \sigma^\prime}}{4} \int \mathrm{d}^d x \langle N(k, \sigma^\prime, j ) |  \frac{\delta^2 W  \left( {\bf{q}}_V, {\bf{q}}_A, {\bf{q}}_W \right)}{ \delta {\bf{q}}^0_{V} \left( x \right) \delta {\bf{q}}^0_{V} \left( 0 \right)}   \Bigg|_{{\bf{q}}=0} | N(k, \sigma, i) \rangle . \label{eq:GVV0}
\end{align}
At tree level and one loop in HBChPT at leading and next-to-leading orders, the correlation functions $G_{VV} |^{\rm HBChPT}$ and $G_{VV,0} |^{\rm HBChPT}$ take the form
\begin{align}
    \left. G_{VV}\right |^{\rm HBChPT} &= e^2 F^2_\pi \left[ f_1 + f_3  +  6 \left( g_A^{(0)}\right)^2 \frac{Z_\pi \pi m_\pi}{\left(  4\pi F_\pi\right)^2}   \right], \\
    \left. G_{VV,0}\right |^{\rm HBChPT} &= 2 e^2 F^2_\pi f_4.
\end{align}

Exploiting the conservation of a vector current in LEFT, $G_{VV} |^{\rm LEFT}$ and $G_{VV,0} |^{\rm LEFT}$ can be expressed in terms of the ultraviolet-finite integrals from the two-current nucleon matrix elements $\tau^{\mu \nu}_{VV}$ and $\tau^{\mu \nu}_{VV,0}$, respectively, as
\begin{align}
   \left. G_{VV} \right|^{\rm LEFT} &= e^2  \int \frac{ \mathrm{d}^4 q}{(2\pi)^4} \frac{ g_{\mu \nu} \tau^{\mu \nu}_{V V} \left(q, v \right)}{q^2 - \lambda_\gamma^2} , \\
   \left. G_{VV,0} \right|^{\rm LEFT} &= e^2  \int \frac{ \mathrm{d}^4 q}{(2\pi)^4} \frac{ g_{\mu \nu} \tau^{\mu \nu}_{V V, 0} \left(q, v \right)}{q^2 - \lambda_\gamma^2} ,
\end{align}
with the small photon mass $\lambda_\gamma$ as an infrared regulator and the nonperturbative hadronic objects $\tau^{\mu \nu}_{VV}$ and $\tau^{\mu \nu}_{VV,0}$:
\begin{align}
    \tau^{\mu \nu}_{VV} \left(q, v \right) &= \frac{\delta^{a b} \delta^{i j} \delta^{\sigma \sigma^\prime}}{48} \int \mathrm{d}^d x e^{ i q \cdot x} \langle N (k, \sigma^\prime, j) | T \left[ \overline{q} \gamma^\mu \tau^b q (x) \, \overline{q} \gamma^\nu \tau^a q(0) \right] | N ( k, \sigma, i) \rangle \label{eq:tauVV}, \\
    \tau^{\mu \nu}_{VV,0} \left(q, v \right) &= \frac{\delta^{i j} \delta^{\sigma \sigma^\prime}}{16}\frac{1}{9} \int \mathrm{d}^d x e^{ i q \cdot x} \langle N (k, \sigma^\prime, j) | T \left[ \overline{q} \gamma^\mu  q (x) \, \overline{q} \gamma^\nu  q(0) \right] | N ( k, \sigma, i) \rangle \label{eq:tauVV0}.
\end{align}

By equating LEFT and HBChPT expressions for the nonperturbative objects in Eqs.~(\ref{eq:GVV}) and~(\ref{eq:GVV0}), i.e., $G_{VV} |^{\rm HBChPT} = G_{VV} |^{\rm LEFT}$ and $G_{VV,0} |^{\rm HBChPT} = G_{VV,0} |^{\rm LEFT}$, we determine the gauge-, scale-, and scheme-independent combinations of the coupling constants of interest,
\begin{align} \label{eq:f1f3_match_result}
     f_1 + f_3 &= \frac{1}{F^2_\pi} \int \frac{ \mathrm{d}^4 q}{(2\pi)^4} \frac{ g_{\mu \nu} \tau^{\mu \nu}_{V V} \left(q, v \right)}{q^2} - 6 \left( g_A^{(0)}\right)^2 \frac{Z_\pi \pi m_\pi}{\left(  4\pi F_\pi\right)^2}  , \\
     2 f_4 &= \frac{1}{F^2_\pi} \int \frac{ \mathrm{d}^4 q}{(2\pi)^4} \frac{ g_{\mu \nu} \tau^{\mu \nu}_{V V,0} \left(q, v \right)}{q^2}. \label{eq:f4_match_result}
\end{align}

\subsection{Semileptonic electroweak coupling constants $A_{1,2,3,4}$}
\label{sec:subsec32}

\begin{figure}[h]
    \centering
    \includegraphics[width=0.9\textwidth]{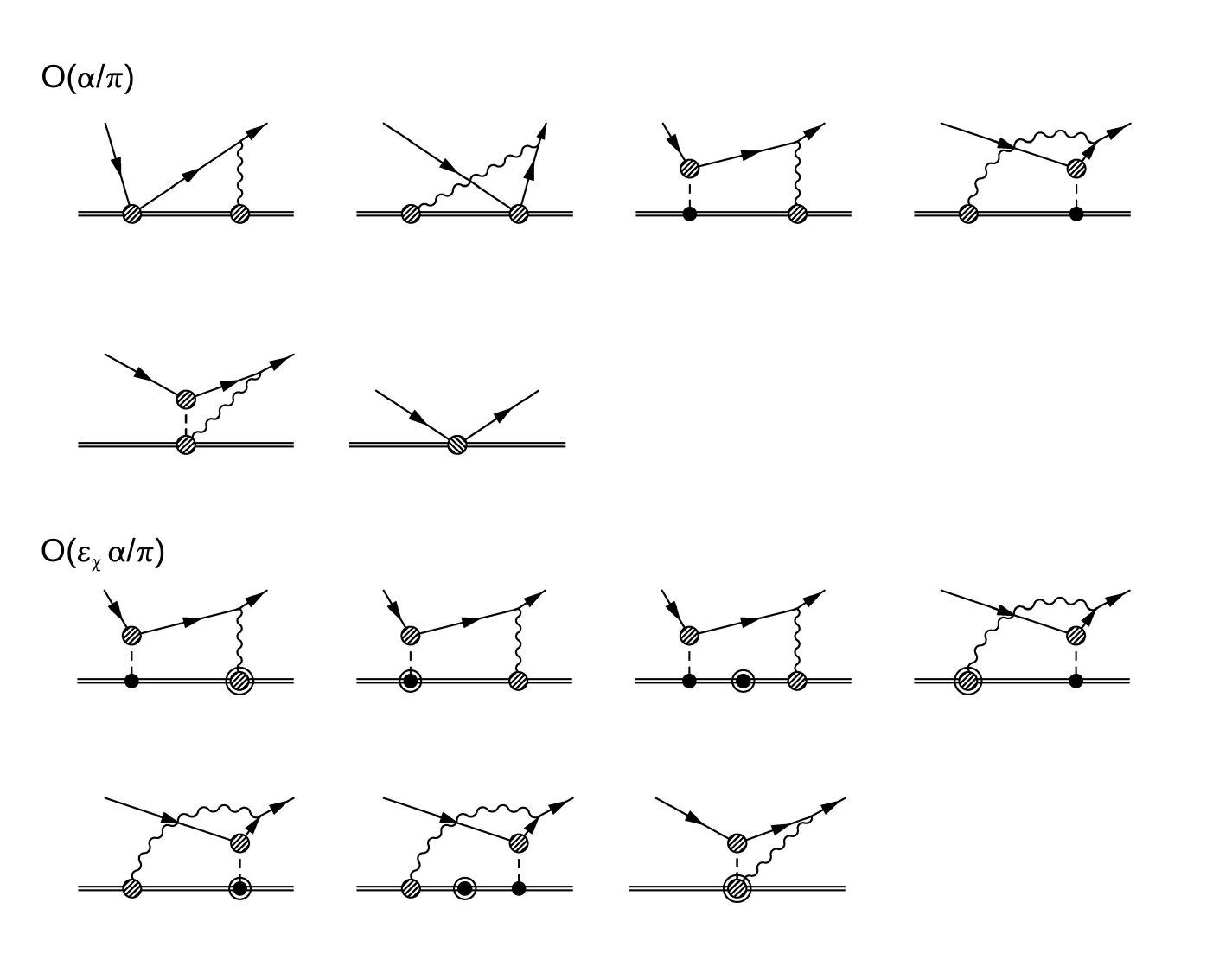}
    \caption{HBChPT diagrams that contribute to the correlation functions $\Gamma^{(0,1)}_{VW}$ are shown. Single line denotes electrons and neutrinos. Double, wiggly, and dashed lines denote nucleons, photons, and pions, respectively. Dashed circles denote insertions of the sources ${\bf q}^{}_V$ or ${\bf q}_W$, while dots denote interactions from the LO chiral Lagrangian, $\mathcal L^{p}_{\pi N}$. Circled dashed circles and circled dots denote the insertions of the sources and pion-nucleon interactions from the NLO chiral Lagrangian $\mathcal L^{p^2}_{\pi N}$. In the NLO diagrams, we omitted corrections to diagrams without pions, which vanish as $\lambda_\gamma \rightarrow 0$.} \label{fig:axialA}
\end{figure}

To determine the relevant combinations of the semileptonic electroweak coupling constants $A_{1,2,3,4}$ from the generating functional $W$, we define the correlators
\begin{align}
    \overline{e}  \slashed{b} P_L \nu_e  \, \Gamma^{(1)}_{VW} &=  b_\rho \frac{ \varepsilon^{a b c} \tau^c_{i j} S^\rho_{\sigma \sigma^\prime}}{6} \frac{i}{2} \int \mathrm{d}^d x \langle e^- \overline{\nu}_e N(k, \sigma^\prime, j ) |  \frac{\delta^2 W  \left( {\bf{q}}_V, {\bf{q}}_A, {\bf{q}}_W \right)}{ \delta {\bf{q}}^b_{V} \left( x \right) \delta {\bf{q}}^a_{W} \left( 0 \right)}   \Bigg|_{{\bf{q}}=0} | N(k, \sigma, i) \rangle,\\
    \overline{e}  \slashed{b} P_L \nu_e  \, \Gamma^{(0)}_{VW} &= b_\rho  \frac{ \tau^a_{i j} S^\rho_{\sigma \sigma^\prime}}{6} \int \mathrm{d}^d x \langle e^- \overline{\nu}_e N(k, \sigma^\prime, j ) | \frac{\delta^2 W  \left( {\bf{q}}_V, {\bf{q}}_A, {\bf{q}}_W \right)}{ \delta {\bf{q}}^0_{V} \left( x \right) \delta {\bf{q}}^a_{W} \left( 0 \right)}   \Bigg|_{{\bf{q}}=0} | N(k, \sigma, i) \rangle,
\end{align}
where $b$ denotes an arbitrary vector, which satisfies $ v \cdot b = 0$ and $b_\rho b^\rho = -1$. We show the HBChPT diagrams that contribute to $\Gamma^{(0,1)}_{VW}$ in Fig.~\ref{fig:axialA}. $\Gamma^{(1)}_{VW}$ receives contributions from both loop diagrams and LECs. The loop corrections to $\Gamma^{(0)}_{VW}$, on the other hand, vanish, and this correlation function only receives contributions from the LECs. At leading and next-to-leading orders in HBChPT, we find
\begin{align}
    \left.   \Gamma^{(1)}_{VW}\right|^{\rm HBChPT} &=  e^2 g^{(0)}_A \left( A_1 + A_2 - \frac{1}{2} \int \frac{i \mathrm{d}^d q}{(2\pi)^d} \frac{1}{q^2 \left(q^2 - \lambda_\gamma^2 \right)} + \frac{1-\xi}{2} \int \frac{i \mathrm{d}^d q}{(2\pi)^d} \frac{1}{\left(q^2 - \lambda_\gamma^2\right) \left(q^2 - \xi \lambda_\gamma^2\right)} \right) \nonumber \\ 
    &- \frac{e^2 g^{(0)}_A }{2} \int \frac{i \mathrm{d}^d q}{(2\pi)^d} \frac{1}{\left( q^2  - m_\pi^2 \right) \left(q^2 - \lambda_\gamma^2 \right)} +  \frac{e^2 g_A^{(0)} (1+\kappa_1) }{4 m_N} \frac{d-2}{d-1} \int \frac{i \mathrm{d}^d q}{(2\pi)^d} \frac{q^2 - \left( v\cdot q \right)^2} {q^2 \left(q^2 - m_\pi^2\right) v\cdot q}\nonumber \\ 
    &+  \frac{e^2 g_A^{(0)}}{4 m_N} \frac{1}{d-1} \int \frac{i \mathrm{d}^d q}{(2\pi)^d} \frac{q^2 - \left( v\cdot q \right)^2} {q^2 \left(q^2 - m_\pi^2\right) v\cdot q}, \label{eq:GammaVW11} \\
    \left. \Gamma^{(0)}_{VW} \right|^{\rm HBChPT} &= e^2 g^{(0)}_A (A^r_3 + A^r_4).\label{eq:GammaVW1}
\end{align}

The matching condition is given by the observation that the expressions for  $\Gamma^{(0,1)}_{VW}$ in HBChPT and LEFT must coincide in the regime in which both EFTs are valid. To express the correlators $\Gamma^{(0,1)}_{VW}$ in LEFT in terms of nonperturbative matrix elements, we introduce the objects
\begin{subequations}
\begin{align}
t^{\mu \nu \rho}_{V V (A)} \left(q, v \right) &= \frac{\varepsilon^{abc} \tau^c_{ij} S^\rho_{\sigma \sigma^\prime}}{6}  \frac{i}{4} \int \mathrm{d}^d x e^{ i q \cdot x} \langle N (k, \sigma^\prime, j) | T \left[ \overline{q} \gamma^\mu \tau^b q (x) \, \overline{q} \gamma^\nu \left( \gamma_5 \right) \tau^a q(0) \right] | N ( k, \sigma, i) \rangle, \label{eq:tAVV}\\
t^{\mu \nu \rho}_{V V (A),0} \left(q, v \right) &= \frac{\tau^a_{i j} S^\rho_{\sigma \sigma^\prime}}{6} \frac{i}{6} \int \mathrm{d}^d x e^{ i q \cdot x} \langle N (k, \sigma^\prime, j) | T \left[ \overline{q} \gamma^\mu q (x) \, \overline{q} \gamma^\nu \left( \gamma_5 \right)  \tau^a q(0) \right] | N ( k, \sigma, i) \rangle,  \label{eq:tAVV0}\\
t^{\mu \rho}_{V P} \left(q, v \right) &= \frac{\varepsilon^{abc} \tau^c_{ij} S^\rho_{\sigma \sigma^\prime}}{6} \frac{i}{4} \int \mathrm{d}^d x e^{ i q \cdot x} \langle N (k, \sigma^\prime, j) | T \left[ \overline{q} \gamma^\mu \tau^b q (x) \, \overline{q} \gamma_5 \tau^a q(0) \right] | N ( k, \sigma, i) \rangle, \label{eq:tPV}\\
t^{\mu \rho}_{VP,0} \left(q, v \right) &= \frac{\tau^a_{i j} S^\rho_{\sigma \sigma^\prime}}{6} \frac{i}{6} \int \mathrm{d}^d x e^{ i q \cdot x} \langle N (k, \sigma^\prime, j) | T \left[ \overline{q} \gamma^\mu q (x) \, \overline{q} \gamma_5  \tau^a q(0) \right] | N ( k, \sigma, i) \rangle, \label{eq:tPV0}
\end{align}
\end{subequations}
which single out the desired spin and isospin components of the two-current matrix elements in the nucleon. These objects satisfy the Ward identities
\begin{subequations}
\begin{align}
q_\mu t^{\nu \mu \rho}_{V A} \left(q, v \right) &= - i g^{(0)}_A \left( g^{\nu \rho} - v^\nu v^\rho \right) + 2 \overline{m}  \, t^{\nu \rho}_{V P} \left(q, v\right), &\qquad q_\mu t^{\mu \nu \rho}_{V A} \left(q, v \right) &= - i g^{(0)}_A \left(g^{\nu \rho} - v^\nu v^\rho  \right), \label{eq:Ward_like_t_VA} \\
q_\mu t^{\nu \mu \rho}_{V A, 0} \left(q, v \right) &=  2 \overline{m}  \, t^{\nu \rho}_{VP, 0} \left(q, v \right), &\qquad q_\mu t^{\mu \nu \rho}_{V A, 0} \left(q, v \right) &= 0, \end{align}
\end{subequations}
with the isospin-averaged quark mass $\overline{m} = \left( m_u + m_d \right)/2$ and assumption of the isospin symmetry, i.e., $m_u = m_d$. For the vector-vector correlators, the Ward identities and the spin projections imply $q_\mu t^{\mu \nu \rho}_{VV} = q_\nu t^{\mu \nu \rho}_{VV} = q_\mu t^{\mu \nu \rho}_{VV, 0} = q_\nu t^{\mu \nu \rho}_{VV, 0} = 0$. We provide details on the hadronic objects $t^{\mu \nu \rho}_{V A}$ and $t_{V P}\sim g_{\mu\rho}t_{VP}^{\mu\rho}$ in Appendix~\ref{app:relations}.

Using the definitions in Eqs.~\eqref{eq:tAVV} and~\eqref{eq:tAVV0}, we relate the correlation functions $\Gamma^{(0,1)}_{VW}$ in LEFT to the hadronic matrix elements as
\begin{align}
\left.  \overline{e}  \slashed{b} P_L \nu_e  \Gamma^{(1)}_{VW}\right|^{\rm LEFT}  &= e^2 g^{(0)}_A \overline{e}  \slashed{b} P_L \nu_e \left ( \frac{g_{02} - g_{03}}{4} + \frac{ 1-\xi}{2} \int \frac{i \mathrm{d}^d q}{(2\pi)^d} \frac{1}{\left(q^2 - \lambda_\gamma^2\right) \left(q^2 - \xi \lambda_\gamma^2\right)} \right) \nonumber \\
&- \frac{e^2}{2} \int \frac{\mathrm{d}^d q}{(2\pi)^d} \frac{b_\rho}{q^2 \left( q^2 - \lambda_\gamma^2 \right) } \overline{e}  \gamma_\mu \slashed{q} \gamma_\nu P_L \nu_e  \left(t^{\mu \nu \rho}_{VV} \left(q, v \right) - t^{\mu \nu \rho}_{VA} \left(q, v \right) \right),  \label{eq:a1a2_match} \\
\overline{e}  \slashed{b} P_L \nu_e  \left.  \Gamma^{(0)}_{VW} \right|^{\rm LEFT}  & = - e^2 g^{(0)}_A \overline{e}  \slashed{b} P_L \nu_e \frac{g_{02} + g_{03}}{12} \nonumber \\ 
&+ \frac{e^2}{2} \int \frac{i \mathrm{d}^d q}{(2\pi)^d} \frac{b_\rho}{q^2 \left( q^2 - \lambda_\gamma^2 \right) } \overline{e}  \gamma_\mu \slashed{q} \gamma_\nu P_L \nu_e  \left(t^{\mu \nu \rho}_{VV, 0} \left(q, v \right) - t^{\mu \nu \rho}_{VA, 0} \left(q, v \right) \right). \label{eq:a3a4_match}
\end{align}
The above matrix elements (and similar ones to be discussed later) may contain ultraviolet (UV) divergences, controlled by the asymptotic short-distance (high-momentum) behavior of the time-ordered product of quark bilinears. To isolate the UV divergences, following Ref.~\cite{Cirigliano:2023fnz} we add and subtract the high-momentum limit of the hadronic tensors and define subtracted objects as
\begin{equation}
    \overline{t} = t - t_{\rm OPE}, 
\end{equation}
where $t_{\rm OPE}$ generically indicates the expression for the hadronic tensors obtained by using the leading terms in the operator product expansion (OPE) of the quark bilinears at short distance. The integrals involving $\overline{t}$ are UV convergent and can be evaluated in $d=4$ so that we can unambiguously perform the Dirac algebra on the leptonic leg, while all the scheme dependence resides in the OPE contribution. The hadronic tensors defined in Eqs.~(\ref{eq:tAVV}) and~(\ref{eq:tAVV0}) have the same OPE as the analogous objects in Ref.~\cite{Cirigliano:2023fnz}:
\begin{align}
    \overline{e} \gamma_\mu \slashed{q} \gamma_\nu P_L \nu_e  \left(\left.  t^{\mu \nu \rho}_{VV} - t^{\mu \nu \rho}_{VA}  \right)\right|_{\rm OPE} & = i g_A^{(0)} \frac{  3 d - 2 }{d}  \frac{ q^2}{q^2 -\mu_0^2} \, \overline{e} \left(\gamma^\rho -  v^\rho  \slashed{v}  \right) P_L \nu_e + ..., \label{eq:ope1} \\
    \overline{e} \gamma_\mu \slashed{q} \gamma_\nu P_L \nu_e \left( \left.  t^{\mu \nu \rho}_{VV,\, 0} - t^{\mu \nu \rho}_{VA, 0} \right) \right|_{\rm OPE} &=  \frac{g_A^{(0)}}{d} \left( \left( 4-d \right) \left( 1 + \frac{4a}{3} \right) - 2 \right) \frac{q^2}{q^2 - \mu_0^2} \, \, \overline{e} \left(\gamma^\rho -  v^\rho  \slashed{v}  \right) P_L \nu_e + ... , \label{eq:ope2}
\end{align}
where we have introduced an arbitrary scale $\mu_0$ to regulate infrared divergences~\cite{Cirigliano:2023fnz} and used the relation $q_{\mu} q_\nu  f(q^2) \to g_{\mu \nu} f(q^2) q^2/d$, which holds only under $d$-dimensional integration over $q$. From the OPE and the Ward identities, it also follows that, for the subtracted tensor,
\begin{align}
    q_\mu \left( \overline{t}^{\mu  \nu \rho}_{VV} - \overline{t}^{\mu  \nu \rho}_{VA}\right) &= - i g_A^{(0)} \left(g^{\nu \rho} - v^\nu v^{\rho}\right) \frac{\mu_0^2}{q^2 - \mu_0^2}, \\ q_\nu \left( \overline{t}^{\mu  \nu \rho}_{VV} - \overline{t}^{\mu  \nu \rho}_{VA}\right) &= - i g_A^{(0)} \left(g^{\mu \rho} - v^\mu v^{\rho}\right) \frac{\mu_0^2}{q^2 - \mu_0^2}  - 2 \overline{m} t^{\mu \rho}_{VP}.
\end{align}
Finally, accounting for time reversal and crossing symmetry~\cite{Sirlin:1967zza,Seng:2018qru}, the hadronic functions in the matching Eqs.~\eqref{eq:a1a2_match} and~\eqref{eq:a3a4_match} are even or odd under $q \rightarrow -q$~\cite{Cirigliano:2023fnz}, such that
\begin{subequations}
\begin{align}
    \overline{t}^{\mu \nu \rho}_{VV}(q^2,v \cdot q) &= \overline{t}^{\mu  \nu \rho}_{VV}(q^2,-v \cdot q), & \qquad {\overline t}^{\mu \nu \rho}_{VA} (q^2, v \cdot q) &= -{\overline t}^{\mu \nu \rho}_{VA} (q^2, -v \cdot q),\label{eq:crossing1} \\
    \overline{t}^{\mu  \nu \rho}_{VV,\, 0}(q^2,v \cdot q) &=  - \overline{t}^{\mu  \nu \rho}_{VV,\,0}(q^2,-v \cdot q), & \qquad {\overline t}^{\mu \nu \rho}_{VA,\, 0} (q^2, v \cdot q) &=  {\overline t}^{\mu \nu \rho}_{VA,\,0 } (q^2, -v \cdot q),\label{eq:crossing2}
\end{align}
\end{subequations}
where we indicated that the corresponding invariant amplitudes depend only on the kinematic invariants
\begin{equation}
    Q^2 = - q^2, \qquad \nu = v\cdot q.
\end{equation}
Since ${\overline t}^{\mu\nu\rho}_{VV} -{\overline t}^{\mu\nu\rho}_{VA}$ and ${\overline t}^{\mu\nu\rho}_{VV,0} - {\overline t}^{\mu\nu\rho}_{VA,0}$ in Eqs.~\eqref{eq:a1a2_match} and~\eqref{eq:a3a4_match} are multiplied by an odd function of $q$, ${\overline t}_{VA,0}$ and ${\overline t}_{VV}$ do not contribute to the matching.

Exploiting the Ward identities, the operator product expansion, and the time-reversal, parity, and crossing properties, and equating the correlation functions in LEFT and HBChPT, i.e., $ \Gamma^{(0,1)}_{VW}|^{\rm HBChPT} =\Gamma^{(0,1)}_{VW}|^{\rm LEFT}$, we can express the LECs in HBChPT in terms of the subtracted hadronic objects in LEFT as
\begin{align}   
2 \left( A^r_1 + A^r_2 \right) \left( \mu_\chi, \mu, \xi \right) &= 2 \overline{m} \int \frac{i \mathrm{d}^4 q}{(2\pi)^4} \frac{{t}_{VP} \left(q, v \right)}{ q^2 - \lambda_\gamma^2 } - \int \frac{i \mathrm{d}^4 q}{(2\pi)^4} \frac{{\overline t}_{VA} \left(q, v \right)}{ q^2 - \lambda_\gamma^2 } - \frac{1}{3} \frac{3 + 2 \kappa_1}{(4\pi)^2}\frac{\pi m_\pi}{m_N} \nonumber \\
&+  \frac{1}{(4\pi)^2}\left[ 2 \ln \frac{\mu^2}{\lambda_\gamma^2} + \frac{1}{2} \ln \frac{\mu^2}{\mu_0^2 } - \ln \frac{\mu_\chi^2}{\lambda_\gamma^2}- \ln \frac{\mu_\chi^2}{m_\pi^2}  + \frac{9}{4} + (1-\xi) \left(\ln \frac{\mu_\chi^2}{\mu^2} - 1\right) \right],\label{eq:a1a2_match_result}  \\
2 \left( A^r_3 + A^r_4 \right) \left( a, \mu \right)  & = \int \frac{i \mathrm{d}^4 q}{(2\pi)^4} \frac{{\overline t}_{VV, 0} \left(q, v \right)}{ q^2 - \lambda_\gamma^2 } +  \frac{1}{(4\pi)^2}\left[ \frac{1}{2} \ln \frac{\mu^2}{\mu^2_0} + \frac{3 - 8 a}{12} \right]. \label{eq:a3a4_match_result}
\end{align}
In writing the expressions above, we have introduced scalar projections of the hadronic correlation functions according to the definitions\footnote{Using the relation $q_{\mu} q_\nu  f(q^2) \to g_{\mu \nu} f(q^2) q^2/4$, which holds under four-dimensional integration over $q$, we can also define the hadronic objects $\overline{t}_{VV, 0} \left(q, v \right), t_{VP} \left(q, v \right),$ and $\overline{t}_{VA} \left(q, v \right)$, as
\begin{align}
b_\rho \varepsilon_{\mu \nu \alpha \beta} q^\alpha \overline{e}  \gamma^\beta P_L \nu_e {\overline t}^{\mu\nu \rho}_{VV, 0} \left(q, v \right) &= i g^{(0)}_A \overline{e}  \slashed{b} P_L \nu_e q^2 {\overline t}_{VV, 0} \left(q, v \right), \label{eq:tVVscalarint} \\
b_\rho \overline{e} \gamma_\mu P_L \nu_e {t}^{\mu \rho}_{VP} \left(q, v \right) &= i g^{(0)}_A \overline{e}  \slashed{b} P_L \nu_e q^2 {t}_{VP} \left(q, v \right), \label{eq:tPVscalarint} \\
g_{\mu \nu} b_\rho \overline{e} \slashed{q} P_L \nu_e {\overline t}^{\mu \nu \rho}_{VA} \left(q, v \right) & = i g^{(0)}_A \overline{e}  \slashed{b} P_L \nu_e q^2 {\overline t}_{VA} \left(q, v \right). \label{eq:tVAscalarint}
\end{align}}
\begin{align}
    \varepsilon_{\mu \nu \alpha \rho} q^\alpha \overline{t}^{\mu \nu \rho}_{VV,0}  &= (d-1)\,i g_A^{(0)} q^2 \overline{t}_{VV, 0} \left(q, v \right), \label{eq:tVVscalar}  \\
    g_{\mu \rho} {t}^{\mu \rho}_{VP} &= (d-1)\, i g_A^{(0)} q^2 {t}_{VP} \left(q, v \right), \label{eq:tPVscalar}  \\
    g_{\mu \nu} q_\rho \overline{t}^{\mu \nu \rho}_{VA} &= (d-1)\, i g_A^{(0)} q^2 \overline{t}_{VA} \left(q, v \right). \label{eq:tVAscalar} 
\end{align}

A few comments are in order: (i) In the expression for $A_1^r+A_2^r$,  the integral involving $t_{VP}$ converges in the ultraviolet and does not need a subtraction, given that the leading power of the OPE at vanishing quark masses scales as $t_{VP} |_\mathrm{OPE} \sim {q_\mathrm{ext}}/{Q^3}$, with the external momentum $q_\mathrm{ext}$. (ii) The combination $A^r_1 + A^r_2$ is scheme independent but depends on the gauge parameter $\xi$. The scheme dependence of $A^r_3 + A^r_4$, which is encoded by $B(a)$, is such as to cancel the scheme dependence of the LEFT matching coefficient $C_{\beta}^r (a, \mu)$. (iii) Finally, we note that the LECs do not depend on $\mu_0$ and $m_\pi$. The explicit dependence on $\mu_0$ in the expressions above is canceled by the implicit dependence on $\mu_0$ in $\overline{t}_{VA}$  and $\overline{t}_{VV,0}$. Similarly, the explicit dependence on $m_\pi$ in the expression for $A_1^r+A_2^r$ is canceled by the infrared behavior of the integrals. After subtracting the infrared behavior ${t}_{VP}^\mathrm{IR}$, which we determine in Appendix~\ref{app:IR}, from the pseudoscalar-vector correlator, we define the subtracted hadronic object under the integral in Eq.~(\ref{eq:a1a2_match_result}) $\tilde{t}_{VP}$:\footnote{The isospin-averaged quark mass $\overline{m}$ is related to the squared pion mass $m^2_\pi$ as $2 B_0 \overline{m} = m^2_\pi$, with the chiral quark condensate parameter $B_0$.}
\begin{align}\label{eq:tPVtilde}
\tilde{t}_{VP} & = {t}_{VP} - {t}_{VP}^\mathrm{IR}, \\
{t}_{VP}^\mathrm{IR} & = - \frac{B_0}{q^2 \left( q^2 - m_\pi^2 \right)} \left( 1 + \frac{3+2\kappa_1}{6 m_N} q^2 i \pi \delta \left( v \cdot q \right) \right). \label{eq:tPVIR}
\end{align}
The corresponding integral over this infrared-subtracted hadronic object $\tilde{t}_{VP}$,
\begin{align}\label{eq:vanishing_tPV}
    2 \overline{m} \int \frac{i \mathrm{d}^4 q}{(2\pi)^4} \frac{{\tilde t}_{VP} \left(q, v \right)}{ q^2 - \lambda_\gamma^2 } = \mathcal{O} \left( \frac{m^2_\pi}{m_N^2}, \frac{m^2_\pi}{\left( 4 \pi F_\pi \right)^2} \right),
\end{align}
vanishes at the leading power in the HBChPT expansion. Accounting for Eq.~(\ref{eq:vanishing_tPV}) and neglecting $m^2_\pi$-suppressed contributions, we rewrite the matching relation for $A^r_1 + A^r_2$ in the infrared-finite form 
\begin{align}
2 \left( A^r_1 + A^r_2 \right) \left(\mu_\chi, \mu, \xi \right) &= \frac{1}{(4\pi)^2}\left[ 2 \ln \frac{\mu^2}{\mu_\chi^2} + \frac{1}{2} \ln \frac{\mu^2}{\mu_0^2 } + \frac{9}{4} + \left( 1-\xi \right) \left(\ln \frac{\mu_\chi^2}{\mu^2} - 1\right) \right] - \int \frac{i \mathrm{d}^4 q}{(2\pi)^4} \frac{{\overline t}_{VA} \left(q, v \right)}{ q^2 }, \label{eq:a1a2_matching_result_no_IR}
\end{align}
where the hadronic physics is encoded in ${\overline t}_{VA}$.

\subsection{The electromagnetic coupling constant $g_{11} + g_{12}$}
\label{sec:subsec33}

\begin{figure}
    \centering
    \includegraphics[width=0.9\textwidth]{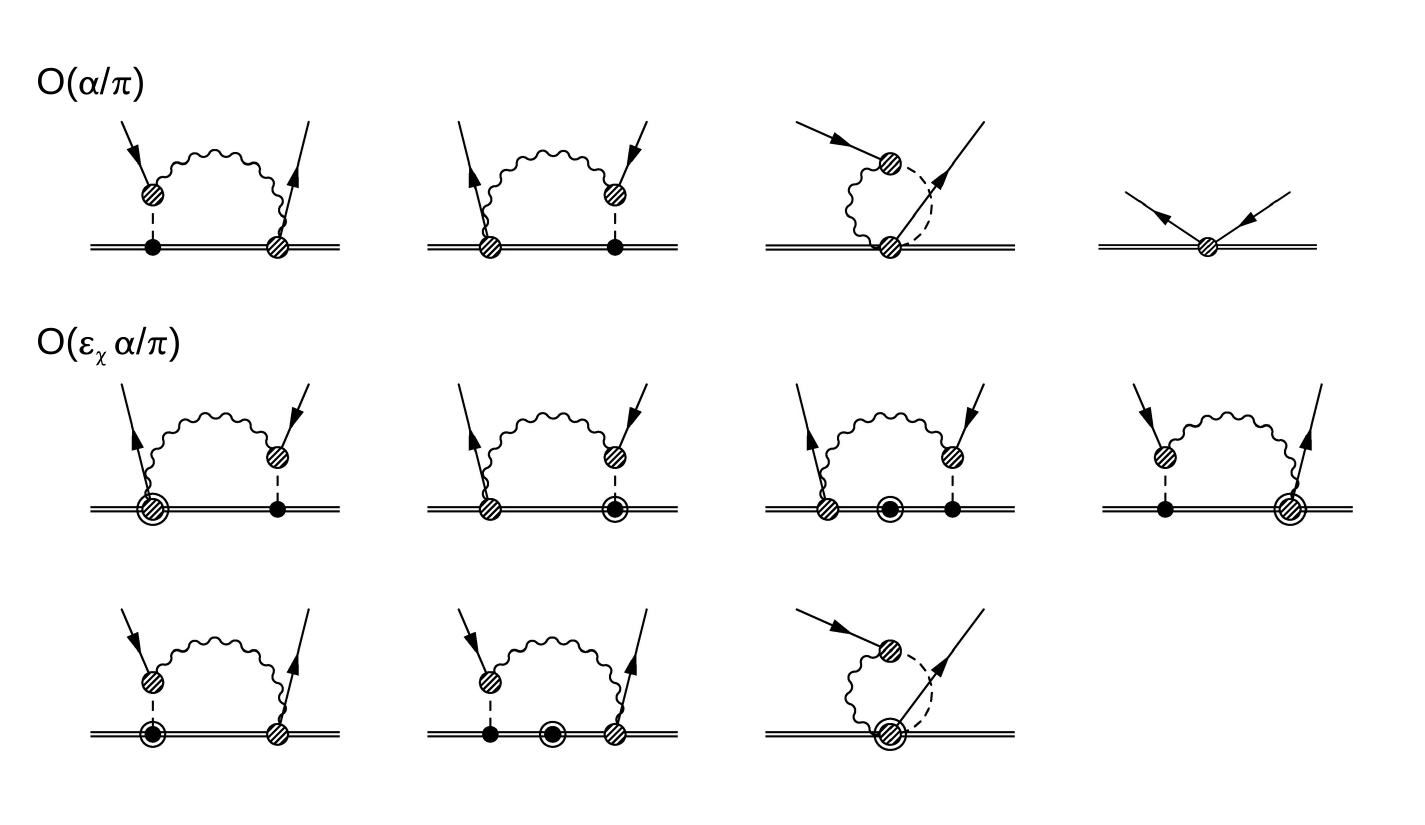}
        \caption{Contributions to the correlation function $\Gamma_{VA}$ in HBChPT are shown as Feynman diagrams. Double, wiggly, and dashed lines denote nucleons, photons, and pions, respectively. Dashed circles represent insertions of the sources ${\bf q}^{a}_V$ and ${\bf q}^{b}_A$. Circled dashed circles denote the insertion of the source from the NLO chiral Lagrangian $\mathcal L^{p^2}_{\pi N}$. The arrows denote the flow of the momentum $r$, which enters the axial-vector source and exits from the vector source.} \label{fig:axial}
\end{figure}

To express the combination of electromagnetic LECs $g_{11} + g_{12}$ in terms of two-point functions, we introduce the correlator
\begin{align}\label{eq:GammaVA}
    \overline{e}  \slashed{b} P_L \nu_e  \, \Gamma_{VA} = b_\rho \frac{ \varepsilon^{a b c} \tau^c_{i j} S^\rho_{\sigma \sigma^\prime}}{3} \overline{e} \gamma_\lambda P_L \nu_e i \frac{\partial}{\partial r_\lambda} \left. \left(  \int \mathrm{d}^d x e^{i r \cdot x} \langle  N (k, \sigma^\prime, j) | \frac{\delta^2 W \left(  {\bf{q}}_V, {\bf{q}}_A  \right)}{\delta {\bf{q}}^a_{V} \left( x \right) \delta {\bf{q}}^b_{A} \left( 0 \right)} \Bigg|_{{\bf{q}}=0} | N (k, \sigma, i) \rangle   \right)  \right|_{r_\lambda=0},
\end{align}
and perform the matching of LEFT to HBChPT at the level of this object.

We present the tree-level and one-loop diagrams in HBChPT at leading and next-to-leading orders, which contribute to this correlation function, in Fig.~\ref{fig:axial}. All one-loop diagrams involve pions, while there are no contributions in which both spurion fields couple directly to the nucleon line. In arbitrary $R_\xi$ gauge, we obtain
\begin{align}
\left. \Gamma_{VA}\right|^{\rm HBChPT} &= e^2 \left( g_{11} + g_{12} - 2 g^{(0)}_A (1-\xi)  \int \frac{i \mathrm{d}^d q}{(2\pi)^d} \frac{q^2}{q^2 - m^2_\pi } \frac{1}{\left( q^2 - \lambda_\gamma^2 \right) \left( q^2 - \xi \lambda_\gamma^2\right)}\right. \nonumber \\ 
& \left. + \frac{g_A^{(0)} (1+\kappa_1) }{m_N} \frac{d-2}{d-1} \int \frac{i \mathrm{d}^d q}{(2\pi)^d} \frac{q^2 - \left( v\cdot q \right)^2}{ \left(q^2 - m_\pi^2\right) v\cdot q}  \left(\frac{1}{q^2 - \lambda_\gamma^2} - \xi \frac{1}{q^2 - \xi \lambda_\gamma^2} \right) \right. \nonumber \\
&\left. + \frac{g_A^{(0)}}{m_N} \frac{1-\xi}{d-1} \int \frac{i \mathrm{d}^d q}{(2\pi)^d} \frac{q^2 - \left( v\cdot q \right)^2}{\left(q^2 - m_\pi^2\right) v\cdot q} \frac{1}{q^2 - \lambda_\gamma^2}\right). \label{eq:HBChPT}
\end{align}
In the Feynman-`t Hooft gauge, i.e., $\xi =1$, the HBChPT loop contributions vanish in agreement with anomalous dimensions in Ref.~\cite{Gasser:2002am} and results in Ref.~\cite{Seng:2024ker}. Upon performing the integrations in Eq.~\eqref{eq:HBChPT}, the HBChPT correlation function $\Gamma_{VA} |^{\rm HBChPT}$ can be explicitly written in terms of the renormalized couplings as
\begin{align}
\left. \Gamma_{VA}\right|^{\rm HBChPT} = e^2 \left( g^r_{11} + g^r_{12} + \frac{2 g_A^{(0)}}{\left( 4 \pi \right)^2} \left( 1 - \xi \right) \ln \frac{\mu^2}{m^2_\pi}  + \frac{2}{3} \frac{3+ 2\kappa_1}{(4\pi)^2} (1-\xi) \frac{\pi m_\pi}{m_N} \right). \label{eq:HBChPT_integrated}
\end{align}

In  LEFT, the correlator $\Gamma_{VA}$ can be expressed in terms of the hadronic tensor $t^{\mu \nu \rho}_{VA}$ from Eq.~(\ref{eq:tAVV}) or equivalently, after using the Ward identities and definitions of Eqs.~(\ref{eq:tPVscalar}) and~(\ref{eq:tVAscalar}), in terms of scalar hadronic amplitudes $\overline{t}_{VA}$ and ${t}_{VP}$ as
\begin{align}
\left. \Gamma_{VA}\right|^{\rm LEFT} &=  4 e^2 g^{(0)}_A \left ( g_{23} + \int \frac{i \mathrm{d}^4 q}{(2\pi)^4} \frac{q^2 \overline{t}_{V A} \left(q, v \right)}{\left[q^2 - \lambda_\gamma^2\right]^2} + \left( 1 - \xi \right)  \overline{m} \int \frac{i \mathrm{d}^4 q}{(2\pi)^4} \frac{q^2 {t}_{VP} \left(q, v \right)}{\left( q^2 - \lambda_\gamma^2 \right) \left( q^2 - \xi \lambda_\gamma^2\right)} \right) \nonumber \\
&+ 4 e^2 g^{(0)}_A \frac{ 1 - \xi}{d} \int \frac{i \mathrm{d}^d q}{(2\pi)^d} \frac{1}{\left( q^2 - \lambda_\gamma^2 \right) \left( q^2 - \xi \lambda_\gamma^2\right)} \left(  \frac{q^2}{q^2 - \lambda_\gamma^2} + \frac{q^2}{q^2 - \xi \lambda_\gamma^2} -d \right) \nonumber \\
&+ 4 e^2 g^{(0)}_A \frac{d-2}{d} \int \frac{i \mathrm{d}^d q}{(2\pi)^d} \frac{1}{\left[q^2 - \lambda_\gamma^2\right]^2} \frac{q^2}{q^2-\mu_0^2}. \label{eq:g11_g12_LEFT}
\end{align}
Performing the integrations in the last two terms of Eq.~\eqref{eq:g11_g12_LEFT}, one obtains 
\begin{align}
\left. \Gamma_{VA}\right|^{\rm LEFT} &=  4 e^2 g^{(0)}_A \left ( g^r_{23} + \int \frac{i \mathrm{d}^4 q}{(2\pi)^4} \frac{q^2 \overline{t}_{V A} \left(q, v \right)}{\left[q^2 - \lambda_\gamma^2\right]^2} + \left( 1 - \xi \right)  \overline{m} \int \frac{i \mathrm{d}^4 q}{(2\pi)^4} \frac{q^2 {t}_{VP} \left(q, v \right)}{\left( q^2 - \lambda_\gamma^2 \right) \left( q^2 - \xi \lambda_\gamma^2\right)} \right) \nonumber \\
& - 4 e^2 \frac{g^{(0)}_A}{\left( 4 \pi \right)^2} \left( \frac{1}{2} \ln \frac{\mu^2}{\mu^2_0} + \frac{1}{4} - \frac{1-\xi}{2} \left( 1 + \ln \frac{\mu^2}{\lambda^2_\gamma}\right) -\frac{\xi}{2} \ln \xi \right). \label{eq:g11_g12_LEFT_explicit}
\end{align}
To derive Eq.~\eqref{eq:g11_g12_LEFT}, we subtracted the OPE by exploiting
\begin{align}\label{eq:OPEtVA}
 \left. g_{\mu \nu} t^{\mu \nu \rho}_{VA}(q,v) \right|_{\rm OPE} = i g_A^{(0)} (q^{\rho} - \left( v \cdot q \right) v^{\rho}) (d-2)  \frac{1}{q^2 - \mu_0^2}.
\end{align}

By equating LEFT and HBChPT Eqs.~(\ref{eq:HBChPT_integrated}) and~(\ref{eq:g11_g12_LEFT_explicit}), i.e., $ \Gamma_{VA}|^{\rm HBChPT} =\Gamma_{VA}|^{\rm LEFT}$, we obtain the representation for the combination of electromagnetic coupling constants $g_{11} + g_{12}$:
\begin{align}
\frac{g^r_{11} + g^r_{12}}{4 g_A^{(0)}} \left( \mu_\chi, \mu, \xi \right) &=  \int \frac{i \mathrm{d}^4 q}{(2\pi)^4} \frac{q^2 \overline{t}_{V A} \left(q, v \right)}{\left[q^2 - \lambda_\gamma^2\right]^2} + \left( 1 - \xi \right)  \overline{m} \int \frac{i \mathrm{d}^4 q}{(2\pi)^4} \frac{q^2 {t}_{VP} \left(q, v \right)}{\left( q^2 - \lambda_\gamma^2 \right) \left( q^2 - \xi \lambda_\gamma^2\right)} \nonumber \\
&- \frac{1}{\left(4 \pi \right)^2} \left[ \frac{1}{2} \ln \frac{\mu^2}{\mu_0^2} + \frac{1-\xi}{2}\ln\frac{\mu^2_\chi}{\mu^2} - \frac{1}{4} + \frac{\xi}{2} \left( 1 - \ln \xi \right)- \frac{1-\xi}{2} \ln \frac{m^2_\pi}{\lambda_\gamma^2} \right] \nonumber \\
& - \frac{1}{6} \frac{3+2\kappa_1}{(4\pi)^2} \left( 1 - \xi \right) \frac{\pi m_\pi}{m_N}. \label{eq:g11_g12_couplings}
\end{align}

Accounting for the infrared behavior of the hadronic object ${t}_{VP}$, cf. Appendix~\ref{app:IR}, and neglecting $m^2_\pi/\Lambda_\chi^2$-suppressed contributions, we rewrite the matching relation for the combination $g^r_{11} + g^r_{12}$ in the infrared-finite form
\begin{align}
\frac{g^r_{11} + g^r_{12}}{4 g_A^{(0)}} \left( \mu_\chi, \mu, \xi \right) &= \int \frac{i \mathrm{d}^4 q}{(2\pi)^4} \frac{\overline{t}_{V A} \left(q, v \right)}{q^2} - \frac{1}{\left(4 \pi \right)^2} \left[ \frac{1}{2} \ln \frac{\mu^2}{\mu_0^2} + \frac{1-\xi}{2} \ln\frac{\mu^2_\chi}{\mu^2} - \frac{1}{4} + \frac{\xi}{2}  \right], \label{eq:g11_g12_couplings_clean}
\end{align}
where the hadronic physics is encoded in ${\overline t}_{VA}$.

\subsection{Representation for the contribution from two-point correlation functions}
\label{sec:subsec34}

Combining the determination of HBChPT coupling constants $X_6$, $A_1 + A_2$, and $g_{11} + g_{12}$ from Eqs.~(\ref{eq:X6_coupling_constant}),~(\ref{eq:a1a2_match_result}), and~(\ref{eq:g11_g12_couplings}), respectively, we determine the relevant low-energy combination:
\begin{align}\label{eq:CA2ptmatch}
& \left( - \frac{X^r_6}{2} + 2 (A^r_1 + A^r_2) + \frac{g^r_{11} + g^r_{12}}{4 g_A^{(0)}}\right) \left( \mu_\chi, \mu \right) = \frac{1}{(4\pi)^2} \left[ \frac{3}{2} \left(1 - \ln \frac{\mu^2_\chi}{\mu^2}\right) + \left(1 + \frac{1-\xi}{2} \right)  \ln \frac{m^2_\pi}{\lambda_\gamma^2} + \frac{\xi}{2} \ln \xi \right] \nonumber\\
& - \frac{1}{3} \frac{3+2\kappa_1}{(4\pi)^2} \left(1 + \frac{1-\xi}{2}\right) \frac{\pi m_\pi}{m_N}  + \int \frac{i \mathrm{d}^4 q}{(2\pi)^4} \frac{\lambda_\gamma^2 \overline{t}_{V A} \left(q, v \right)}{\left[q^2\right]^2} + 2  \overline{m} \int \frac{i \mathrm{d}^4 q}{(2\pi)^4} \frac{{t}_{VP} \left(q, v \right)}{ q^2 - \lambda_\gamma^2} \left( 1 + \frac{1 - \xi}{2} \frac{q^2}{ q^2 - \xi \lambda_\gamma^2} \right).
\end{align}
Again accounting for the infrared behavior of the hadronic object ${t}_{VP}$, cf. Appendix~\ref{app:IR}, exploiting infrared behavior for $\overline{t}_{V A}$, and neglecting $m^2_\pi/\Lambda_\chi^2$-suppressed contributions, we obtain the $m_\pi,~\lambda_\gamma,~\kappa_1,$ and $\xi$-independent relations
\begin{align}
\left( - \frac{X^r_6}{2} + 2 \left( A^r_1 + A^r_2 \right) + \frac{g^r_{11} + g^r_{12}}{4 g_A^{(0)}}\right) \left( \mu_\chi, \mu \right) &= \frac{1}{(4\pi)^2} \frac{3}{2} \left(1 - \ln \frac{\mu^2_\chi}{\mu^2}\right), \label{eq:A1A2}\\
2 \left( A^r_3 + A^r_4 \right) \left( a, \mu \right) &= \frac{1}{(4\pi)^2}\left[ \frac{1}{2} \ln \frac{\mu^2}{\mu^2_0} + \frac{3 - 8 a}{12} \right] + \int \frac{i \mathrm{d}^4 q}{(2\pi)^4} \frac{{\overline t}_{VV, 0} \left(q, v \right)}{ q^2}, \label{eq:A3A4}
\end{align}
from which one can immediately obtain the two-point function of Eq.~\eqref{eq:split} as
\begin{equation}\label{eq:CA2pt}
   \left[ \hat C_A \right]_{\rm 2pt} =    8 \pi^2 \left \{ - \frac{X_6}{2}  + 2 \left( A_1 + A_2 + A_3 + A_4 \right) + \frac{g_{11} + g_{12}}{4 g_A^{(0)}} \right\}. 
\end{equation}

The r.h.s. of Eq.~\eqref{eq:A3A4} is conveniently expressed in terms of the difference between proton and neutron spin structure functions, $S_{1,2} = S^p_{1,2} - S^n_{1,2}$, as
\begin{align}\label{eq:A34S12}
2 \left( A^r_3 + A^r_4 \right) \left( a, \mu \right) &= \frac{1}{(4\pi)^2}\left[ \frac{1}{2} \ln \frac{\mu^2}{\mu^2_0} + \frac{3 - 8 a}{12} \right] + \frac{2}{g^{(0)}_A} \int \frac{i \mathrm{d}^4 q}{\left( 2 \pi \right)^4}  \left( \frac{ \nu^2 - 2 Q^2}{3 Q^2} \frac{\overline{S}_1 (\nu, Q^2)}{Q^2} - \frac{\nu^2}{Q^2 } \frac{\overline{S}_2 (\nu, Q^2)}{m_N \nu} \right),
\end{align}
where we defined the amplitudes $S_1$ and $S_2$ from the tensor decomposition of the antisymmetric part of $t^{\mu \nu \rho}_{V V,0}$ as
\begin{align}
t^{\mu \nu \rho}_{V V,0} &= i \left[ \varepsilon^{\mu \nu \rho \alpha} q_\alpha S_1 (\nu, Q^2) + \varepsilon^{\mu \nu \beta \alpha} q_\alpha \left( \delta^\rho_\beta \frac{\nu}{m_N} - v_\beta \frac{q^\rho}{m_N}  \right) S_2\left(\nu, Q^2\right) \right],  \label{eq:tVV0_decomposition}
\end{align}
with the OPE-subtracted expressions
\begin{align}
\overline{S}_1 \left( \nu, Q^2 \right) &= {S}_1 \left( \nu, Q^2 \right) + \frac{1}{3}\frac{g^{(0)}_A}{Q^2 + \mu_0^2}, \\
\overline{S}_2 \left( \nu, Q^2 \right) &= {S}_2 \left( \nu, Q^2 \right).
\end{align}
Contrary to hadronic contributions to the vector coupling constant~\cite{Marciano:2005ec,Seng:2018qru,Seng:2018yzq,Cirigliano:2023fnz} that do not correspond to physical processes in the same isospin channel, the spin structure functions $S_1$ and $S_2$ in Eq.~(\ref{eq:A34S12}) are directly related to experimentally measured proton and neutron spin structure by unsubtracted dispersion relations~\cite{Drell:1966kk,Hayen:2020cxh,Gorchtein:2021fce}.

Our resulting expression in terms of the spin structure functions is in agreement with known results~\cite{Hayen:2020cxh,Gorchtein:2021fce} in the limit of a large mass of the $W$ boson $M_W \to \infty$.

\subsection{Electromagnetic coupling constants: three-point correlation functions}
\label{sec:subsec35}

In this Subsection, we evaluate the combination of HBChPT electromagnetic LECs that determine $\left[ \hat C_A \right]_{\rm 3pt}$ in Eq.~\eqref{eq:split} in terms of the LEFT three-point correlation functions by two alternative ways and verify the consistency between the two results. We perform the matching at the level of correlation functions with two vector currents and either a pseudoscalar density or an axial-vector current in Subsections~\ref{sec:subsec351} and~\ref{sec:subsec352}, respectively.

\subsubsection{Pseudoscalar density  and two vector currents}
\label{sec:subsec351}

\begin{figure}
\includegraphics[width=\textwidth]{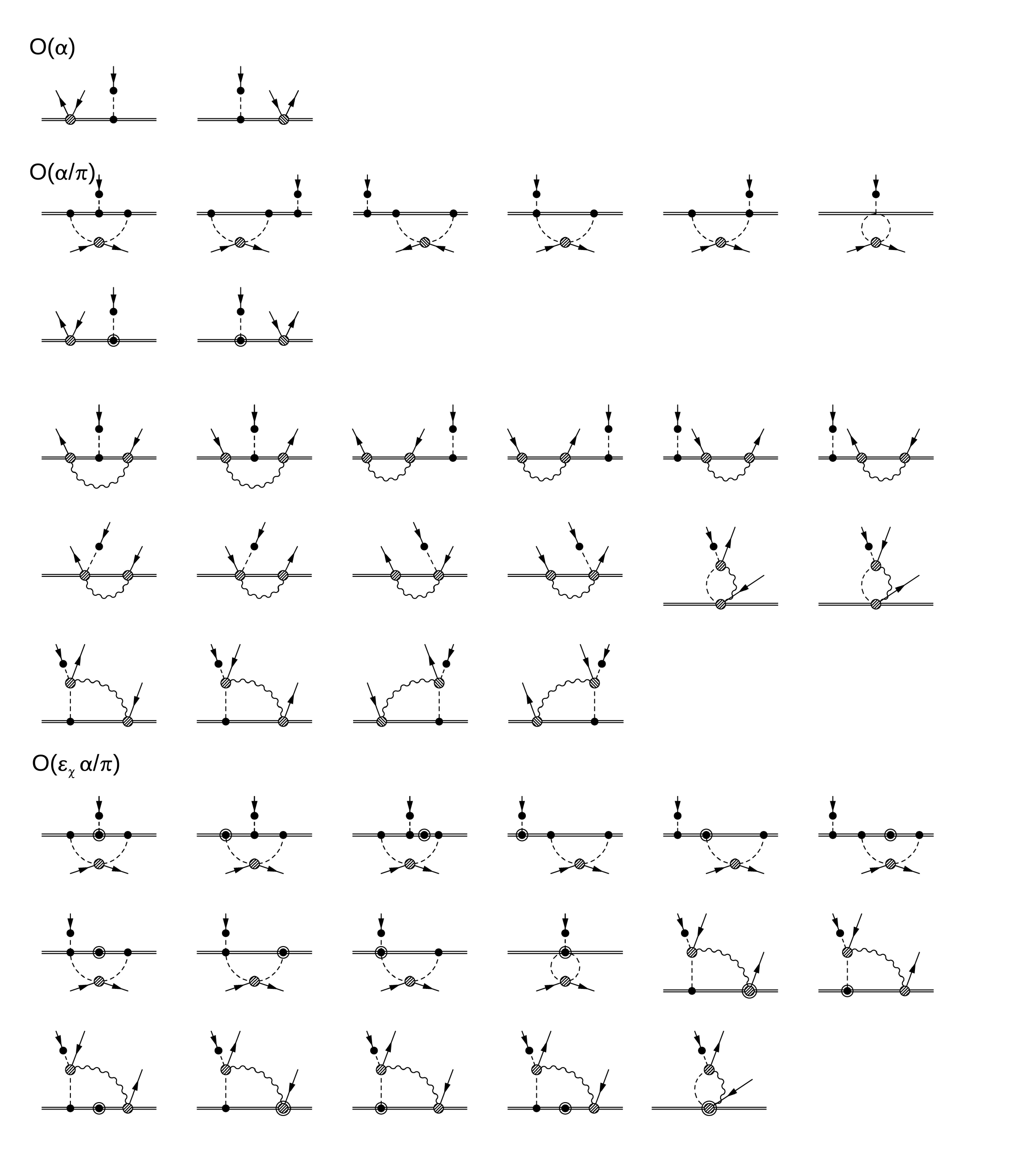}
\caption{HBChPT one-loop contributions to the pseudoscalar correlator $W_P$ are represented as Feynman diagrams. The incoming plain lines denote a vector source, which injects zero momentum, the outgoing plain line denotes a vector source with outgoing momentum $r$. The plain line that ends with a dot indicates a pseudoscalar source, which injects the momentum $r$. The remaining notations are as in Fig. \ref{fig:axialA}. Notice that only a few possible orderings of the $\mathcal O \left( \epsilon_\chi \alpha/\pi \right)$ diagrams are shown, e.g., the mirror images of diagrams in the last three rows, while the pseudoscalar current to the right of the loop are not shown.}\label{fig:PS}
\end{figure}

To express the combination of LECs that enter $\left[\hat{C}_A \right]_{\rm 3pt}$ in~\eqref{eq:split}, we  perform the matching at the level of the following correlation function, which involves derivatives w.r.t. the pseudoscalar source $\overline{p}$,
\begin{align}
W_P &= b_\rho  b_\lambda \frac{ \overline{m} \tau^c_{ij} S^\rho_{\sigma \sigma^\prime}}{6 i} \frac{\partial}{\partial r_\lambda } \int \mathrm{d}^d x \mathrm{d}^d y e^{i r \cdot (x- y)}  \nonumber \\
&\left. \langle N (k, \sigma^\prime, j)| P^{a b c d} \frac{\delta^3 W \left( \overline{p}, {\bf q}_V \right)}{ \delta {\overline{p}}^d(y) \delta {\bf q}_V^{b}(x) \delta {\bf q}_V^{a}(0)}  +  \frac{\delta^3 W \left( \overline{p}, {\bf q}_V \right)}{\delta {\overline{p}}^c(y) \delta {\bf q}_V^{0}(x)  \delta {\bf q}_V^{0}(0)} | N (k, \sigma, i) \rangle \right|_{r_\lambda = 0}, \label{eq:WP3PT_definition}
\end{align}
where the projector $P^{a b c d}$ acts in the isospin space as
\begin{equation}
P^{a b c d} = \frac{2}{5} \left( \delta^{a b} \delta^{c d} - \frac{\delta^{a c} \delta^{bd} + \delta^{ad } \delta^{bc}}{4} \right).
\end{equation}

The tree-level contribution of HBChPT low-energy constants to the correlator of Eq.~(\ref{eq:WP3PT_definition}) $W_{P,  \rm LEC}^{\rm HBChPT}$ can be expressed as
\begin{align}
    W_{P,\, \rm LEC}^{\rm HBChPT}=  e^2 \left[ \frac{1}{2}\left(g_{1} + g_{13} + \frac{g_{11}-g_{12}}{2}\right) + \left(g_2 + \frac{g_{1} + g_{13}}{2} \right) \right] = \frac{e^2 g_A^{(0)}}{8\pi^2} \left[\hat{C}_A\right]_{\rm 3pt},
\end{align}
where we separated the LECs contributions to the isovector and isoscalar correlators, respectively. One-loop contributions at leading and next-to-leading orders in HBChPT are illustrated in Fig.~\ref{fig:PS}, where the pseudoscalar source attaches to the pion field. These diagrams are UV divergent and $W_P^{\rm HBChPT}$ is rendered finite by renormalizing the LECs. In the Feynman-`t Hooft gauge, one finds~\cite{Gasser:2002am}\footnote{Ref.~\cite{Gasser:2002am} works in the Feynman-`t Hooft gauge. Here, we extend the analysis to the arbitrary $R_\xi$ gauge.}
\begin{align}
    g_1 + g_2 + g_{13} + \frac{g_{11}-g_{12}}{4} &= g^r_1 + g^r_2 + g^r_{13} + \frac{g^r_{11} - g^r_{12}}{4} - \frac{1}{2} \frac{1}{(4\pi)^2} \left(\frac{1}{\varepsilon} + 1\right) 2 g^{(0)}_A  Z_\pi \left[ 1 + 3 \left( g^{(0)}_A \right)^2 \right].
\end{align}
The resulting expression for $W^{\rm HBChPT}_P$ to NLO in HBChPT can be written as
\begin{align} \label{eq:coup2}
W_P^{\rm HBChPT} &= \frac{e^2 g_A^{(0)}}{2} \frac{1}{(4\pi)^2} \left\{ 2 Z_\pi \left(  \left[ 1 + 3 \left( g_A^{(0)} \right)^2 \right] \left(  \ln \frac{\mu^2_\chi}{m_\pi^2}  - 1\right) + 8 \pi m_\pi \left[ c_4 - c_3 + \frac{3}{8 m_N} + \frac{9}{16} \frac{\left( g_A^{(0)} \right)^2}{m_N} \right] \right)   \nonumber \right. \\
& \left. - 4 Z_\pi \left( g_A^{(0)} \right)^2 +  2 \ln \frac{m_\pi^2}{\lambda_\gamma^2} - 2 + \left( 1 - \xi \right) \ln \frac{m_\pi^2}{\lambda_\gamma^2} + \xi \ln \xi - 2 \frac{3 + 2 \kappa_1}{3} \left(1 + \frac{1-\xi}{2} \right) \frac{\pi m_\pi}{m_N}   \right\} \nonumber \\ 
& +\frac{e^2 g_A^{\left( 0 \right)}}{2 m_N} F^2_\pi \left[ f_1 + f_3 + 2 f_4 + 6 \left( g_A^{(0)}\right)^2 \frac{Z_\pi \pi m_\pi}{\left(  4\pi F_\pi\right)^2} \right] + e^2 \left( g^r_1 + g^r_2 + g^r_{13} + \frac{g^r_{11}-g^r_{12}}{4} \right)  \nonumber \\
& +\left(\frac{1}{- v \cdot r + i \eta} + \frac{1}{v \cdot r + i \eta} \right)  g^{(0)}_A \Delta_{\rm em} m_N.
\end{align}
The last line vanishes if the energy inserted by the pseudoscalar current, $v \cdot r$, is different from zero. For $v \cdot r \rightarrow 0$, the nucleon propagator between the insertion of the pseudoscalar current and the two vector currents goes on shell, and $W_P$ develops a pole proportional to the electromagnetic correction to the averaged nucleon mass $\Delta_{\rm em} m_N$:
\begin{equation}
   \Delta_{\rm em} m_N = - \frac{e^2 F_\pi^2}{2} \left[ f_1 + f_3 + 2 f_4 + 6 \left( g_A^{(0)}\right)^2 \frac{Z_\pi \pi m_\pi}{\left(  4\pi F_\pi\right)^2}   + \ldots \right] = - \frac{1}{2}\left( G^{\rm HBChPT}_{VV} + G^{\rm HBChPT}_{VV,0} \right).  
\end{equation}

In LEFT, $W^{\rm LEFT}_P$ receives no contributions from the counterterms $g_{23}$, $g_{02}$, and $g_{03}$. It can be expressed as
\begin{align}
W_P^{\rm LEFT} &= e^2 \overline{m} b_\rho b_\lambda \frac{\partial}{\partial r_\lambda} \left. \left( \int \frac{i \mathrm{d}^d q}{(2\pi)^d} \frac{t^{\mu \nu \rho}_{PVV} \left( r, q, v \right) + t^{\mu \nu \rho}_{PVV,0}\left( r, q, v \right)}{q^2 - \lambda_\gamma^2} \left( g_{\mu \nu} - \left( 1-\xi \right) \frac{q_\mu q_\nu}{q^2 - \lambda_\gamma^2 \xi }\right) \right) \right|_{r_\lambda = 0}, 
\end{align}
where we introduced the nonperturbative objects in LEFT $t^{\mu \nu \rho}_{PVV}$ and $t^{\mu \nu \rho}_{PVV,0}$ as
\begin{align}
    t^{\mu \nu \rho}_{PVV} \left(r, q, v\right) &=  \frac{\tau^c_{i j} S^\rho_{\sigma \sigma^\prime}}{24} P^{abcd} \int \mathrm{d}^d x \mathrm{d}^d y e^{i r \cdot \left( x - y \right) + i q \cdot x} \nonumber \\
    &\langle N (k, \sigma^\prime, j) | T \left[ \overline{q} \gamma^5 \tau^d q(y)\, \overline{q} \gamma^\mu \tau^b q (x) \, \overline{q} \gamma^\nu \tau^a q(0) \right] | N ( k, \sigma, i) \rangle, \\ 
    t^{\mu \nu \rho}_{PVV,\, 0} \left(r, q, v \right) &= \frac{\tau^c_{i j} S^\rho_{\sigma \sigma^\prime}}{24} \frac{1}{9} \int \mathrm{d}^d x \mathrm{d}^d y e^{i r \cdot \left( x - y \right) + i q \cdot x} \langle N (k, \sigma^\prime, j) | T \left[ \overline{q} \gamma^5 \tau^c q(y)\, \overline{q} \gamma^\mu q (x) \, \overline{q} \gamma^\nu q(0) \right] | N ( k, \sigma, i) \rangle.
\end{align}
$t_{PVV}^{\mu \nu \rho}$ and $t^{\mu \nu \rho}_{PVV,\, 0}$ satisfy the following Ward identities:
\begin{align}
    q_\nu t^{\mu \nu \rho}_{PVV} \left( r, q, v \right) &= \, i t^{\mu \rho}_{VP} \left( r + q, v \right), \\
    q_\nu t^{\mu \nu \rho}_{PVV,\, 0} \left( r, q, v \right) &=  0,  \\
    q_\mu t^{\mu \rho}_{VP} \left(r + q, v \right) &= \left( r + q \right)_\mu t^{\mu \rho}_{VP} \left( r + q, v\right) - r_\mu t^{\mu \rho}_{VP} \left(r + q, v\right) = - r_\mu t^{\mu \rho}_{VP} \left(r + q, v\right),
\end{align}
resulting in
\begin{align}
    W_P^{\rm LEFT} &=  e^2 \overline{m} b_\rho b_\lambda \frac{\partial}{\partial r_\lambda} \left. \left( \int \frac{i \mathrm{d}^d q}{(2\pi)^d} \frac{g_{\mu \nu} \left( t^{\mu \nu \rho}_{PVV} \left( r, q, v \right) + t^{\mu \nu \rho}_{PVV,0}\left( r, q, v \right) \right)}{q^2 - \lambda_\gamma^2}  \right) \right|_{r_\lambda = 0} \nonumber \\
    &+ e^2 \overline{m} b_\rho b_\lambda \int \frac{\mathrm{d}^d q }{(2\pi)^d}  \frac{  \left( 1-\xi \right) t^{\lambda \rho}_{VP} \left( q, v\right)}{\left( q^2 - \lambda_\gamma^2 \right) \left( q^2 - \xi \lambda_\gamma^2 \right)}.
\end{align}
The only counterterm operator that could contribute to the correlation function with quarks corresponds to the coupling constant $g_{23}$. It enters in an antisymmetric combination w.r.t. the indexes of the vector spurions and, consequently, vanishes both in the isoscalar correlator and in the isovector correlator after the contraction with the symmetric projection operator $P^{abcd}$. As a consistency check, we verified that the leading OPE for the hadronic objects with pseudoscalar densities ensures convergence of the integrals, and thus no counterterm is needed to absorb UV divergences.

Equating the LEFT and HBChPT expressions for $W_P$, i.e., $W^{\rm HBChPT}_P=W^{\rm LEFT}_P$, using the notation of Eq.~(\ref{eq:tPVscalar}) and the infrared expression for ${t}_{VP}$ of Eq.~(\ref{eq:tPVIR}), we determine the gauge-independent  combination of coupling constants $\left[\hat{C}_{A}\right]_{\rm 3pt}$ in Eq.~(\ref{eq:split})
\begin{align}\label{eq:matchingPS3pt}
    & \left[\hat{C}_{A}\right]_{\rm 3pt} \left( \mu_\chi \right) = \frac{Z_\pi}{2} \left( \left[ 1 + 3 \left( g_A^{(0)} \right)^2 \right] \left( 1 -  \ln \frac{\mu^2_\chi}{m_\pi^2}  \right) + 2 \left( g_A^{(0)} \right)^2 + 8 \pi m_\pi \left[ c_3 - c_4 - \frac{3}{8 m_N} -  \frac{9}{16} \frac{\left( g_A^{(0)} \right)^2}{m_N}\right] \right)   \nonumber \\
    & + \frac{1}{2} + \frac{1}{2} \ln \frac{\lambda_\gamma^2}{m_\pi^2} + \frac{3 + 2 \kappa_1}{6} \frac{\pi m_\pi}{m_N} - \frac{4 \pi^2}{m_N} \int \frac{ \mathrm{d}^4 q}{(2\pi)^4} \frac{ g_{\mu \nu} \left( \tau^{\mu \nu}_{V V} \left(q, v \right) + \tau^{\mu \nu}_{V V, 0} \left(q, v \right) \right)}{q^2 - \lambda_\gamma^2} \nonumber \\ 
    &+ \frac{8 \pi^2}{g_A^{(0)}} \left.\left[ \overline{m} b_\rho b_\lambda \frac{\partial}{\partial r_\lambda} \left( \int \frac{i \mathrm{d}^4 q}{(2\pi)^4} \frac{g_{\mu \nu} \left( {t}^{\mu \nu \rho}_{PVV} \left( r, q, v \right) + {t}^{\mu \nu \rho}_{PVV,0}\left( r, q, v \right) \right)}{q^2 - \lambda_\gamma^2} \right) + 2 i \pi \delta(v \cdot r) \frac{g_A^{(0)}\Delta_{\rm em} m_N}{e^2} \right]\right|_{r_\lambda =0}. 
\end{align}
The last line is finite in the limit $v\cdot r \rightarrow 0$, so that the terms in square brackets correspond to the subtracted correlation function. This is analogous to the subtraction prescription used in the traditional approach \cite{Brown:1969lsx,Sirlin:1977sv}.

After combining the separately gauge-independent pieces into Eq.~(\ref{eq:split}), we obtain an exact cancellation of the contributions from the magnetic moment and infrared regions between two-point and three-point functions:
\begin{align}\label{eq:matchingPS}
    &    \hat{C}_{A} \left( a, \mu_\chi, \mu \right) =  \frac{Z_\pi}{2} \left( \left[ 1 + 3 \left( g_A^{(0)} \right)^2 \right] \left( 1 -  \ln \frac{\mu^2_\chi}{m_\pi^2}  \right) + 2 \left( g_A^{(0)} \right)^2 + 8 \pi m_\pi \left[ c_3 - c_4 - \frac{3}{8 m_N} -  \frac{9}{16} \frac{\left( g_A^{(0)} \right)^2}{m_N}\right] \right)  \nonumber \\
    &+ \frac{1}{2}\left[ \frac{1}{2} \ln \frac{\mu^2}{\mu^2_0} - \frac{3}{2} \ln \frac{\mu^2_\chi}{\mu^2} - 4 B \left(a \right)- \frac{1}{4} \right] + \frac{16 \pi^2}{g^{(0)}_A} \int \frac{i \mathrm{d}^4 q}{\left( 2 \pi \right)^4}  \left( \frac{ \nu^2 - 2 Q^2}{3 Q^2} \frac{\overline{S}_1 (\nu, Q^2)}{Q^2} - \frac{\nu^2}{Q^2 } \frac{\overline{S}_2 (\nu, Q^2)}{m_N \nu} \right) \nonumber \\
    &+ \frac{8 \pi^2}{g_A^{(0)}} \left.\left[ \overline{m} b_\rho b_\lambda \frac{\partial}{\partial r_\lambda} \left( \int \frac{i \mathrm{d}^4 q}{(2\pi)^4} \frac{g_{\mu \nu} \left( {t}^{\mu \nu \rho}_{PVV} \left( r, q, v \right) + {t}^{\mu \nu \rho}_{PVV,0}\left( r, q, v \right) \right)}{q^2 - \lambda_\gamma^2} \right) + 2 i \pi \delta(v \cdot r) \frac{g_A^{(0)}\Delta_{\rm em} m_N}{e^2} \right]\right|_{r_\lambda =0} \nonumber \\ 
    & + 16 \pi^2  \overline{m} \int \frac{i \mathrm{d}^4 q}{(2\pi)^4} \frac{{t}_{VP} \left(q, v \right)}{ q^2 - \lambda_\gamma^2} -  \frac{4 \pi^2}{m_N} \int \frac{ \mathrm{d}^4 q}{(2\pi)^4} \frac{ g_{\mu \nu} \left( \tau^{\mu \nu}_{V V} \left(q, v \right) + \tau^{\mu \nu}_{V V, 0} \left(q, v \right) \right)}{q^2}.
\end{align}
The combination of $PVV$, $VP$, and $VV$ hadronic objects in the last two lines of Eq.~(\ref{eq:matchingPS}) is infrared finite for $\lambda_\gamma \to 0$ and depends on the pion mass in such a way as to cancel the explicit pion-mass dependence in the first line of Eq.~\eqref{eq:matchingPS}. This can be shown explicitly by making use of the fact that pion-mass dependence can arise in the LEFT correlators whenever intermediate states involving pions contribute. Such intermediate states can lead to single-particle poles or multiparticle cuts in the correlation functions, which, when integrated over, result in $m_\pi$ dependence. The $m_\pi$ dependence without the pion mass splitting coupling $Z_\pi$ results from intermediate states involving a single pion, while the $m_\pi$ dependence of the terms $\sim Z_\pi$ in Eq.~\eqref{eq:matchingPS} is canceled by contributions of $2\pi$ intermediate states to the ${PVV}$ correlator. As the $m_\pi$ dependence of the correlation functions can be determined by studying their poles (or cuts), we loosely refer to these techniques as ``polology''~\cite{Weinberg:1995mt}, which is discussed in more detail in Appendix \ref{app:IR}.

\subsubsection{Axial-vector current and two vector currents}
\label{sec:subsec352}

An alternative representation for the electromagnetic LECs can be obtained by matching at the level of the following correlation function that involves the derivative w.r.t. the axial-vector external source
\begin{align}
W_A &= - b_\rho  b_\lambda \frac{\tau^c_{ij} S^\rho_{\sigma \sigma^\prime}}{6} \nonumber \\
& \int \mathrm{d}^d x \mathrm{d}^d y \langle N (k, \sigma^\prime, j)| P^{a b c d} \frac{\delta^3 W \left( \overline{a}, {\bf q}_V \right)}{ \delta {\overline{a}}_\lambda^d \left( y \right) \delta {\bf q}_V^{b} \left( x \right) \delta {\bf q}_V^{a} \left( 0 \right)}  +  \frac{\delta^3 W \left( \overline{a}, {\bf q}_V \right)}{\delta {\overline{a}}_\lambda^d \left( y \right) \delta {\bf q}_V^{0} \left( x \right)  \delta {\bf q}_V^{0} \left( 0 \right)} | N (k, \sigma, i) \rangle. \label{eq:WA3PT_definition}
\end{align}
As for the pseudoscalar correlator, we calculate $W_A$ by injecting some nonzero energy $v\cdot r$ in the axial-vector current. In the limit $v\cdot r \rightarrow 0$, a nucleon pole arises on both the HBChPT and LEFT sides of the matching equation, which therefore cancels in the matching and does not affect the determination of LECs.

The LEC contribution to this correlator is given by
\begin{align}\label{eq:WAchi}
    W_{A,\, {\rm LEC}}^{\rm HBChPT} &=  e^2 \left[ \frac{1}{2}\left(g_{1} + g_{13} + g_{11} \right) + g_2 + \frac{g_{1} + g_{13}}{2}  \right] = e^2 \left( \frac{g_A^{(0)}}{8 \pi^2}\left[\hat{C}_A\right]_{\rm 3pt}  + \frac{g_{11} + g_{12}}{4}\right).
\end{align}
The HBChPT loops contributing at leading and next-to-leading orders to $W^{\rm HBChPT}_A$ are similar to the ones that give $W^{\rm HBChPT}_P$. The Feynman diagrams for the axial-vector current insertion are exactly the same as in Fig.~\ref{fig:PS}, with the pseudoscalar source attached to a pion propagator replaced by an axial-vector source. After explicit evaluation, we find
\begin{align}\label{eq:coup22}
W_A^{\rm HBChPT} &= \frac{e^2 g_A^{(0)}}{2} \frac{1}{(4\pi)^2} \left\{ 2 Z_\pi \left(  \left[ 1 + 3 \left( g_A^{(0)} \right)^2 \right] \left(  \ln \frac{\mu^2_\chi}{m_\pi^2}  - 1\right) + 8 \pi m_\pi \left[ c_4 - c_3 + \frac{3}{8 m_N} + \frac{9}{16} \frac{\left( g_A^{(0)} \right)^2}{m_N} \right]  \right)   \nonumber \right. \\
& \left.  - 4 Z_\pi \left( g_A^{(0)} \right)^2 - 2 \frac{3 + 2 \kappa_1}{3} \frac{\pi m_\pi}{m_N} +  2 \ln \frac{m_\pi^2}{\lambda_\gamma^2} - 2 + \left( 1 - \xi \right) \ln \frac{\mu_\chi^2}{\lambda_\gamma^2} + \xi \ln \xi \right\} \nonumber \\
&+ e^2 \left( g^r_{1} + g^r_2 + g^r_{13} + \frac{g^r_{11}}{2} \right) +\left(\frac{1}{- v \cdot r + i \eta} + \frac{1}{v \cdot r + i \eta} \right)  g^{(0)}_A \Delta_{\rm em} m_N.
\end{align}

In LEFT, we obtain
\begin{align}
    W_A^{\rm LEFT} &= e^2 \left[ g_A^{(0)} g_{23} - b_\rho b_\lambda \int \frac{i \mathrm{d}^d q}{(2\pi)^d} \frac{t^{\mu \nu \lambda \rho}_{AVV} \left( q, v \right) + t^{\mu \nu \lambda \rho}_{AVV,0}\left( q, v \right)}{q^2 - \lambda_\gamma^2} \left( -g_{\mu \nu} + \left( 1-\xi \right) \frac{q_\mu q_\nu}{q^2 - \lambda_\gamma^2 \xi }\right) \right], 
\end{align}
where we introduced the nonperturbative QCD objects $t^{\mu \nu \lambda \rho}_{AVV}$ and $t^{\mu \nu \lambda \rho}_{AVV,0}$ as
\begin{align}\
    t^{\mu \nu \lambda \rho}_{AVV} \left(q, v\right) &=  \frac{\tau^c_{i j} S^\rho_{\sigma \sigma^\prime}}{48} P^{abcd} \int \mathrm{d}^d x \mathrm{d}^d y e^{i q \cdot x} \langle N (k, \sigma^\prime, j) | T \left[ \overline{q} \gamma^\lambda \gamma^5 \tau^d q(y)\, \overline{q} \gamma^\mu \tau^b q (x) \, \overline{q} \gamma^\nu \tau^a q(0) \right] | N ( k, \sigma, i) \rangle, \label{eq:tAVVdef} \\ 
    t^{\mu \nu \lambda \rho}_{AVV,\, 0} \left(q, v \right) &= \frac{\tau^c_{i j} S^\rho_{\sigma \sigma^\prime}}{48} \frac{1}{9} \int \mathrm{d}^d x \mathrm{d}^d y e^{i q \cdot x} \langle N (k, \sigma^\prime, j) | T \left[ \overline{q} \gamma^\lambda \gamma^5 \tau^c q(y)\, \overline{q} \gamma^\mu q (x) \, \overline{q} \gamma^\nu q(0) \right] | N ( k, \sigma, i) \rangle, \label{eq:tAVV0def}
\end{align}
which satisfy the Ward identities 
\begin{align}
    q_\mu q_\nu t^{\mu \nu \lambda \rho}_{AVV} \left( q, v \right) &= \, \frac{i}{2} q_\mu t^{\mu \lambda \rho}_{VA} \left( q, v \right) = \frac{i}{2} q_\mu t^{\mu \lambda \rho}_{VA} \left( -q, v \right)  =  - \frac{g_A^{(0)}}{2} \left( g^{\lambda \rho} - v^\lambda v^\rho \right), \label{eq:Ward_like_t_AVV} \\
    q_\nu t^{\mu \nu \lambda \rho}_{AVV,\, 0} \left( q, v \right) &=  0.
\end{align}
The above results imply 
\begin{align}
    W_A^{\rm LEFT} &= e^2 g_A^{(0)} \left( g_{23} -  \frac{1 - \xi}{2} \int \frac{i \mathrm{d}^d q}{(2\pi)^d}  \frac{1}{\left( q^2 - \lambda_\gamma^2 \right) \left( q^2 - \xi \lambda_\gamma^2 \right)}  \right) \nonumber \\
    &+ e^2 b_\rho b_\lambda \int \frac{i \mathrm{d}^d q}{(2\pi)^d} \frac{g_{\mu \nu} \left( t^{\mu \nu \lambda \rho}_{AVV} \left( q, v \right) + t^{\mu \nu \lambda \rho}_{AVV,0}\left( q, v \right) \right)}{q^2 - \lambda_\gamma^2}.
\end{align}
To avoid integration in $d$ dimensions, we add and subtract the OPE of the hadronic tensors $t^{\mu \nu \lambda \rho}_{AVV}$ and $t^{\mu \nu \lambda \rho}_{AVV,0}$ and define the subtracted objects $\overline{t}^{\mu \nu \lambda \rho}_{AVV}$ and $\overline{t}^{\mu \nu \lambda \rho}_{AVV,0}$ as
\begin{align}
    g_{\mu \nu}\overline{t}^{\mu \nu \lambda \rho}_{AVV} \left(q, v\right) &= g_{\mu \nu} t^{\mu \nu \lambda \rho}_{AVV} + g_A^{(0)} \left( g^{\lambda \rho} - v^\lambda v^\rho \right)   \frac{\left( 2 - d \right)^2}{2 d} \frac{1}{q^2-\mu_0^2},\label{eq:OPEtAVV} \\ 
    g_{\mu \nu} \overline{t}^{\mu \nu \lambda \rho}_{AVV,\, 0} \left(q, v \right) &= g_{\mu \nu} t^{\mu \nu \lambda \rho}_{AVV,0}.\label{eq:OPEtAVV0}
\end{align}
The resulting correlation function $W_A^{\rm LEFT}$ is expressed in terms of the momentum-space integrals as
\begin{align}\label{eq:WArenormalized}
    W_A^{\rm LEFT} &= e^2 g_A^{(0)} \left( g_{23} - \frac{1 - \xi}{2} \int \frac{i \mathrm{d}^d q}{(2\pi)^d}  \frac{1}{\left( q^2 - \lambda_\gamma^2 \right) \left( q^2 - \xi \lambda_\gamma^2 \right)} + \frac{\left( 2 - d \right)^2}{2 d}\int \frac{i \mathrm{d}^d q}{(2\pi)^d}  \frac{1}{\left( q^2 - \lambda_\gamma^2 \right) \left( q^2 - \mu_0^2 \right)} \right) \nonumber \\
    &+ e^2 b_\rho b_\lambda \int \frac{i \mathrm{d}^4 q}{(2\pi)^4} \frac{g_{\mu \nu} \left( \overline{t}^{\mu \nu \lambda \rho}_{AVV} \left( q, v \right) + \overline{t}^{\mu \nu \lambda \rho}_{AVV,0}\left( q, v \right) \right)}{q^2 - \lambda_\gamma^2}, 
\end{align}
or alternatively, after performing the integrals,
\begin{align}\label{eq:WALEFT}
    W_A^{\rm LEFT} &= \frac{e^2 g_A^{(0)}}{ \left( 4 \pi \right)^2} \left[ \frac{1 - \xi}{2} \left(1 + \ln \frac{\mu^2}{\lambda_\gamma^2} \right) + \frac{\xi}{2} \ln \xi - \frac{1}{2} \ln \frac{\mu^2}{\mu_0^2} + \frac{1}{4} \right]  \nonumber \\
    &+ e^2 b_\rho b_\lambda \int \frac{i \mathrm{d}^4 q}{(2\pi)^4} \frac{g_{\mu \nu} \left( \overline{t}^{\mu \nu \lambda \rho}_{AVV} \left( q, v \right) + \overline{t}^{\mu \nu \lambda \rho}_{AVV,0}\left( q, v \right) \right)}{q^2 - \lambda_\gamma^2}.
\end{align}
Equating Eqs.~\eqref{eq:coup22} and~\eqref{eq:WALEFT}, i.e., $W^{\rm HBChPT}_A=W^{\rm LEFT}_A$, we determine the relevant combination of electromagnetic LECs as
\begin{align}\label{eq:matchingAS3pt}
& \left( g^r_{1}+ g^r_2 + g^r_{13} + \frac{g^r_{11}}{2} \right) \left( \mu_\chi, \mu \right) = \nonumber \\
&\left. \left[  b_\rho b_\lambda \int \frac{i \mathrm{d}^4 q}{(2\pi)^4} \frac{g_{\mu \nu} \left( \overline{t}^{\mu \nu \lambda \rho}_{AVV} \left( q, v \right) + \overline{t}^{\mu \nu \lambda \rho}_{AVV,0}\left( q, v \right) \right)}{q^2 - \lambda_\gamma^2} + 2 i \pi  \delta(v\cdot r) \, \frac{g_A^{(0)} \Delta_{\rm em} m_N}{e^2} \right] \right|_{r_\lambda =0} \nonumber \\
&+\frac{Z_\pi g_A^{(0)}}{(4\pi)^2} \left( \left[ 1 + 3 \left( g_A^{(0)} \right)^2 \right] \left( 1 -  \ln \frac{\mu^2_\chi}{m_\pi^2}  \right) + 8 \pi m_\pi \left[ c_3 - c_4 - \frac{3}{8 m_N}- \frac{9}{16} \frac{\left( g_A^{(0)} \right)^2}{m_N} \right] \right)  \nonumber \\
&+ \frac{g_A^{(0)}}{(4\pi)^2} \left( 2 Z_\pi \left( g_A^{(0)} \right)^2 + \frac{5}{4} - \ln \frac{m^2_\pi}{\lambda^2_\gamma} - \frac{1}{2} \ln \frac{\mu^2}{\mu_0^2} + \frac{1 - \xi}{2} \left( 1 + \ln \frac{\mu^2}{\mu_\chi^2}\right)  +  \frac{3 + 2 \kappa_1}{3} \frac{\pi m_\pi}{m_N} \right).
\end{align}
As in the case of the correlation function $W_P$, the term in brackets in the second line of Eq.~\eqref{eq:matchingAS3pt} is finite in the limit $v\cdot r \rightarrow 0$. The matching at the level of the axial-vector coupling constant is expressed now in terms of the gauge-independent expression,
\begin{align}\label{eq:matchingAS}
&    \hat{C}_{A} \left( a, \mu_\chi, \mu \right) =  \frac{Z_\pi}{2} \left( \left[ 1 + 3 \left( g_A^{(0)} \right)^2 \right] \left( 1 -  \ln \frac{\mu^2_\chi}{m_\pi^2}  \right) + 2 \left( g_A^{(0)} \right)^2 + 8 \pi m_\pi\left[ c_3 - c_4 - \frac{3}{8 m_N} -  \frac{9}{16} \frac{\left( g_A^{(0)} \right)^2}{m_N}\right] \right) \nonumber \\
&+ \frac{1}{2}\left[\frac{1}{2} \ln \frac{\mu^2}{\mu^2_0} - \frac{3}{2} \ln \frac{\mu_\chi^2}{\mu^2}  - 4 B(a) + \frac{1}{4}  \right]  + \frac{16 \pi^2}{g^{(0)}_A} \int \frac{i \mathrm{d}^4 q}{\left( 2 \pi \right)^4}  \left( \frac{ \nu^2 - 2 Q^2}{3 Q^2} \frac{\overline{S}_1 (\nu, Q^2)}{Q^2} - \frac{\nu^2}{Q^2 } \frac{\overline{S}_2 (\nu, Q^2)}{m_N \nu} \right) \nonumber \\
& + \frac{8 \pi^2}{g_A^{(0)}} \left. \left[  b_\rho b_\lambda \int \frac{i \mathrm{d}^4 q}{(2\pi)^4} \frac{g_{\mu \nu} \left( \overline{t}^{\mu \nu \lambda \rho}_{AVV} \left( q, v \right) + \overline{t}^{\mu \nu \lambda \rho}_{AVV,0}\left( q, v \right) \right)}{q^2 - \lambda_\gamma^2} + 2 i \pi  \delta(v\cdot r) \, g_A^{(0)}\frac{\Delta_{\rm em} m_N}{e^2} \right] \right|_{r_\lambda =0} \nonumber \\ 
& + 8 \pi^2  \int \frac{i \mathrm{d}^4 q}{(2\pi)^4} \frac{2 \overline{m}{t}_{VP} \left(q, v \right) - {\overline t}_{VA} \left(q, v \right)}{ q^2 - \lambda_\gamma^2 }.
\end{align}
Note that the dependence on the nucleon magnetic moment and photon mass regulator cancels between two terms in the last line of Eq.~(\ref{eq:matchingAS}). Comparing the above expression to Eq.~\eqref{eq:matchingPS}, we deduce the following scale-independent relation between nonperturbative LEFT objects:
\begin{align}
& b_\rho b_\lambda \int \frac{i \mathrm{d}^4 q}{(2\pi)^4} \frac{g_{\mu \nu} \left( \overline{t}^{\mu \nu \lambda \rho}_{AVV} \left( q, v \right) + \overline{t}^{\mu \nu \lambda \rho}_{AVV,0}\left( q, v \right) \right)}{q^2 - \lambda_\gamma^2}  + \frac{g_A^{\left( 0 \right)}}{2 m_N} \int \frac{ \mathrm{d}^4 q}{(2\pi)^4} \frac{ g_{\mu \nu} \left( \tau^{\mu \nu}_{V V} \left(q, v \right) + \tau^{\mu \nu}_{V V, 0} \left(q, v \right) \right)}{q^2} + \frac{1}{2} \frac{g_A^{(0)}}{(4\pi)^2} \nonumber \\
&=\overline{m} b_\rho b_\lambda \frac{\partial}{\partial r_\lambda} \left. \left( \int \frac{i \mathrm{d}^4 q}{(2\pi)^4} \frac{g_{\mu \nu} \left( {t}^{\mu \nu \rho}_{PVV} \left( r, q, v \right) + {t}^{\mu \nu \rho}_{PVV,0}\left( r, q, v \right) \right)}{q^2 - \lambda_\gamma^2} \right) \right|_{r_\lambda = 0} + g_A^{(0)} \int \frac{i \mathrm{d}^4 q}{(2\pi)^4} \frac{{\overline t}_{VA} \left(q, v \right)}{q^2}. \label{eq:correlation_functions_matching}
\end{align}
In Appendix~\ref{app:Ward_three_point}, we derive the expression above by using the Ward identities.

The nonperturbative objects $W_P$ and $W_A$, which are defined in Eqs.~(\ref{eq:WP3PT_definition}) and~(\ref{eq:WA3PT_definition}), respectively, are symmetric w.r.t. the isospin of two vector spurions. Therefore, the nucleon pole terms of the three-point functions contain only the isoscalar component of the nucleon mass shift. In accordance with the analysis of Refs.~\cite{Brown:1969lsx,Sirlin:1977sv} and later developments in Refs.~\cite{Hayen:2020cxh,Seng:2024ker}, in Eqs.~(\ref{eq:matchingPS3pt}),~(\ref{eq:matchingPS}),~(\ref{eq:matchingAS3pt}), and~(\ref{eq:matchingAS}), the explicit nucleon-pole contribution from $\Delta_{\rm em} m_N$ cancels with the pole term implicit in the integrals of the LEFT three-point correlation functions $PVV$ and $AVV$.

\section{QED evolution of the axial-vector coupling constant between the baryon and pion-mass scales}
\label{sec4}

In this Section, we present the QED evolution of the HBChPT axial-vector coupling constant. After the nonperturbative input needed for the matching has been determined, the renormalization group equations can be used to evolve $\hat C_A$ from the hadronic scale, $\mu\sim m_N$, to low energies, $\mu\sim q_{\rm ext}$. The QED evolution of the LEC $\hat{C}_A\left( \mu_\chi \right)$ from Eq.~(\ref{eq:axial_contribution_LECs_v1}) for scales $\mu_\chi \lesssim m_\pi$ is governed by the same renormalization group equations (RGEs) as the evolution of $g_V$ in Ref.~\cite{Cirigliano:2023fnz}. Exploiting the results of this work and Ref.~\cite{Cirigliano:2022hob}, we account for virtual pion fields in the QED running between baryon and pion-mass scales and generalize the RGEs of Ref.~\cite{Cirigliano:2023fnz} by including the leading- and next-to-leading order HBChPT contribution to the QED anomalous dimension $\tilde{\gamma}^\pi_0$. The renormalization group equation for $\hat{C}_A\left( \mu_\chi \right)$ can be expressed as
\begin{subequations} \label{eq:RGE_LE}
\begin{align}
    \mu_\chi \frac{\mathrm{d} \hat{C}_A\left( \mu_\chi \right)}{\mathrm{d} \mu_\chi} &=  \gamma \left( \alpha \right),  \\
    \gamma (\alpha) &= \left( \tilde{\gamma}^\pi_0 + \tilde{\gamma}_0 \right) \, \frac{\alpha}{\pi} + \tilde{\gamma}_1 \left( \frac{\alpha}{\pi} \right)^2 \  +  \   \cdots, \\
    \tilde{\gamma}^\pi_0 &= - Z_\pi \frac{1 + 3 \left( g^{(0)}_A \right)^2}{2},  \\
    \tilde{\gamma}_0 &= -\frac{3}{4},  \\
    \tilde{\gamma}_1 &=  \frac{5 \tilde{n}}{24} + \frac{5}{32} - \frac{\pi^2}{6},
\end{align}
\end{subequations}
with the effective number of particles $\tilde n$, as described in Appendix A.2 of Ref.~\cite{Cirigliano:2023fnz}, and QED anomalous dimensions from Ref.~\cite{Cirigliano:2023fnz} and references therein. The change of the axial-vector coupling $g_A$ from the HBChPT running associated with $\tilde{\gamma}^\pi_0$ exceeds the effects of pure QED running by a factor larger than $3$.

\section{Discussion}
\label{sec:discussion}

In this Section, we provide a summary of our results (Subsection~\ref{sec:subsec51}), a comparison with the recent approach presented in Ref.~\cite{Seng:2024ker} (Subsection~\ref{sec:subsec52}), and a version of our results that is more suitable for interfacing with nonperturbative methods, such as lattice-QCD or dispersive calculations (Subsection~\ref{sec:subsec53}).

\subsection{Summary of results}
\label{sec:subsec51}

In this paper, we have derived two equivalent representations for the LECs that shift the axial-vector coupling constant to $O(\alpha)$, both in terms of QCD correlation functions of quark currents. The first representation splits $\hat C_A$, the LEC correction to $g_A$, see Eqs.~(\ref{eq:axial_coupling_expansion})-(\ref{eq:axial_contribution_LECs_v1}), into the sum of gauge-independent contributions from two-point and three-point correlation functions as
\begin{align}\label{eq:splitPS2}
\hat C_A &=  8 \pi^2 \left[ -\frac{X_6}{2} + 2 \left(A_1 + A_2 + A_3 + A_4 \right) + \frac{g_{11}+ g_{12}}{4 g_A^{(0)}}  \right] + \frac{8 \pi^2}{g_A^{(0)}} \left( g_1 + g_2 + g_{13} + \frac{g_{11} - g_{12}}{4} \right),
\end{align}
and relates the three-point contribution, the second term in Eq.~(\ref{eq:splitPS2}), to the correlators of pseudoscalar and two vector quark bilinears $t_{PVV}$ with the resulting expression for the axial-vector coupling constant in Eq.~(\ref{eq:matchingPS}).

In the alternative determination, we split $\hat C_A$ into the gauge-dependent contributions from electromagnetic and electroweak HBChPT LECs as
\begin{align}\label{eq:splitAS2}
\hat C_A &=  8 \pi^2 \left[ -\frac{X_6}{2} + 2 \left(A_1 + A_2 + A_3 + A_4 \right) \right] + \frac{8 \pi^2}{g_A^{(0)}} \left( g_1 + g_2 + g_{13} + \frac{g_{11}}{2} \right),
\end{align}
and relate the second term in Eq.~(\ref{eq:splitAS2}) to the QCD correlation functions of the axial-vector and two vector currents $t_{AVV}$ in Eq.~(\ref{eq:matchingAS}). The two representations are shown to be equivalent by using the Ward identities.

Besides these main results, we clarified the amplitude decomposition for the vector-axial-vector two-point correlation functions and derived a new sum rule for the corresponding nucleon amplitudes in Appendix~\ref{app:relations}, we provided explicitly the low-energy behavior of all the relevant hadronic objects in Appendix~\ref{app:IR}, and derived and verified Ward identities between two- and three-point correlation functions in Appendix~\ref{app:Ward}. Along the way, we have obtained representations for the following combinations of the HBChPT LECs: $A_1 + A_2$, $A_3 + A_4$, $g_{11}+g_{12}$, $g_1 + g_2 + g_{13} + \frac{g_{11}}{2}$, $f_1 + f_3$, and $f_4$, as well as for all next-to-leading order HBChPT LECs in Appendix~\ref{app:NLOChPT}.

\subsection{Comparison to the approach of Ref.~\cite{Seng:2024ker}}
\label{sec:subsec52}

In this Subsection, we compare our work with the recent analysis of radiative corrections to the axial-vector coupling performed in Ref.~\cite{Seng:2024ker}.

Ref.~\cite{Seng:2024ker} uses a hybrid approach that copes both with effective field theory, HBChPT, and with a traditional ``Sirlin representation." The calculation of HBChPT diagrams is therefore very similar to our approach, while an obvious difference is that Ref.~\cite{Seng:2024ker} does not match onto the LEFT. Their expressions at quark level are always rendered finite due to the $W$-boson propagator that appears within the full Standard Model. In contrast, the matching to the LEFT in this paper allows us to consistently resum leading and next-to-leading large logarithms of the ratio between the electroweak and hadronic scales, e.g., the ratio $M_W/m_N$. In our paper, the matching of LEFT to HBChPT is performed at the level of derivatives of the generating functional w.r.t. the spurion fields, while the matching in Ref.~\cite{Seng:2024ker} is performed by equating the HBChPT expression for the matrix element to the Sirlin representation.

Apart from the ability to resum large logarithms, the two approaches lead to several differences in the representations for the two- and three-point functions. Our representation involving the ${AVV}$ correlator in Eq.~\eqref{eq:matchingAS} agrees with Ref.~\cite{Seng:2024ker} on the required nonperturbative input in the three-point function that contributes to the combination of HBChPT LECs $g_1+g_2+g_{13}+g_{11}/2$, but we find an additional term proportional to $t_{VA}$ in the two-point function. In both approaches, one can show that such $VA$ contributions cancel between two- and three-point functions {\it if} one chooses to rewrite the $AVV$ correlator in terms of $PVV$ and $VA$ correlators by using the Ward identity, see Eq.~\eqref{eq:correlation_functions_matching} or Refs.~\cite{Hayen:2020cxh,Seng:2024ker} for the analog in the traditional approach. One thus has a choice on whether to represent the LECs in terms of a $PVV$ correlator, in which case the contribution from the $t_{VA}$ correlator cancels,  or in terms of $AVV$ and $VA$ correlators, see Eqs.~\eqref{eq:matchingPS} and~\eqref{eq:matchingAS}, respectively. The discrepancy with Ref.~\cite{Seng:2024ker} results from their assumption that the $VA$ contributions cancel,\footnote{This cancellation is correct only for the loop part of $\Gamma_{VA}|^{\rm HBChPT}$ in the Feynman-`t Hooft gauge~\cite{Seng:2024ker}.} which is inconsistent with their use of the $AVV$ representation. We discuss the comparison to Ref.~\cite{Seng:2024ker} in more detail below.

\subsubsection{Three-point function}
\label{sec:subsubsec521}

Setting our representation for $g_1+g_2+g_{13}+g_{11}/2$ (the combination of electromagnetic LECs that enters the three-point function) equal to the representation of the same quantity  given in Ref.~\cite{Seng:2024ker} leads to (in the Feynman-`t Hooft gauge, i.e., $\xi=1$),
\begin{align}\label{eq:CY3pt}
W^{\rm LEFT}_A &\supset e^2 b_\rho b_\lambda \int\frac{i \mathrm{d}^dq}{\left(2\pi \right)^d}\frac{g_{\mu\nu} \left( t_{AVV}^{\mu\nu\lambda\rho}\left(q,v \right)+t_{AVV,0}^{\mu\nu\lambda\rho}\left(q,v \right) \right)}{q^2-\lambda_\gamma^2} + 2 i \pi  \delta(v\cdot r) g_A^{(0)} \Delta_{\rm em} m_N\nonumber\\
&\stackrel{\rm ?}{=} b_\rho b_\lambda  S^\rho_{\sigma\sigma^\prime}\left[\langle p \left(\sigma^\prime \right)|J_A^\lambda |n \left(\sigma \right)\rangle_{\rm QCD+QED}-\langle p \left(\sigma' \right)|J_A^\lambda |n \left( \sigma \right)\rangle_{\rm QCD}\right],
\end{align}
where $J_A^\lambda = \overline{q} \gamma^\lambda \gamma_5\tau^+q$ and ``QCD(+QED)" denotes the matrix element without(with) QED effects. The first line results from the expressions obtained here, while the second arises in the approach of Ref.~\cite{Seng:2024ker}. One can check that the two sides are equivalent by making use of the isospin invariance. The representation obtained in this work is proportional to the correlators in Eqs.~\eqref{eq:tAVVdef} and~\eqref{eq:tAVV0def}, with the isospin structure inside the correlator $t_{AVV}^{\mu\nu\lambda\rho}\left(q,v \right)$
\begin{align}
 \tau^c_{ij} \langle N (k, \sigma^\prime, j) | T \left[ \overline{q} \gamma^\lambda \gamma^5 \tau^d q(y)\, \overline{q} \gamma^\mu \tau^b q (x) \, \overline{q} \gamma^\nu \tau^a q(0) \right] | N ( k, \sigma, i) \rangle &=c_1^{\mu\nu\lambda} \delta ^{ab}\delta ^{cd}+c_2^{\mu\nu\lambda} \delta ^{ac}\delta ^{bd}+c_3^{\mu\nu\lambda} \delta ^{ad}\delta ^{bc}.
\end{align}
As this equation is contracted with $P^{abcd}$ in $t_{AVV}^{\mu\nu\lambda \rho}$, the contributions from $c_{2,3}^{\mu\nu\lambda}$ terms vanish and only $c_1^{\mu\nu\lambda}$ contributes with a factor $6$. This allows us to project onto the neutron and proton states that are used in the representation of Ref.~\cite{Seng:2024ker}. Up to the nucleon-pole contribution, the first line in Eq.~\eqref{eq:CY3pt} is given by
\begin{align}
W^{\rm LEFT}_A \supset \frac{e^2}{2} b_\rho b_\lambda S^\rho_{\sigma\sigma'} \int\frac{i \mathrm{d}^dq}{(2\pi)^d}\int \mathrm{d}^d x \mathrm{d}^d y e^{i q \cdot x}  \frac{ g_{\mu\nu} \langle p \left( \sigma^\prime \right)|  T\left[ J_A^\lambda \left(y \right) \left( J^\mu_3 \left(x \right) J^\nu_3 \left(0 \right) + J^\mu_0 \left(x \right) J^\nu_0 \left(0 \right) \right)  \right] |n(\sigma )\rangle}{q^2-\lambda_\gamma^2},
\end{align}
where $J_3^\mu = \overline{q} \gamma^\mu \tau^3 q/2$ and $J_0^\mu = \overline{q} \gamma^\mu q/6$ are the isovector and isoscalar components of the electromagnetic current $J_{\rm em}^\mu$: $J_{\rm em}^\mu = J_3^\mu + J_0^\mu$. As possible mixed terms $\sim J_0\times J_3$ do not contribute due to their time-reversal properties, the above is proportional to $T \left[ J_A^\lambda \left(y \right) J_{\rm em}^\mu \left(x \right) J_{\rm em}^\nu \left(0 \right)  \right]$. This implies that the first line in Eq.~\eqref{eq:CY3pt} captures the leading, $\mathcal{O} \left( e^2 \right)$, QED contributions to the matrix element of the axial-vector current, and Eq.~\eqref{eq:CY3pt} is satisfied implying an agreement with Ref.~\cite{Seng:2024ker} on the nonperturbative inputs in the determination of $g_1+g_2+g_{13}+g_{11}/2$.

\subsubsection{Two-point function}
\label{sec:subsubsec522}

Besides the determination of the three-point function, Ref.~\cite{Seng:2024ker} provides a representation for the combination of LECs that enters the two-point function of Eq.~\eqref{eq:split}. In our conventions, this combination of LECs corresponds to
\begin{align}
A_{\rm 2pt}^{\rm \left(Ref.~[1] \right)} &= -\frac{X^r_6}{2} + \frac{\tilde X^r_3}{g^{(0)}_A}, \\
A_{\rm 2pt}^{\rm \left(this~work \right)}&=-\frac{X^r_6}{2}+2\left(A^r_1+A^r_2+A^r_3+A^r_4\right)+\frac{C^r_\beta-1}{e^2}.
\end{align}
Given the correspondence between different notations for LECs, we should have $A_{\rm 2pt}^{\rm (this~work)}=A_{\rm 2pt}^{\rm (Ref.~[1])}$. However, using Eqs.~\eqref{eq:a1a2_matching_result_no_IR} and~\eqref{eq:A34S12}, along with the fact that the appearing combination of the ${S}_{1,2}$ structure functions match onto $\Box_{\gamma W}^A$ of Ref.~\cite{Seng:2024ker} for $\mu_0\simeq M_W$, we find
\begin{align}
A_{\rm 2pt}^{\rm \left(this~work\right)}&= \frac{1}{\left(4\pi \right)^2}\left[3\ln \frac{M_Z}{\mu_\chi}+\ln\frac{M_Z}{M_W}-\frac{5}{4}\right]+\frac{1}{e^2}\Box_{\gamma W}^A+\left[\frac{1}{\left(4\pi \right)^2}\left(\frac{1}{4}+\ln\frac{\mu}{M_W}\right)- \int\frac{i \mathrm{d}^4 q}{\left(2\pi \right)^4}\frac{\overline{t}_{VA}}{q^2}\right]\nonumber\\
&\neq \nonumber\\
A_{\rm 2pt}^{\rm \left(Ref.~[1]\right)}&= \frac{1}{\left( 4\pi \right)^2}\left[ 3\ln \frac{M_Z}{\mu_\chi}+\ln\frac{M_Z}{M_W}-\frac{5}{4}+\tilde a_g\right]+\frac{1}{e^2}\Box_{\gamma W}^{A}\,. \label{eq:comp1}
\end{align} 
Neglecting the QCD correction, $\tilde a_g$, the two representations match up to the last term in square brackets in the first line of Eq.~\eqref{eq:comp1}. Crucially, our representation of the LECs involves an additional nonperturbative object, $\overline{t}_{VA}$, which needs to be modeled or computed on the lattice. In terms of LECs, this additional correlation function is related to $g_{11}+g_{12}$, see e.g.\ Eq.\ \eqref{eq:g11_g12_couplings_clean}. Thus, while the structure functions $ S_{1,2}$ are the only nonperturbative input needed for $\left[\hat C_A\right]_{\rm 2pt}\sim A_{\rm 2pt} - \frac{C^r_\beta-1}{e^2} + \frac{g^r_{11}+g^r_{12}}{4g_A^{(0)}}$, see Eqs.\ \eqref{eq:A1A2}-\eqref{eq:CA2pt}, the combination of LECs in $A_{\rm 2pt}$ involves the additional nonperturbative object $\overline{t}_{VA}$.

\subsubsection{Mapping to Sirlin's representation}
\label{sec:subsubsec523}

As discussed above, the missing $t_{VA}$ contribution would cancel with terms arising from the three-point function if one chooses to represent it in terms of the matrix element of $\partial \cdot J_A$ instead of $J_A$ itself. The cancellation can be seen after rewriting $g_{\mu \nu} \langle p|J^\lambda_A J^\mu_V J^\nu_V|n\rangle = - g_{\mu \nu} \overline{q}_\kappa \frac{\partial}{\partial \overline{q}_\lambda} \langle p|J_A^\kappa J^\mu_V J^\nu_V|n\rangle + g_{\mu \nu} \frac{\partial}{\partial \overline{q}_\lambda} \left(\overline{q}_\kappa \langle p|J^\kappa_A J^\mu_V J^\nu_V|n\rangle \right)$, where $\bar q$ is the momentum injected in the axial-vector current. Now taking $\overline{q}$ to be of order of the neutron decay kinematics or below it, one can neglect the kinematically suppressed first term and simplify the second by using the Ward identities. The remainder is a three-point function of the form $g_{\mu \nu} \frac{\partial}{\partial \overline{q}_\lambda} \langle p|\partial\cdot J_A  J^\mu_V J^\nu_V|n\rangle$ together with two-point functions. The latter, which arise from use of the Ward identities, then cancel terms with $t_{VA}$ in the combination of two- and three-point functions. This cancellation of the two-point correlator, $t_{VA}$, is similar to our representation of LECs in terms of $t_{PVV}$, which does not depend on $t_{VA}$ as can be seen from Eq.~\eqref{eq:matchingPS}. 

Indeed, we can perform a similar comparison of the two results for $C^r_\beta \left( 1+\frac{\alpha}{2\pi}\hat{C}_A \right)$, but using the $PVV$ representation of Eq.~\eqref{eq:matchingPS} instead of the one for $AVV$. In doing so we find (using $\mu_0\simeq M_W$) that the difference between the two results now resides purely in the three-point function, for which we would obtain
\begin{align}\label{eq:PVVcomparison}
\frac{e^2 g_A^{(0)}}{8\pi^2} \left[\hat{C}_A\right]_{\rm 3pt}&\stackrel{\rm this~work}{=} 
e^2 \overline{m} b_\rho b_\lambda \frac{\partial}{\partial r_\lambda} \left. \left( \int\frac{i \mathrm{d}^dq}{(2\pi)^d}\frac{g_{\mu\nu}  \left( {t}_{PVV}^{\mu\nu\rho}\left( r,q,v \right)+{t}_{PVV,0}^{\mu\nu\rho}\left( r,q,v \right) \right)}{q^2-\lambda_\gamma^2} \right) \right|_{r_\lambda = 0} \nonumber\\
&+ 2 i \pi  \delta(v\cdot r) g_A^{(0)} \Delta_{\rm em} m_N+\dots\,,
\end{align}
where the second line subtracts the pole contribution. Instead, the three-point function obtained in Ref.\cite{Seng:2024ker} can be written as
\begin{align}\label{eq:PVVcomparisonCY}
\frac{e^2 g_A^{(0)}}{8\pi^2} \left[\hat{C}_A\right]_{\rm 3pt}&\stackrel{\rm Ref.~\left[1 \right]}{=}b_\rho b_\lambda S^\rho_{\sigma\sigma^\prime}\left[\langle p \left(\sigma^\prime \right)|J_A^\lambda |n \left(\sigma \right)\rangle_{\rm QCD+QED}-\langle p \left(\sigma^\prime \right)|J_A^\lambda |n \left(\sigma \right)\rangle_{\rm QCD}\right]+\dots\,.
\end{align}
After canceling terms that are common to both approaches, the only difference is the mismatch between the explicitly shown terms in Eqs.\ \eqref{eq:PVVcomparison} and \eqref{eq:PVVcomparisonCY}.

The matrix element involving $PVV$ in the first line of Eq.~\eqref{eq:PVVcomparison} can be written in a form that  more closely resembles the matrix element $g_{\mu \nu} \frac{\partial}{\partial \overline{q}_\lambda} \langle p|\partial\cdot J_A  J^\mu_V J^\nu_V|n\rangle$  that appears in the traditional approach~\cite{Hayen:2020cxh,Seng:2024ker}. In the following, we use $i \overline{m}t_{PVV}^{\mu\nu\rho}= t^{\mu \nu  \rho}_{\partial A VV}$ and similar relation for the isoscalar combination with
\begin{align}
    t^{\mu \nu  \rho}_{\partial A VV} \left(q, v\right) &=  \frac{\tau^c_{i j} S^\rho_{\sigma \sigma^\prime}}{48} P^{abcd} \int \mathrm{d}^d x \mathrm{d}^d y e^{ir\cdot(x-y)+i q \cdot x} \nonumber \\
    &\langle N (k, \sigma^\prime, j) | T \left[\partial_\lambda \left[\overline{q} \gamma^\lambda \gamma^5 \tau^d q(y) \right]\, \overline{q} \gamma^\mu \tau^b q (x) \, \overline{q} \gamma^\nu \tau^a q(0) \right] | N ( k, \sigma, i) \rangle, \label{eq:derivative_AVV} \\ 
    t^{\mu \nu  \rho}_{\partial A VV,\, 0} \left(q, v \right) &= \frac{\tau^c_{i j} S^\rho_{\sigma \sigma^\prime}}{48} \frac{1}{9} \int \mathrm{d}^d x \mathrm{d}^d y e^{ir\cdot(x-y)+i q \cdot x} \nonumber \\
    &\langle N (k, \sigma^\prime, j) | T \left[ \partial_\lambda \left[\overline{q} \gamma^\lambda \gamma^5 \tau^c q(y) \right]\, \overline{q} \gamma^\mu q (x) \, \overline{q} \gamma^\nu q(0) \right] | N ( k, \sigma, i) \rangle. \label{eq:derivative_AVV0} 
\end{align}
The first line in Eq.~\eqref{eq:PVVcomparison} can be written in a form similar to Refs.~\cite{Hayen:2020cxh,Seng:2024ker}, after introducing an additional UV regulator $\frac{M_W^2}{M_W^2-q^2}$, as
\begin{align} \label{eq:PVVway_D}
W^{\rm LEFT}_P \supset e^2 b_\rho b_\lambda \frac{\partial}{\partial r_\lambda} \left. \left( \int\frac{ \mathrm{d}^dq}{(2\pi)^d}\frac{ g_{\mu\nu}  \left( {t}_{\partial A VV}^{\mu\nu\rho}\left( r,q,v \right)+{t}_{\partial A VV,0}^{\mu\nu\rho} \left( r,q,v \right) \right)}{q^2-\lambda_\gamma^2} \right) \right|_{r_\lambda = 0} \simeq - i  b_\rho b_\lambda S_{\sigma\sigma^\prime}^\rho \left. \frac{\partial D_{\sigma^\prime \sigma}}{\partial \overline{q}_\lambda} \right|_{ \overline{q}_\lambda \to 0} \,,
\end{align}
with the nonperturbative object $D_{\sigma^\prime \sigma}$:
\begin{align}
D_{\sigma^\prime \sigma} &= \frac{e^2}{2}\int\frac{ i\mathrm{d}^dq}{(2\pi)^d}\frac{M_W^2}{M_W^2-q^2}\frac{g_{\mu \nu}}{q^2-\lambda_\gamma^2}\int \mathrm{d}^d x \mathrm{d}^d y e^{i \overline{q} \cdot x+i q\cdot y}\langle p \left(\sigma^\prime \right)|T\left[ \partial \cdot J_A \left(x \right) J^\mu_{\rm em} \left(y \right)J^\nu_{\rm em} \left(0 \right)\right]|n \left(\sigma \right)\rangle.
\end{align}
This evaluation becomes equivalent to the forms used in Refs.~\cite{Hayen:2020cxh,Seng:2024ker} if we specify $\overline q = p_n - p_p $ as a difference of the neutron and proton momenta $p_n$ and $p_p$, respectively. Consequently, evaluating the derivative of $D_{\sigma^\prime \sigma}$ in Eq.~(\ref{eq:PVVway_D}) and the nonperturbative object $t_{VP}$ is equivalent to the evaluation of Eq.~(\ref{eq:CY3pt}) and the objects $t_{VA}$ and $t_{VP}$. To directly compare Eq.~(\ref{eq:PVVway_D}) with Ref.~\cite{Seng:2024ker}, we take the limit $\overline{q} \to 0$ and  subtract the isoscalar contribution from the nucleon poles. The difference to Ref.~\cite{Seng:2024ker} comes from the missing $t_{VA}$ contribution and a difference in the HBChPT recoil terms between $W_A$ and $W_P$, which can be expressed in terms of the $\tau^{\mu \nu}_{VV}$ and $\tau^{\mu \nu}_{VV,0}$ nonperturbative objects in LEFT.

\subsection{Connection to lattice QCD}
\label{sec:subsec53}

In this Subsection, we recast the results of our work in a form more suitable to interface with the nonperturbative input from lattice QCD and phenomenological methods.

We find it convenient to split $\hat{C}_A$ as in Eq.~\eqref{eq:split}: $\hat{C}_A = \left[\hat C_A \right]_{\rm 2pt} +\left[\hat C_A \right]_{\rm 3pt}$, into $ \left[\hat C_A \right]_{\rm 2pt}$, which only depends on one-nucleon matrix elements of two quark bilinears, and $\left[\hat C_A \right]_{\rm 3pt}$, which involves one-nucleon matrix elements with three quark currents. As we already remarked, $\left[\hat C_A\right]_{\rm 2pt}$ and $\left[\hat C_A\right]_{\rm 3pt}$ are separately gauge-independent. $\left[\hat C_A \right]_{\rm 2pt}$ depends on the LEFT renormalization scale $\mu$ and on the parameter $a$ specifying the renormalization scheme in such a way as to compensate the dependence of the LEFT matching coefficient $C_{\beta}^r$. $\left[\hat C_A \right]_{\rm 3pt}$, on the other hand, is scale and scheme independent. The representation for $\left[\hat C_A \right]_{\rm 2pt}$ was derived in Section~\ref{sec:subsec34}, and it depends on the ``axial" $\gamma W$ box diagram
\begin{equation}
    \left[ \hat{C}_A\right]_{\rm 2 pt} = \frac{1}{2} \left[ \frac{1}{2} \ln \frac{\mu^2}{\mu_0^2} - \frac{3}{2}\ln \frac{\mu^2_\chi}{\mu^2} - \frac{5}{4} - 4 B(a) \right]  + \frac{8\pi^2}{e^2} \overline{\Box}_{\rm Had}^{A},
\end{equation}
where we defined the ``subtracted" hadronic contribution $\overline{\Box}_{\rm Had}^{A}$ as 
\begin{equation}
     \overline{\Box}_{\rm Had}^{A} = e^2 \frac{2}{g^{(0)}_A} \int \frac{i \mathrm{d}^4 q}{\left( 2 \pi \right)^4}  \left( \frac{ \nu^2 - 2 Q^2}{3 Q^2} \frac{\overline{S}_1 (\nu, Q^2)}{Q^2} - \frac{\nu^2}{Q^2 } \frac{\overline{S}_2 (\nu, Q^2)}{m_N \nu} \right).
\end{equation}
The difference between $\left[\hat C_A \right]_{\rm 2pt}$ and the correction to the vector coupling $\hat C_V$ given in Ref.~\cite{Cirigliano:2023fnz} is scale and scheme independent. Up to higher-order corrections in $\alpha_s$,
\begin{equation} \label{eq:axial_vector_difference}
  \frac{\alpha}{2\pi} \left( \left[\hat C_A\right]_{\rm 2pt} - \hat C_V \right) = \overline{\Box}_{\rm Had}^{A} - \overline{\Box}_{\rm Had}^{V}.
\end{equation}  
The nonperturbative input for $\overline{\Box}_{\rm Had}^{A}$ is well determined from the measurement of the proton and neutron spin structure functions $g_{1}$ and $g_{2}$~\cite{Gorchtein:2021fce}. A lattice calculation of $\overline{\Box}_{\rm Had}^{A}$ as well as other physical quantities that are based on the nucleon two-point functions would thus mainly serve as a validation of the methods for the evaluation of $\overline{\Box}_{\rm Had}^{V}$~\cite{Feng:2020zdc,Ma:2023kfr}. At the order in $\alpha_s$ we are working at, the difference $\overline{\Box}_{\rm Had}^{A} - \overline{\Box}_{\rm Had}^{V}$ is equal to the difference of traditional axial and vector $\gamma W$ boxes, ${\Box}^{A}_{\gamma W} - \Box^V_{\gamma W}$. We can thus use the extraction of Ref. \cite{Gorchtein:2021fce}. Interestingly, the first data-driven evaluation finds the result for Eq.~(\ref{eq:axial_vector_difference}) to be very close to zero: $\overline{\Box}_{\rm Had}^{A} - \overline{\Box}_{\rm Had}^{V} = 0.13(11)_V(6)_A \times 10^{-3}$~\cite{Gorchtein:2021fce}.

The representations for $\hat C_A$, cf. Eqs.~\eqref{eq:matchingPS} and~\eqref{eq:matchingAS}, involve integrals of the nonperturbative nucleon matrix elements of two- and three-point functions that run over all momenta. The low-momentum (long-distance) contribution of the integrals is sensitive to infrared physics (through $m_\pi$ and $\lambda_\gamma$) that is already captured in the HBChPT loops. Since the LECs cannot depend on infrared (IR) scales, the IR sensitivity in the integrals is canceled by the explicit $m_\pi$-dependent terms in  Eqs.~\eqref{eq:matchingPS} and~\eqref{eq:matchingAS}. While this is perfectly fine, the final expression obscures the fact that the nonperturbative input in $\left[\hat C_A\right]_{\rm 3pt}$ comes from the intermediate momentum region of the integrals in Eqs.~\eqref{eq:matchingPS} and~\eqref{eq:matchingAS}. Therefore, a lattice-QCD calculation of the current matrix elements will have the largest impact on the intermediate momentum region. To make this point more explicit, we introduce an intermediate scale $\Lambda_1$ such that $m_\pi \ll \Lambda_1 \ll m_N$. We then split the integrals in two regions $Q^2 < \Lambda_1^2$ and $Q^2 > \Lambda_1^2$ and use the polology expressions from Appendix~\ref{app:IR} for the evaluation of the long-distance region of the momentum-space integrals, i.e., $Q^2 \le \Lambda_1^2$. The region $Q^2 \ge \Lambda_1^2$ can be accessed with future phenomenological evaluations or lattice-QCD calculations, as it is realized for the hadronic contributions to the nucleon vector coupling constant~\cite{Marciano:2005ec,Seng:2018qru,Czarnecki:2019mwq,Hayen:2020cxh,Shiells:2020fqp,Seng:2020wjq,Yoo:2022lmt,Ma:2023kfr}. The expression for $\left[\hat{C}_{A} \right]_{\rm 3pt}$  involving the $PVV$ correlation functions, see Eqs.~\eqref{eq:matchingPS3pt} and~\eqref{eq:matchingPS}, results in
\begin{align}\label{eq:matchingPSptlattice_3pt}
&\left[\hat{C}_{A} \right]_{\rm 3 pt} \left( \mu_\chi\right) = \frac{1}{2}
+ \frac{Z_\pi}{2} \left( \left[ 1 + 3 \left( g_A^{(0)} \right)^2 \right] \ln \frac{\Lambda_1^2}{\mu^2_\chi} +  2\left( g_A^{(0)} \right)^2   + \frac{32}{3} \Lambda_1\left[ c_3 - c_4 - \frac{3}{8 m_N} -  \frac{9}{16} \frac{\left( g_A^{(0)} \right)^2}{m_N}\right] \right)  \nonumber \\
&+ \frac{8 \pi^2}{g_A^{(0)}} \overline{m} b_\rho b_\lambda \frac{\partial}{\partial r_\lambda} \left. \left( \int \frac{i \mathrm{d}^4 q}{(2\pi)^4} \frac{g_{\mu \nu} \left( {t}^{\mu \nu \rho}_{PVV} \left( r, q, v \right) + {t}^{\mu \nu \rho}_{PVV,0}\left( r, q, v \right) \right) - 2 g_A^{(0)} {t}_{VP} \left(q, v \right) r^{\rho} }{q^2} \Theta \left( Q^2 - \Lambda_1^2\right) \right) \right|_{r_\lambda = 0}\nonumber \\
&-  8 \pi^2 \left[ \frac{1}{2 m_N} + i \pi \delta \left( v \cdot r \right) \right] \int \frac{ \mathrm{d}^4 q}{(2\pi)^4} \frac{ g_{\mu \nu} \left( \tau^{\mu \nu}_{V V} \left(q, v \right) + \tau^{\mu \nu}_{VV,0} \left(q, v \right) \right)}{q^2} \Theta \left( Q^2 - \Lambda_1^2\right) + R_A \left( \frac{m_\pi}{\Lambda_1}, \frac{\Lambda_1}{m_N} \right),
\end{align}
where $R_A \left( m_\pi/\Lambda_1, \Lambda_1/m_N \right)$ is a function that goes to zero as both its arguments tend to zero:
\begin{align}
&   R_A \left( \frac{m_\pi}{\Lambda_1}, \frac{\Lambda_1}{m_N} \right) = \frac{Z_\pi}{2} \left[ 1 + 3 \left( g_A^{(0)} \right)^2 \right] \left( \ln \left( 1 + \frac{m_\pi^2}{\Lambda_1^2} \right)  + \frac{m_\pi^2}{m_\pi^2 + \Lambda_1^2}     \right) \nonumber \\
&+ 8 Z_\pi m_\pi  \left[ c_3 - c_4 - \frac{3}{8 m_N} -  \frac{9}{16} \frac{\left( g_A^{(0)} \right)^2}{m_N}\right] \left( \frac{\pi}{2} - \arctan\frac{\Lambda_1}{m_\pi}  + \frac{1}{3} \frac{m_\pi \Lambda_1}{m_\pi^2 + \Lambda_1^2} \right) + \mathcal{O} \left( \frac{\Lambda_1^2}{m_N^2} \right).
\end{align}
The representation for $\left[ \hat C_A \right]_{\rm 3pt}$ involving the $AVV$ correlation functions, see Eq.~\eqref{eq:matchingAS}, is obtained by replacing the nonperturbative inputs, besides the function $R_A$, in the second and third lines in Eq.~\eqref{eq:matchingPSptlattice_3pt} with
\begin{align}
&\frac{8 \pi^2}{g_A^{(0)}} b_\rho b_\lambda \int \frac{i \mathrm{d}^4 q}{(2\pi)^4} \frac{g_{\mu \nu} \left( \overline{t}^{\mu \nu \lambda \rho}_{AVV} \left( q, v \right) + \overline{t}^{\mu \nu \lambda \rho}_{AVV,0}\left( q, v \right) \right) - g_A^{(0)} g^{\lambda \rho} \left( 2 \overline{m}{t}_{VP} \left(q, v \right) - {\overline t}_{VA} \left(q, v \right) \right) }{q^2} \Theta \left( Q^2 - \Lambda_1^2\right) \nonumber \\
&-  8 \pi^3 \delta \left( v \cdot r \right) \int \frac{ i \mathrm{d}^4 q}{(2\pi)^4} \frac{ g_{\mu \nu} \left( \tau^{\mu \nu}_{V V} \left(q, v \right) + \tau^{\mu \nu}_{VV,0} \left(q, v \right) \right)}{q^2} \Theta \left( Q^2 - \Lambda_1^2\right),
\end{align}
and by replacing $1/2$ with $3/4$ in the constant term in the first line of Eq.~\eqref{eq:matchingPSptlattice_3pt}.

Alternatively, the contribution to $\left[ \hat C_A \right]_{\rm 3pt}$ could be evaluated by lattice-QCD simulations with dynamical QED, as it is suggested in Ref.~\cite{Seng:2024ker}. Several formulations of lattice QCD+QED exist, with different approaches to deal with the complications stemming from the long-range nature of photon interactions. QED$_{\rm M}$~\cite{Endres:2015gda}, which works with a massive photon, is perhaps the closest approach to the formalism of this paper. As shown in Eq.~\eqref{eq:CY3pt}, we can identify the shift to the axial-vector form factor at zero squared momentum transfer, in the presence of QED and QCD, as
\begin{align}\label{eq:gAQED}
\Delta g_A^{\rm QCD + QED}(\xi, \lambda_\gamma) &= e^2 b_\rho b_\lambda \int\frac{i \mathrm{d}^dq}{\left(2\pi \right)^d} \frac{\left( t_{AVV}^{\mu\nu\lambda\rho}\left(q,v \right)+t_{AVV,0}^{\mu\nu\lambda\rho}\left(q,v \right) \right)}{q^2-\lambda_\gamma^2} \left(g_{\mu\nu} - (1-\xi) \frac{q_\mu q_\nu}{q^2 - \xi \lambda_\gamma^2} \right) \nonumber \\
& + 2 i \pi  \delta(v\cdot r) g_A^{(0)} \Delta_{\rm em} m_N,
\end{align}
where  in the first line we inject nonzero energy $v\cdot r$ in the axial-vector current (see discussion following~\eqref{eq:WA3PT_definition}). In general, $\Delta g_A^{\rm QCD + QED} \equiv  g_A^{\rm QCD + QED}  - g_A^{\rm QCD}$ depends on the gauge and on the IR regulator. 
We also notice that the r.h.s. of Eq.~\eqref{eq:gAQED} is UV divergent, and the divergence is canceled by the LEFT counterterm $g_{23}$, see Eq.~\eqref{eq:WArenormalized}. Therefore, $\Delta g_A^{\rm QCD + QED}$ needs to be calculated in a given nonperturbative renormalization scheme, and then matched to the $\overline{\rm{MS}}$ scheme. For the sake of illustration, we define a renormalized $\Delta \overline{g}_A^{\rm QCD + QED}$ by using the OPE-subtracted axial-vector three-point function
\begin{align}\label{eq:gAQEDr}
\Delta \overline{g}_A^{\rm QCD + QED} \left( \xi=1, \lambda_\gamma, \mu_0 \right) \equiv e^2 b_\rho b_\lambda \int\frac{i \mathrm{d}^dq}{\left(2\pi \right)^d} \frac{g_{\mu\nu} \left( \overline{t}_{AVV}^{\mu\nu\lambda\rho}\left(q,v \right)+\overline{t}_{AVV,0}^{\mu\nu\lambda\rho}\left(q,v \right) \right)}{q^2-\lambda_\gamma^2} + 2 i \pi  \delta(v\cdot r) g_A^{(0)} \Delta_{\rm em} m_N.
\end{align}
With this definition,
\begin{align}
& \left[\hat C_A\right]_{\rm 3pt} 
= \frac{Z_\pi}{2} \left( \left[ 1 + 3 \left( g_A^{(0)} \right)^2 \right] \left( 1 -  \ln \frac{\mu^2_\chi}{m_\pi^2}  \right) + 2 \left( g_A^{(0)} \right)^2 + 8 \pi m_\pi \left[ c_3 - c_4 - \frac{3}{8 m_N} -  \frac{9}{16} \frac{\left( g_A^{(0)} \right)^2}{m_N}\right] \right) \nonumber\\
&+ \frac{8\pi^2}{e^2 g_A^{(0)}}   \Delta\overline{g}_A^{\rm QCD + QED} \left(\xi=1, \lambda_\gamma, \mu_0 \right) + 8 \pi^2  \int \frac{i \mathrm{d}^4 q}{(2\pi)^4} \frac{2 \overline{m}{t}_{VP} \left(q, v \right) - {\overline t}_{VA} \left(q, v \right)}{ q^2 - \lambda_\gamma^2 }.
\label{eq:gAQEDr2}
\end{align}
In this scheme, $\Delta\overline g_A$ depends on the subtraction scale $\mu_0$ but $\left[ \hat C_A \right]_{\rm 3pt}$ does not, as the $\mu_0$ dependence cancels between $\Delta\overline g_A$ and $\overline{t}_{VA}$. We conclude by reiterating that \eqref{eq:gAQEDr} and \eqref{eq:gAQEDr2} are valid in our specific renormalization scheme and need to be generalized to a scheme suitable for lattice calculations, such as RI-MOM~\cite{Martinelli:1994ty,Sturm:2009kb}.

\section{Conclusions and Outlook}
\label{sec6}

In this paper, we extended the systematically improvable effective field theory framework for radiative corrections to charged-current processes to the determination of the nucleon axial-vector coupling  $g_A$. The matching of the Standard Model to the LEFT with effective four-fermion interactions and running from the electroweak scale down to the hadronic scale is encoded in the same LEFT coupling constant $C^r_\beta \left( a, \mu \right)$ that enters the determination of the HBChPT vector coupling  $g_V$. We performed the matching to the heavy-baryon chiral perturbation theory in two different ways: by connection to the vector-vector-axial-vector three-point correlation functions and by connection to the pseudoscalar-vector-vector three-point correlation functions. We verified the consistency of our calculations between these two objects through the Ward identities. As a result, we expressed the nucleon axial-vector coupling in terms of the correlation functions of quark currents that can be evaluated in lattice QCD or with dispersive methods and hadronic models. As discussed in Section \ref{sec:subsec53}, these expressions are amenable to both lattice-QCD computations that use perturbative insertions of quark bilinears as well as those using dynamical QED. We pointed out the difference in the QED running of the axial-vector coupling constant $g_A$ compared to the running of the vector coupling constant $g_V$. Compared to previous formulations, we 
(1) add a missing nonperturbative contribution from the $VA$ correlation functions,
(2) present three-point and two-point QCD correlation functions, which are not always discussed in the first-principles determinations of the axial-vector coupling constant, and 
(3) provide a first determination for physically relevant combinations of electromagnetic and electroweak as well as all next-to-leading order HBChPT LECs.

\section*{Acknowledgments}
We thank C.-Y.~Seng for several insightful discussions and for comments on the manuscript. We acknowledge conversations with T.~Bhattacharya, J.~de~Vries, R.~Gupta, L.~Hayen, M.~Hoferichter, M.~Wagman, and A.~Walker-Loud and thank them for their feedback on this work. V.C. and W.D. acknowledge support by the U.S. DOE under Grant No. DE-FG02-00ER41132. E.M. and O.T. acknowledge support by the US Department of Energy through the Los Alamos National Laboratory and by LANL’s Laboratory Directed Research and Development (LDRD/PRD) program under projects 20210968PRD4 and 20210190ER. Los Alamos National Laboratory is operated by Triad National Security, LLC, for the National Nuclear Security Administration of U.S. Department of Energy (Contract No. 89233218CNA000001). We acknowledge support from the DOE Topical Collaboration ``Nuclear Theory for New Physics,'' award No.\ DE-SC0023663. For facilitating portions of this research, O.T. wishes to acknowledge the Center for Theoretical Underground Physics and Related Areas (CETUP*), The Institute for Underground Science at Sanford Underground Research Facility (SURF), and the South Dakota Science and Technology Authority for hospitality and financial support, and the Theory Division at Fermilab for hospitality and financial support. Wolfram Mathematica~\cite{Mathematica} was extremely useful in this work.

\appendix

\section{Invariant amplitude decomposition for the nucleon two-point correlation functions}
\label{app:relations}

In this Appendix, we present the invariant amplitude decomposition for the two-point correlation functions of the axial-vector and vector currents and discuss the crossing properties of the invariant amplitudes. We derive novel relations between two-point invariant amplitudes and a new sum rule for these amplitudes. We also present expressions for the correlation functions of the pseudoscalar-vector and axial-vector-vector currents.

Exploiting the tensor decomposition for the part of $t^{\mu \nu \rho}_{V A}$ that is symmetric in $\mu\leftrightarrow \nu$ in terms of the amplitudes $A_1,~A_2,~A_4,~A_5,~B_1,$ and $B_2$~\cite{Ji:1993ey}
\begin{align}
- i t^{\mu \nu \rho}_{V A, S} &= \left( - g^{\mu \nu} + \frac{q^\mu q^\nu}{q^2}\right) q^\rho A_1 (\nu, Q^2) + \left( v^\mu - \frac{\nu}{q^2} q^\mu \right) \left(  v^\nu - \frac{\nu}{q^2} q^\nu \right) q^\rho A_2 (\nu, Q^2) + \frac{q^\mu q^\nu}{m_N^2} q^\rho A_4 (\nu, Q^2) \nonumber \\
&+ \frac{v^\mu q^\nu + v^\nu q^\mu}{2 m_N} q^\rho A_5 (\nu, Q^2) + \frac{m_N}{2} \left[ \left( g^{\mu \rho} - \frac{q^\mu q^\rho}{q^2} \right) v^\nu + \left( g^{\nu \rho} - \frac{q^\nu q^\rho}{q^2} \right) v^\mu  \right] B_2 (\nu, Q^2) \nonumber \\
&+ \frac{m_N}{2} \left[ \left( g^{\mu \rho} - \frac{q^\mu q^\rho}{q^2} \right) \left( v^\nu  - \frac{\nu q^\nu}{q^2}\right) + \left( g^{\nu \rho} - \frac{q^\nu q^\rho}{q^2} \right) \left( v^\mu  - \frac{\nu q^\mu}{q^2}\right)  \right] B_1 (\nu, Q^2), \label{eq:tVA_decomposition}
\end{align}
we express the hadronic contribution from $\overline{t}_{VA}$ and ${t}_{VP}$ in terms of the nucleon two-point amplitudes by making a Lorentz contraction:
\begin{align}
& b_\rho \int \frac{\mathrm{d}^4 q}{(2\pi)^4} \frac{\overline{e} \gamma_\mu \slashed{q} \gamma_\nu P_L \nu_e {\overline t}^{\mu \nu \rho}_{VA,S} \left(q, v \right)}{ q^2 \left( q^2 - \lambda_\gamma^2 \right)} = \frac{1}{g^{(0)}_A} \int \frac{i \mathrm{d}^4 q}{\left( 2 \pi \right)^4} \frac{ \left( \nu^2 + Q^2 \right) \overline{A}_1 (\nu, Q^2) + m_N \nu \overline{B}_2 (\nu, Q^2)}{ Q^2 \left( Q^2 + \lambda_\gamma^2 \right)} \nonumber \\
&- \frac{1}{g^{(0)}_A} \frac{1}{d-1} \int \frac{i \mathrm{d}^4 q}{\left( 2 \pi \right)^4} \frac{ \nu^2 + Q^2}{ Q^2 \left( Q^2 + \lambda_\gamma^2 \right)} \left[ \frac{m_N \nu}{Q^2} \overline{B}_1 (\nu, Q^2) + \frac{Q^2}{m^2_N} \overline{A}_4 (\nu, Q^2) + \frac{\nu^2 + Q^2}{Q^2} \overline{A}_2 (\nu, Q^2) \right] \nonumber \\
&=2 \overline{m} \int \frac{i \mathrm{d}^4 q}{(2\pi)^4} \frac{{t}_{VP} \left(q, v \right)}{ q^2 - \lambda_\gamma^2 } - \int \frac{i \mathrm{d}^4 q}{(2\pi)^4} \frac{{\overline t}_{VA} \left(q, v \right) }{ q^2} -  \int \frac{i \mathrm{d}^4 q}{(2\pi)^4} \frac{2}{q^2 \left( q^2 - \lambda^2_\gamma \right)} \left( 1 - \frac{q^2}{q^2-\mu^2_0} \right),
\end{align}
with resulting relations
\begin{align}
g_A^{(0)} {\overline t}_{VA} \left(q, v \right) &= \frac{\nu^2 + Q^2}{Q^2} \left( \overline{A}_1 \left(\nu, Q^2\right) - \frac{1}{d-1} \left[ \frac{\nu^2 + Q^2}{Q^2} \overline{A}_2 \left(\nu, Q^2\right) - \frac{Q^2}{m^2_N} \overline{A}_4 \left(\nu, Q^2\right) + \frac{\nu}{m_N} \overline{A}_5 \left(\nu, Q^2\right) \right] \right) \nonumber \\
&- \frac{1}{d-1} \frac{\nu^2 + Q^2}{Q^2} \frac{m_N \nu}{Q^2} \left( \overline{B}_1 \left(\nu, Q^2\right) + \overline{B}_2 \left(\nu, Q^2\right) \right), \\
2 \overline{m} g_A^{(0)} {t}_{VP} \left(q, v \right) &= - \frac{1}{d - 1} \frac{\nu^2 + Q^2}{Q^2} \left[ \frac{\nu}{m_N} \overline{A}_5 \left(\nu, Q^2\right) - 2 \frac{Q^2}{m^2_N} \overline{A}_4 \left(\nu, Q^2\right) + \frac{m_N \nu}{Q^2} \overline{B}_2 \left(\nu, Q^2\right) \right] \nonumber \\
& - \frac{m_N \nu}{Q^2}\overline{B}_2 \left(\nu, Q^2\right) + \frac{2 g_A^{(0)}}{q^2} \left( 1 - \frac{q^2}{q^2-\mu^2_0} \right). \label{eq:PV_symmetric}
\end{align}
The invariant amplitudes have the following crossing properties:
\begin{align}
    A_{1,5} \left( - \nu, Q^2 \right) &= A_{1,5} \left( \nu, Q^2 \right), \\
    A_{2,4} \left( - \nu, Q^2 \right) &= - A_{2,4} \left( \nu, Q^2 \right), \\
    B_{1,2} \left( - \nu, Q^2 \right) &= - B_{1,2} \left( \nu, Q^2 \right).   
\end{align}
The amplitudes $A_4,~A_5,$ and $B_2$ vanish in the massless quark limit. The OPE-subtracted expressions for the amplitudes in $t^{\mu \nu \rho}_{V A, S}$ are given by
\begin{align}
\overline{A}_1 \left( \nu, Q^2 \right) &= {A}_1 \left( \nu, Q^2 \right) - \frac{g^{(0)}_A}{Q^2 + \mu_0^2}, \\
\overline{A}_{2, 4, 5} \left( \nu, Q^2 \right) &= {A}_{2, 4, 5} \left( \nu, Q^2 \right), \\
\overline{B}_1 \left( \nu, Q^2 \right) &= {B}_1 \left( \nu, Q^2 \right) - \frac{2 g^{(0)}_A}{m_N \nu} \frac{Q^2}{Q^2+\mu_0^2}, \\
\overline{B}_2 \left( \nu, Q^2 \right) &= {B}_2 \left( \nu, Q^2 \right) + \frac{2 g^{(0)}_A}{m_N \nu} \frac{Q^2}{Q^2+\mu_0^2}.
\end{align}

The $VA$ correlation functions enter physical processes with the coupling of a $W^\pm$ boson and photon to the nucleon line and appear linearly as an integrand in interference terms of virtual radiative corrections to charged-current neutrino-, hadron-, and quark-induced interactions with nucleons and nuclei, and quadratically in the same processes with radiation of the hard photon. The symmetric piece of the tensor $t^{\mu \nu \rho}_{VA}$ satisfies the following identity:
\begin{align} \label{eq:symmetry_identity_t_VA}
  \int \frac{\mathrm{d}^d q}{(2\pi)^4} \frac{\overline{e} q_\mu \gamma_\nu P_L \nu_e}{q^2 \left( q^2 - \lambda^2_\gamma \right)} t^{\mu \nu \rho}_{VA,S} \left(q, v \right) = \int \frac{\mathrm{d}^d q}{(2\pi)^4} \frac{\overline{e} q_\nu \gamma_\mu P_L \nu_e}{q^2 \left( q^2 - \lambda^2_\gamma \right)} t^{\mu \nu \rho}_{VA, S} \left(q, v \right).
\end{align}
Comparing this relation with the Ward identity on the left side in Eqs.~(\ref{eq:Ward_like_t_VA}), we see that $t^{\mu \nu \rho}_{VA}$ must have an antisymmetric part unless $t_{VP}=0$, which is in contradiction with nonvanishing infrared behavior from Appendix~\ref{app:IR}. To preserve the nonvanishing infrared behavior of the integrand with $t_{VP}$, we introduce the antisymmetric part for the $VA$ correlator $t^{\mu \nu \rho}_{VA, A}$: $t^{\mu \nu \rho}_{VA} = t^{\mu \nu \rho}_{VA, S} + t^{\mu \nu \rho}_{VA, A}$ by writing the most general representation,\footnote{We omit the term proportional to $\varepsilon^{\mu \nu \alpha \beta}$, that does not contribute neither to Eq.~(\ref{eq:symmetry_identity_t_VA}) nor to the scalar hadronic amplitudes $t_{VA}$ and $t_{VP}$.}
\begin{align}
- i t^{\mu \nu \rho}_{VA, A} \left(q, v \right) &= \frac{v^\mu q^\nu - v^\nu q^\mu}{2 m_N}  q^\rho C_5 \left( \nu, Q^2 \right)  + \frac{m_N}{2} \left[ \left( g^{\mu \rho} - \frac{q^\mu q^\rho}{q^2} \right) v^\nu - \left( g^{\nu \rho} - \frac{q^\nu q^\rho}{q^2} \right) v^\mu  \right] D_2 (\nu, Q^2) \nonumber \\
&+ \frac{m_N}{2} \left[ \left( g^{\mu \rho} - \frac{q^\mu q^\rho}{q^2} \right) \left( v^\nu  - \frac{\nu q^\nu}{q^2}\right) - \left( g^{\nu \rho} - \frac{q^\nu q^\rho}{q^2} \right) \left( v^\mu  - \frac{\nu q^\mu}{q^2}\right)  \right] D_1 (\nu, Q^2). \label{eq:tVA_decomposition_antisymmetric}
\end{align}
The crossing properties for $C_5,~D_1,$ and $D_2$ are opposite to the crossing properties of $A_5,~B_1,$ and $B_2$, respectively. The Ward identity on the left side in Eqs.~(\ref{eq:Ward_like_t_VA}) together with decomposition of Eqs.~\eqref{eq:tVA_decomposition} and~(\ref{eq:tVA_decomposition_antisymmetric}) completely determine the hadronic object ${t}_{VP}$:
\begin{align}
2 \overline{m} g_A^{(0)} {t}_{VP} \left(q, v \right) &=  \frac{1}{d-1} \frac{\nu^2 + Q^2}{Q^2} \left[ \frac{\nu}{m_N} C_5 \left(\nu, Q^2\right) - \frac{m_N \nu}{Q^2} D_2 \left(\nu, Q^2\right) \right] - \frac{m_N \nu}{Q^2} D_2 \left(\nu, Q^2\right). \label{eq:PV_antisymmetric}
\end{align}

Contracting $ t^{\mu \nu \rho}_{VA}$ with two momenta, i.e., $q_\mu q_\nu \left( t^{\mu \nu \rho}_{VA, S} + t^{\mu \nu \rho}_{VA, A} \right)$, we obtain a relation between the amplitudes $A_4$ and $A_5$:
\begin{align}
\frac{Q^2}{m^2_N} A_4 \left( \nu, Q^2 \right) - \frac{ \nu}{m_N} A_5 \left( \nu, Q^2 \right) = - \frac{g_A^{\left( 0 \right)}}{Q^2}.
\end{align}

Contracting $ t^{\mu \nu \rho}_{VA}$ with one momentum, i.e., $q_\mu \left( t^{\mu \nu \rho}_{VA, S} + t^{\mu \nu \rho}_{VA, A} \right)$ or $q_\nu \left( t^{\mu \nu \rho}_{VA, S} + t^{\mu \nu \rho}_{VA, A} \right)$, we obtain three more relations between the amplitudes $A_4,~A_5$, $B_2,~D_2$, and $C_5$:\footnote{Model-independent relations for the Born contributions to the $VV$ correlator were derived in Ref.~\cite{Bjorken:1966jh} by exploiting a similar idea.}
\begin{align}
\frac{Q^2}{m^2_N} A_4 \left( \nu, Q^2 \right) &=  \frac{1}{2} \frac{ \nu}{m_N} \left[ A_5 \left( \nu, Q^2 \right) + C_5 \left( \nu, Q^2 \right) \right] + \frac{1}{2} \frac{m_N \nu}{Q^2} \left[ B_2 \left( \nu, Q^2 \right) - D_2 \left( \nu, Q^2 \right) \right], \\
B_2 \left( \nu, Q^2 \right) - D_2 \left( \nu, Q^2 \right) &= - \frac{2 g_A^{\left( 0 \right)}}{m_N \nu}, \\
A_5 \left( \nu, Q^2 \right) - C_5 \left( \nu, Q^2 \right) &= 0,
\end{align}
in agreement with the leading power of the OPE. As a result, we derived three independent equations. Exploiting these results, we find agreement for the hadronic object ${t}_{VP}$ expressed in terms of symmetric amplitudes in Eq.~(\ref{eq:PV_symmetric}) and antisymmetric amplitudes in Eq.~(\ref{eq:PV_antisymmetric}).

Accounting for the derived relations, the $VA$ hadronic object can be expressed as
\begin{align}
- i \left( t^{\mu \nu \rho}_{VA, S} + t^{\mu \nu \rho}_{VA, A} \right) &= \left( - g^{\mu \nu} + \frac{q^\mu q^\nu}{q^2}\right) q^\rho A_1 (\nu, Q^2) + \left( v^\mu - \frac{\nu}{q^2} q^\mu \right) \left(  v^\nu - \frac{\nu}{q^2} q^\nu \right) q^\rho A_2 (\nu, Q^2) \nonumber \\
&+ \frac{m_N}{2} \left[ \left( g^{\mu \rho} - \frac{q^\mu q^\rho}{q^2} \right) \left( v^\nu  - \frac{\nu q^\nu}{q^2}\right) + \left( g^{\nu \rho} - \frac{q^\nu q^\rho}{q^2} \right) \left( v^\mu  - \frac{\nu q^\mu}{q^2}\right)  \right] B_1 (\nu, Q^2) \nonumber \\
&+ \frac{m_N}{2} \left[ \left( g^{\mu \rho} - \frac{q^\mu q^\rho}{q^2} \right) \left( v^\nu  - \frac{\nu q^\nu}{q^2}\right) - \left( g^{\nu \rho} - \frac{q^\nu q^\rho}{q^2} \right) \left( v^\mu  - \frac{\nu q^\mu}{q^2}\right)  \right] D_1 (\nu, Q^2)\nonumber \\
& + \frac{g_A^{(0)}}{v \cdot q} \left[ \left( g^{\mu \rho} - \frac{q^\mu q^\rho}{q^2} \right) v^\nu - \left( g^{\nu \rho} - \frac{q^\nu q^\rho}{q^2} \right) v^\mu  \right] - g_A^{(0)} \frac{q^\mu q^\nu q^\rho}{q^2 q^2} \nonumber \\
& + m_N\left( g^{\mu \rho} - \frac{q^\mu q^\rho}{q^2} \right) v^\nu B_2 (\nu, Q^2) + \left( v^\mu -  \frac{\nu q^\mu}{q^2} \right) \frac{q^\nu}{m_N} q^\rho A_5 (\nu, Q^2). \label{eq:tVA_decomposition_corrected}
\end{align}

Moreover, by evaluating the integral with ${t}_{VP}$ in Eqs.~(\ref{eq:a1a2_match_result}),~(\ref{eq:g11_g12_couplings}), and~(\ref{eq:CA2ptmatch})
\begin{align}
   2 \overline{m} \int \frac{i \mathrm{d}^4 q}{4 \pi^2} \frac{{t}_{VP} \left(q, v \right)}{ q^2 - \lambda_\gamma^2 } = \ln \frac{\lambda_\gamma^2}{m^2_\pi} + \frac{3 + 2 \kappa_1}{3} \frac{\pi m_\pi}{m_N} + \mathcal{O} \left( \frac{m^2_\pi}{m_N^2}, \frac{m^2_\pi}{\left( 4 \pi F_\pi \right)^2} \right), 
\end{align}
we provide new sum rules for the combination of the structure functions $C_5$ and $D_2$, cf. Eq.~(\ref{eq:PV_antisymmetric}), or, equivalently, for the combination of the structure functions $A_5$ and $B_2$, cf. Eq.~(\ref{eq:PV_symmetric}),
\begin{align}
&\frac{1}{3 g_A^{(0)}}  \frac{2}{i \pi} \int \frac{\sqrt{Q^2 + \nu^2} \mathrm{d} \nu \mathrm{d} Q^2}{ Q^2 + \lambda_\gamma^2} \frac{\nu}{m_N} \left( \frac{\nu^2 + Q^2}{Q^2} C_5 \left(\nu, Q^2\right) - \frac{\nu^2 + 4 Q^2}{Q^2} \frac{m_N^2}{Q^2} D_2 \left(\nu, Q^2\right) \right) \nonumber \\
&= \ln \frac{\lambda_\gamma^2}{m^2_\pi} + \frac{3 + 2 \kappa_1}{3} \frac{\pi m_\pi}{m_N} + \mathcal{O} \left( \frac{m^2_\pi}{m_N^2}, \frac{m^2_\pi}{\left( 4 \pi F_\pi \right)^2} \right), \\
&\frac{1}{3 g_A^{(0)}}  \frac{2}{i \pi} \int \frac{\sqrt{Q^2 + \nu^2} \mathrm{d} \nu \mathrm{d} Q^2}{ Q^2 + \lambda_\gamma^2} \frac{\nu}{m_N} \left( \frac{\nu^2 + Q^2}{Q^2} {A}_5 \left(\nu, Q^2\right) - \frac{\nu^2 + 4 Q^2}{Q^2} \frac{m_N^2}{Q^2} \left[ {B}_2 \left(\nu, Q^2\right ) + \frac{2 g_A^{(0)}}{m_N \nu}\right] \right) \nonumber \\
&= \ln \frac{\lambda_\gamma^2}{m^2_\pi} +\frac{3 + 2 \kappa_1}{3} \frac{\pi m_\pi}{m_N} + \mathcal{O} \left( \frac{m^2_\pi}{m_N^2}, \frac{m^2_\pi}{\left( 4 \pi F_\pi \right)^2} \right).
\end{align}

The explicit infrared contributions, cf. Appendix~\ref{app:IR} for the derivation of the IR behavior of the $VA$ amplitudes, are given by
\begin{align}
{A}^\mathrm{IR}_1 \left(\nu, Q^2\right) &= \frac{\nu^2 + Q^2}{Q^2} {A}^\mathrm{IR}_2 \left(\nu, Q^2\right) , \\
{B}^\mathrm{IR}_2 \left(\nu, Q^2\right) &= {D}^\mathrm{IR}_2 \left(\nu, Q^2\right) = - 2 {B}^\mathrm{IR}_1 \left(\nu, Q^2\right)= - 2 {D}^\mathrm{IR}_1 \left(\nu, Q^2\right) , \\
{C}^\mathrm{IR}_5 \left(\nu, Q^2\right) &= - \frac{m_N}{\nu} \frac{g_A^{(0)}}{Q^2 + m^2_\pi} \left[ 1 -  \nu^2  \frac{3+2\kappa_1}{4 m_N} i \pi \delta \left( v \cdot q \right) \right], \\
{B}^\mathrm{IR}_1 \left(\nu, Q^2\right) &= \frac{Q^2}{m_N \nu} \frac{g_A^{(0)}}{Q^2 + m^2_\pi} \left[ 1 - \left( \nu^2 + Q^2 \right) \frac{3+2\kappa_1}{4 m_N} i \pi \delta \left( v \cdot q \right) \right].
\end{align}
Defining the IR-subtracted amplitudes $\tilde{C}_5 \left(\nu, Q^2\right)$ and $\tilde{D}_2 \left(\nu, Q^2\right)$
\begin{align}
\tilde{C}_5 \left(\nu, Q^2\right) &= {C}_5 \left(\nu, Q^2\right) - {C}^\mathrm{IR}_5 \left(\nu, Q^2\right) , \\
\tilde{D}_2 \left(\nu, Q^2\right) &= {D}_2 \left(\nu, Q^2\right) - {D}^\mathrm{IR}_2 \left(\nu, Q^2\right),
\end{align}
we obtain the  sum rule
\begin{align}
&  \int \frac{\sqrt{Q^2 + \nu^2} \mathrm{d} \nu \mathrm{d} Q^2}{ Q^2} \frac{\nu}{m_N} \left( \frac{\nu^2 + Q^2}{Q^2}  \tilde{C}_5 \left(\nu, Q^2\right) - \frac{\nu^2 + 4 Q^2}{Q^2} \frac{m_N^2}{Q^2} \tilde{D}_2 \left(\nu, Q^2\right) \right) = \mathcal{O} \left( \frac{m^2_\pi}{m_N^2}, \frac{m^2_\pi}{\left( 4 \pi F_\pi \right)^2} \right).
\end{align}

\section{Infrared properties of correlation functions in LEFT}
\label{app:IR}

In this Appendix, we investigate the infrared properties of all correlation functions encountered in the main text. We discuss the dependence on the pion mass $m_\pi$ and the photon mass regulator $\lambda_\gamma$. The results collected here are essential in proving that our representation leads to LECs that do not depend on $m_\pi$ and $\lambda_\gamma$.

With the matching relations, we express the HBChPT LECs in terms of correlators in LEFT and loops in HBChPT. From the independence of LECs on $m_\pi$, any appearance of $m_\pi$ should cancel between the HBChPT loops and the LEFT correlators. Although factors and logarithms of $m_\pi$ are explicit in the HBChPT loops, the same dependence is not as easily seen from the expressions of the LEFT correlators. In this Appendix, we derive the $m_\pi$ dependence of the LEFT correlators and verify the expected cancellations explicitly.

To isolate the infrared behavior, we study various correlation functions using the methods presented for example in Chapter 10 of Ref.~\cite{Weinberg:1995mt}, which we loosely refer to as ``polology." These methods allow one to determine the behavior of correlation functions near their singularities, such as single-particle poles and multiparticle cuts. In particular, any correlation function in  momentum space of the form
\begin{align}
G(q_1, \dots, q_n)=\int_{x_1\dots x_n}e^{-i( q_1\cdot x_1+\dots + q_n\cdot x_n)  }\langle 0|T\left[ O_1(x_1)\dots O_n(x_n)\right]|0\rangle\,,
\end{align}
with $\int_{x_1\dots x_n} = \int \mathrm{d}^d x_1 \dots \mathrm{d}^d x_n$, has a pole at $Q^2\to M^2$, with $Q^\mu\equiv q_{r+1}^\mu+\dots + q_n^\mu$, if there is a state $|X(\vec Q)\rangle$ with mass $M$ that has nonvanishing matrix elements $\langle 0|T\left[ O_1(0)\dots O_r(x_r)\right]|X(\vec Q)\rangle$ and $\langle X(\vec Q)| T\left[ O_{r+1}(0)\dots O_n(x_n)\right]|0\rangle$. In the vicinity of this pole, the correlation function $G$ can be approximated as
\begin{align}\label{eq:polology}
G(q_1, \dots, q_n)&\underset{Q^2 \to M^2}{\longrightarrow} \frac{2i Q_0 (2\pi)^d \delta^d (q_1+\dots + q_r+Q)}{Q^2-M^2} \int_{x_2\dots x_r}e^{-i( q_2\cdot x_2+\dots + q_r\cdot x_r)  }\langle 0|T\left[ O_1(0)\dots O_r(x_r)\right]|X(\vec Q)\rangle\nonumber\\
&\times \int_{x_{r+2}\dots x_n}e^{-i( q_{r+2}\cdot x_{r+2}+\dots q_n\cdot x_n)  }\langle X(\vec Q)|T\left[ O_{r+1}(0)\dots O_n(x_n)\right]|0\rangle+\dots \,,
\end{align}
with $Q_0 = \sqrt{\vec Q^2+M^2}$.

\subsection{Two-point functions}
\label{app:a1}

Starting with the two-point functions, we find that the HBChPT loops in Eqs.~\eqref{eq:a1a2_match_result} and~(\ref{eq:g11_g12_couplings}) depend on $m_\pi$. This dependence enters the matching relation in Eq.~\eqref{eq:CA2ptmatch}. To obtain $m_\pi$-independent HBChPT LECs, we used the infrared behavior of ${t}_{VP}$ given in Eq.~\eqref{eq:tPVIR}. Here, we provide a proof of this result. We rewrite the $VP$ correlator in the vicinity of the pion pole in terms of a vacuum matrix element as follows,
\begin{align}
t^{\mu \rho}_{V P} \left(q, v \right) &\to \lim_{k^{\prime}\to k} \int \frac{\mathrm{d}^d q^\prime}{\left(2\pi\right)^d} \frac{\varepsilon^{abc} \tau^c_{ij} S^\rho_{\sigma \sigma^\prime}}{6} \frac{i}{4} \int \mathrm{d}^d x \mathrm{d}^d y \mathrm{d}^d z \mathrm{d}^d x^\prime e^{ i q \cdot x-ik \cdot y+ik'\cdot z + i q^\prime \cdot x^\prime} \nonumber\\
& \times i \left( k^2 - m_N^2 \right) i \left( k^{\prime \,2} - m_N^2 \right) \langle 0 | T \left[ \overline{N}_{i,\sigma }(y) N_{j,\sigma' }(z)\overline{q} \gamma^\mu \tau^b q (x) \, \overline{q} \gamma_5 \tau^a q(x^\prime) \right] | 0 \rangle \,,
\end{align}
where $\overline{N}$ and $N$ are interpolating fields for the nucleons. To establish the $m_\pi$ dependence, we identify pion poles by employing Eq.~\eqref{eq:polology} with the pion intermediate state $|X(\vec Q)\rangle = |\pi^d(\vec Q)\rangle $ with $Q = q - k + k^\prime$. The following contribution to the time-ordered product gives rise to a pion pole,
\begin{align}
t^{\mu \rho}_{V P} \left(q, v \right) &\to \lim_{k^2,k^{\prime2}\to m_N^2} \frac{\varepsilon^{abc} \tau^c_{ij} S^\rho_{\sigma \sigma^\prime}}{6} \frac{i}{4}\frac{2i Q_0}{Q^2-m_\pi^2} i \left( k^2 - m_N^2 \right) i \left( k^{\prime \,2} - m_N^2 \right) \nonumber \\
&\times\int \mathrm{d}^d y \mathrm{d}^d z e^{ -ik \cdot y+ik'\cdot z} \langle 0 | T \left[ \overline{N}_{i,\sigma }(y) N_{j,\sigma' }(z)\overline{q} \gamma^\mu \tau^b q (0) \right]|\pi^d(\vec Q)\rangle\, \langle \pi^d (\vec Q)|\overline{q} \gamma_5 \tau^a q(0)  | 0 \rangle \nonumber\\
&=\frac{\varepsilon^{abc} \tau^c_{ij} S^\rho_{\sigma \sigma^\prime}}{6} \frac{i}{4}\frac{2i q_0}{q^2-m_\pi^2}  \langle N \left(k,\sigma',j \right) | \overline{q} \gamma^\mu \tau^b q (0)|N \left(k,\sigma, i \right)\pi^d(\vec q)\rangle\, \langle \pi^d (\vec q)|\overline{q} \gamma_5 \tau^a q(0)  | 0 \rangle\,.
\end{align}
To compare the above expression with the correlator in the heavy-baryon chiral perturbation theory, we identify the above vacuum to $\pi$ and $\pi \to \pi N$ matrix elements with their representation implied by chiral symmetry (which in practice can be read off the HBChPT Lagrangian~\cite{Scherer:2002tk})
\begin{align}
t^{\mu \rho}_{V P} \to \frac{ig_A^{(0)}}{q^2-m_\pi^2} {B_0} & \left[ \left(v^\mu v^\rho-g^{\mu\rho} \right) \left( 1  - \vec q^2 \frac{1+\kappa_1}{2 m_N} i \pi \delta \left( v \cdot q \right) \right) + \overline{q}^\mu \overline{q}^\rho \frac{\kappa_1}{2 m_N} i \pi \delta \left( v \cdot q \right) \right. \nonumber \\
&\left. - \left( v^\mu v^\rho - \frac{v^\mu q^\rho}{v \cdot q} \right) \left( 1 + \frac{\vec{q}^2}{2 m_N v \cdot q} \right) \right]\,, \label{eq:tPVpolology}
\end{align}
where $\overline{q}^\mu = q^\mu-(v\cdot q) v^\mu $. Note that we took into account a factor of $1/(2 q_0)$, which arises from the normalization of states. The terms involving the magnetic moment arise from the insertion of intermediate nucleon states, schematically, $\langle N | \overline{q} \gamma^\mu \tau^b q (0)|N\pi^d(\vec q)\rangle \supset \int \mathrm{d}^3\vec p\langle N | \overline{q} \gamma^\mu \tau^b q (0)|N(\vec p)\rangle \langle N(\vec p)|N \pi^d(\vec q) \rangle$. Using Eq.~\eqref{eq:tPVpolology}, we determine the following expression for ${t}_{VP}$ in $d=4$,
\begin{align}
{t}_{VP}=-\frac{B_0}{q^2(q^2-m_\pi^2)} \left( 1 - \frac{3+2 \kappa_1}{6 m_N} \vec q^2  i \pi \delta \left( v \cdot q \right) \right)\,, \label{eq:tPVscalarpolology}
\end{align}
which proves Eq.~\eqref{eq:tPVIR}.

For the $VA$ correlator in the vicinity of the pion pole $q^2\to m_\pi^2 $, we obtain the following expression in terms of the matrix elements:
\begin{align}\label{eq:tVApolology}
t^{\mu \nu \rho}_{V A}&\to \frac{\varepsilon^{abc} \tau^c_{ij} S^\rho_{\sigma \sigma^\prime}}{6} \frac{i}{4}\frac{2i q_0}{q^2-m_\pi^2}  \langle N \left(k,\sigma',j \right) | \overline{q} \gamma^\mu \tau^b q (0)|N \left(k,\sigma, i \right)\pi^d(\vec q)\rangle\, \langle \pi^d (\vec q)| \overline{q} \gamma^\nu \gamma_5 \tau^a q (0)  | 0 \rangle\,,
\end{align}
which leads to
\begin{align}
t^{\mu \nu \rho}_{V A}\to \frac{i g^{(0)}_A q^\nu}{q^2-m_\pi^2} &\left[ \left(v^\mu v^\rho-g^{\mu\rho} \right) \left( 1  - \vec q^2 \frac{1+\kappa_1}{2 m_N} i \pi \delta \left( v \cdot q \right) \right) + \overline{q}^\mu \overline{q}^\rho \frac{\kappa_1}{2 m_N} i \pi \delta \left( v \cdot q \right) \right. \nonumber \\
&\left. - \left( v^\mu v^\rho - \frac{v^\mu q^\rho}{v \cdot q} \right) \left( 1 + \frac{\vec{q}^2}{2 m_N v \cdot q} \right) \right].
\end{align}
In the contraction $t_{V A} \sim g_{\mu \nu} t^{\mu \nu \rho}_{V A}$, the leading-order contribution and the contribution from the magnetic moment vanish, which was used in Eqs.~\eqref{eq:a1a2_match_result} and~\eqref{eq:g11_g12_couplings}. As a consistency check, using the Ward identities Eqs.~\eqref{eq:Ward_like_t_VA} and~\eqref{eq:tPVscalar} with the infrared behavior of $t^{\mu \nu \rho}_{V A}$, we also reproduce the infrared behavior for the $VP$ correlator from Eq.~(\ref{eq:tPVscalarpolology}).

\subsection{Three-point functions}
\label{app:a2}

In this Appendix, we study the infrared behavior of the three-point correlation functions $W_P$ and $W_A$,  defined in Eqs.~(\ref{eq:WP3PT_definition}) and~(\ref{eq:WA3PT_definition}), respectively. In particular, we demonstrate the cancellation of the $m_\pi$ dependence of the HBChPT loops by the LEFT correlation functions at the level of the HBChPT coupling constant $[\hat{C}_A]_{\rm 3 pt}$ in Eqs.~(\ref{eq:matchingPS3pt}) and~(\ref{eq:matchingAS3pt}).

First, we determine the leading $Z_\pi$-independent infrared behavior of the hadronic objects $t^{\mu \nu\lambda  \rho}_{AVV}$ and $t^{\mu \nu \rho}_{P V V}$. Following the derivation from Appendix~\ref{app:a1}, we rewrite the correlation function $AVV$ in the vicinity of the pion pole in terms of the matrix elements of quark currents with an intermediate one-pion state as
\begin{align}\label{eq:tVAApolology}
   t^{\mu \nu\lambda  \rho}_{AVV}&\to \frac{ \tau^c_{ij} S^\rho_{\sigma \sigma^\prime}}{48}  \frac{2i q_0 P^{a b c d}}{q^2-m_\pi^2} \int \mathrm{d}^d y \langle N \left(k,\sigma',j \right) |  \overline{q} \gamma^\mu \tau^b q (0)|N \left(k,\sigma, i \right)\pi^e(\vec q)\rangle\, \langle \pi^e (\vec q)| T \left[ \overline{q} \gamma^\nu \tau^a q (0)  \overline{q} \gamma^\lambda \gamma_5 \tau^d q (y) \right] | 0 \rangle\, \nonumber \\
   &+ \frac{ \tau^c_{ij} S^\rho_{\sigma \sigma^\prime}}{48}  \frac{2i q_0 P^{a b c d}}{q^2-m_\pi^2} \int \mathrm{d}^d y \langle N \left(k,\sigma',j \right) |  \overline{q} \gamma^\nu \tau^a q (0)|N \left(k,\sigma, i \right)\pi^e(\vec q)\rangle\, \langle \pi^e (\vec q)| T \left[ \overline{q} \gamma^\mu \tau^b q (0)  \overline{q} \gamma^\lambda \gamma_5 \tau^d q (y) \right] | 0 \rangle\ \nonumber \\
   &= - \frac{g^{(0)}_A g^{\nu \lambda}}{q^2-m_\pi^2} \left[ \left(v^\mu v^\rho-g^{\mu\rho} \right) \left( 1  - \vec q^2 \frac{1+\kappa_1}{2 m_N} i \pi \delta \left( v \cdot q \right) \right) + \overline{q}^\mu \overline{q}^\rho \frac{\kappa_1}{2 m_N} i \pi \delta \left( v \cdot q \right) \right. \nonumber \\
    &\left. \qquad \qquad \qquad - \left( v^\mu v^\rho - \frac{v^\mu q^\rho}{v \cdot q} \right) \left( 1 + \frac{\vec{q}^2}{2 m_N v \cdot q} \right) \right]\,.
\end{align}
Analogously, in the vicinity of the pion pole, we rewrite the correlation function $PVV$ in terms of the matrix elements of quark bilinears with an intermediate one-pion state as
\begin{align}\label{eq:tPVVpolology}
   t^{\mu \nu \rho}_{P V V}&\to \frac{ \tau^c_{ij} S^\rho_{\sigma \sigma^\prime}}{24} \frac{2i q_0 P^{a b c d}}{q^2-m_\pi^2} \int \mathrm{d}^d y e^{-i r \cdot y} \langle N \left(k,\sigma',j \right) | \overline{q} \gamma^\mu \tau^b q (0)|N \left(k,\sigma, i \right)\pi^e(\vec q)\rangle\, \langle \pi^e (\vec q)| \overline{q} \gamma^\nu \tau^a q (0)  \overline{q} \gamma_5 \tau^d q (y)  | 0 \rangle\, \nonumber \\
   &+ \frac{ \tau^c_{ij} S^\rho_{\sigma \sigma^\prime}}{24}  \frac{2 i q_0 P^{a b c d}}{q^2-m_\pi^2} \int \mathrm{d}^d y e^{-i r \cdot y} \langle N \left(k,\sigma',j \right) | \overline{q} \gamma^\nu \tau^a q (0)|N \left(k,\sigma, i \right)\pi^e(\vec q)\rangle\, \langle \pi^e (\vec q)| \overline{q} \gamma^\mu \tau^b q (0)  \overline{q} \gamma_5 \tau^d q (y)  | 0 \rangle\, \nonumber \\
   &=  \frac{g^{(0)}_A}{q^2-m_\pi^2}  \frac{m^2_\pi}{\overline{m}} \frac{q^\nu + r^\nu}{r^2 - m^2_\pi} \left[ \left(v^\mu v^\rho-g^{\mu\rho} \right) \left( 1  - \vec q^2 \frac{1+\kappa_1}{2 m_N} i \pi \delta \left( v \cdot q \right) \right) + \overline{q}^\mu \overline{q}^\rho \frac{\kappa_1}{2 m_N} i \pi \delta \left( v \cdot q \right)  \right] \nonumber \\
   &- \frac{g^{(0)}_A}{q^2-m_\pi^2}  \frac{m^2_\pi}{\overline{m}} \frac{q^\nu + r^\nu}{r^2 - m^2_\pi} \left( v^\mu v^\rho - \frac{v^\mu q^\rho}{v \cdot q} \right)  \left( 1 + \frac{\vec{q}^2}{2 m_N v \cdot q} \right).
\end{align}
The expressions above reproduce the $Z_\pi$-independent terms that involve $m_\pi$ and $\kappa_1$, which describe the whole infrared contribution that does not depend on $Z_\pi$.

In the vicinity of the nucleon pole, an intermediate one-nucleon state, with a heavy nucleon field, contributes the following terms to the $AVV$ correlation function:
\begin{align}\label{eq:tVAApolology_nucleon}
   b_\lambda t^{\mu \nu\lambda  \rho}_{AVV}&\to \frac{ \tau^c_{ij} S^\rho_{\sigma \sigma^\prime} b_\lambda}{48}  \frac{i}{v \cdot \tilde{q} + i \varepsilon} P^{a b c d} \int \mathrm{d}^d x e^{i q \cdot x} \nonumber \\ 
   & \langle N \left(k,\sigma',j \right) |  \overline{q} \gamma^\lambda \gamma_5 \tau^d q (0) | N \left( \tilde{q}, \sigma'', l \right) \rangle\, \langle N \left(  \tilde{q}, \sigma'', l \right)| T \left[ \overline{q} \gamma^\nu \tau^a q (0) \overline{q} \gamma^\mu \tau^b q (x)  \right] | N \left(k,\sigma, i \right) \rangle\, \nonumber \\
   &+ \frac{ \tau^c_{ij} S^\rho_{\sigma \sigma^\prime} b_\lambda}{48}  \frac{i}{v \cdot  \tilde{q} + i \varepsilon} P^{a b c d} \int \mathrm{d}^d x e^{i q \cdot x} \nonumber \\ 
   &\langle N \left(k,\sigma',j \right) |  T \left[ \overline{q} \gamma^\nu \tau^a q (0) \overline{q} \gamma^\mu \tau^b q (x)  \right] | N \left(  \tilde{q}, \sigma'', l \right) \rangle\, \langle N \left(  \tilde{q}, \sigma'', l \right)|  \overline{q} \gamma^\lambda \gamma_5 \tau^d q (0) | N \left(k,\sigma, i \right) \rangle\ \nonumber \\
   &= - \pi \delta \left( v \cdot r \right) g_A^{\left( 0 \right)} \frac{b^\rho  \tau^{\mu \nu}_{VV} \left( q, v \right)}{e^2}, \\
   b_\lambda t^{\mu \nu\lambda  \rho}_{AVV, 0}&\to \frac{ \tau^c_{ij} S^\rho_{\sigma \sigma^\prime} b_\lambda}{48} \frac{1}{9} \frac{i}{v \cdot  \tilde{q} + i \varepsilon} \int \mathrm{d}^d x e^{i q \cdot x} \nonumber \\ 
   & \langle N \left(k,\sigma',j \right) |  \overline{q} \gamma^\lambda \gamma_5 \tau^c q (0) | N \left(  \tilde{q}, \sigma'', l \right) \rangle\, \langle N \left(  \tilde{q}, \sigma'', l \right)| T \left[ \overline{q} \gamma^\nu q (0) \overline{q} \gamma^\mu q (x)  \right] | N \left(k,\sigma, i \right) \rangle\, \nonumber \\
   &+ \frac{ \tau^c_{ij} S^\rho_{\sigma \sigma^\prime} b_\lambda}{48} \frac{1}{9}  \frac{i}{v \cdot \tilde{q} + i \varepsilon} \int \mathrm{d}^d x e^{i q \cdot x} \nonumber \\ 
   &\langle N \left(k,\sigma',j \right) |  T \left[ \overline{q} \gamma^\nu q (0) \overline{q} \gamma^\mu  q (x)  \right] | N \left(  \tilde{q}, \sigma'', l \right) \rangle\, \langle N \left(  \tilde{q}, \sigma'', l \right)|  \overline{q} \gamma^\lambda \gamma_5 \tau^c q (0) | N \left(k,\sigma, i \right) \rangle\ \nonumber \\
   &= - \pi \delta \left( v \cdot r \right) g_A^{\left( 0 \right)} \frac{b^\rho  \tau^{\mu \nu}_{VV,0} \left( q, v \right)}{e^2},
\end{align}
where we injected a nonzero energy $v \cdot r$ in the axial-vector current.

Analogously, we rewrite the correlation function $PVV$ in terms of the matrix elements of quark bilinears with an intermediate one-nucleon state as
\begin{align}\label{eq:tPVVpolology_nucleon}
   b_\lambda \frac{\partial t^{\mu \nu \rho}_{P V V}}{\partial r_\lambda} &\to \frac{ \tau^c_{ij} S^\rho_{\sigma \sigma^\prime} b_\lambda}{24} \frac{\partial}{\partial r_\lambda}  \left( \frac{i}{v \cdot \tilde{q} + i \varepsilon} P^{a b c d} \int \mathrm{d}^d x e^{i r \cdot x + i q \cdot x} \right. \nonumber \\ 
   &  \left. \langle N \left(k,\sigma',j \right) |  \overline{q} \gamma_5 \tau^d q (0) | N \left( \tilde{q}, \sigma'', l \right) \rangle\, \langle N \left( \tilde{q}, \sigma'', l \right)| T \left[ \overline{q} \gamma^\nu \tau^a q (0) \overline{q} \gamma^\mu \tau^b q (x)  \right] | N \left(k,\sigma, i \right) \rangle \right)\, \nonumber \\
   &+ \frac{ \tau^c_{ij} S^\rho_{\sigma \sigma^\prime} b_\lambda}{24} \frac{\partial}{\partial r_\lambda} \left(  \frac{i}{v \cdot \tilde{q} + i \varepsilon} P^{a b c d} \int \mathrm{d}^d x e^{-i r \cdot x + i q \cdot x} \right. \nonumber \\
   &  \left. \langle N \left(k,\sigma',j \right) |  T \left[ \overline{q} \gamma^\nu \tau^a q (0) \overline{q} \gamma^\mu \tau^b q (x)  \right] | N \left( \tilde{q}, \sigma'', l \right) \rangle\, \langle N \left( \tilde{q}, \sigma'', l \right)|  \overline{q} \gamma_5 \tau^d q (0) | N \left(k,\sigma, i \right) \rangle \right)  \nonumber \\
   &= - \pi \delta \left( v \cdot r \right) g_A^{\left( 0 \right)} \frac{b^\rho  \tau^{\mu \nu}_{VV} \left( q, v \right)}{e^2 \overline{m} }, \\
   b_\lambda \frac{\partial t^{\mu \nu \rho}_{P V V, 0}}{\partial r_\lambda} &\to \frac{ \tau^c_{ij} S^\rho_{\sigma \sigma^\prime} b_\lambda}{24} \frac{1}{9} \frac{\partial}{\partial r_\lambda}  \left(  \frac{i}{v \cdot  \tilde{q} + i \varepsilon} \int \mathrm{d}^d x e^{i r \cdot x + i q \cdot x} \right. \nonumber \\
   & \left. \langle N \left(k,\sigma',j \right) |  \overline{q} \gamma_5 \tau^c q (0) | N \left( \tilde{q}, \sigma'', l \right) \rangle\, \langle N \left(  \tilde{q}, \sigma'', l \right)| T \left[ \overline{q} \gamma^\nu  q (0) \overline{q} \gamma^\mu  q (x)  \right] | N \left(k,\sigma, i \right) \rangle \right) \, \nonumber \\
   &+ \frac{ \tau^c_{ij} S^\rho_{\sigma \sigma^\prime} b_\lambda}{24} \frac{1}{9} \frac{\partial}{\partial r_\lambda}  \left( \frac{i}{v \cdot \tilde{q} + i \varepsilon} \int \mathrm{d}^d x e^{-i r \cdot x + i q \cdot x} \right. \nonumber \\
   & \left. \langle N \left(k,\sigma',j \right) |  T \left[ \overline{q} \gamma^\nu q (0) \overline{q} \gamma^\mu q (x)  \right] | N \left( \tilde{q}, \sigma'', l \right) \rangle\, \langle N \left( \tilde{q}, \sigma'', l \right)|  \overline{q} \gamma_5 \tau^c q (0) | N \left(k,\sigma, i \right) \rangle\ \right) \nonumber \\
   &= - \pi \delta \left( v \cdot r \right) g_A^{\left( 0 \right)} \frac{b^\rho  \tau^{\mu \nu}_{VV,0} \left( q, v \right)}{e^2 \overline{m} }.
\end{align}
The polology of three-point correlation functions for the nucleon intermediate state reproduces the expected nucleon pole contribution in Eqs.~(\ref{eq:matchingPS3pt}),~(\ref{eq:matchingPS}), and~(\ref{eq:matchingAS}).

In addition, nonanalytic quark-mass dependence, which is proportional to $Z_\pi$, arises from terms with pion matrix elements of the two vector currents. Such a contribution arises from a two-pion intermediate state. Using similar manipulations as above, the two-pion intermediate state contribution can be written as
\begin{eqnarray}
t^{\mu \nu\lambda  \rho}_{AVV}&\supset &  \frac{ \tau^c_{ij} S^\rho_{\sigma \sigma^\prime}P^{abcd}}{48}\int \frac{\mathrm{d}^{d-1} \vec q_1}{\left( 2\pi \right)^{d-1}} \frac{\mathrm{d}^{d-1} \vec q_2}{\left( 2\pi \right)^{d-1}}   \mathrm{d}^d x\, \mathrm{d}^d y\, e^{i x\cdot q} \langle 0| T[\overline{q}  \gamma^\mu \tau^b q(x) \overline{q}  \gamma^\nu \tau^a q(0)] |\pi^A(\vec q_1)\pi^B(\vec q_2)\rangle\nonumber \\
&&\times \langle \pi^A(\vec q_1)\pi^B( \vec q_2) N \left(k,\sigma',j \right)|  \overline{q}  \gamma^\lambda \gamma_5\tau^d q(y)| N \left(k,\sigma, i \right)\rangle \theta \left( {\rm min} \left( x_0, 0 \right)-y_0 \right)\,.
\end{eqnarray}
The expression above can be simplified by exploiting an integral representation for the step function, $\theta(x) = \frac{i}{2\pi}\int \mathrm{d} w \frac{e^{-iw x}}{w+i\epsilon}$. After using translation invariance to remove the $y$ dependence from the axial-vector current, the integrals over $y$ result into delta functions, $\sim (2\pi)^d \delta (v\cdot (q_1+q_2)+w)\delta^{d-1} (\vec q_1+\vec q_2)$, where $q_{1,2}^0\equiv \sqrt{\vec q_{1,2}^2+m_\pi^2}$. After performing the $\vec q_2$ integral and replacing $q_1 \to q$, this gives rise to
\begin{eqnarray}\label{eq:AVVZpi}
t^{\mu \nu\lambda  \rho}_{AVV}&\supset &  \frac{ \tau^c_{ij} S^\rho_{\sigma \sigma^\prime}P^{abcd}}{48}\int \frac{\mathrm{d}^{d-1} \vec q}{\left( 2\pi \right)^{d-1}}  \mathrm{d}^d x\, e^{i x\cdot q} \langle 0| T[\overline{q}  \gamma^\mu \tau^b q(x) \overline{q}  \gamma^\nu \tau^a q(0)] |\pi^A(\vec q)\pi^B(-\vec q)\rangle \nonumber\\
&&\times \langle \pi^A(\vec q)\pi^B( -\vec q) N \left(k,\sigma',j \right)|  \overline{q}  \gamma^\lambda \gamma_5\tau^d q(0)| N \left(k,\sigma, i \right)\rangle \frac{i}{-2\sqrt{\vec q^2+m_\pi^2}+i\epsilon}\,.
\end{eqnarray}
Combined with the $VV$ matrix element in the first line of Eq.~\eqref{eq:AVVZpi}, this reproduces the Cottingham representation of $Z_\pi$~\cite{Ecker:1988te,Donoghue:1996zn,Cirigliano:2021qko},
\begin{eqnarray}\label{eq:AVVZpi2}
\int \frac{i \mathrm{d}^d q}{\left( 2\pi \right)^d}\frac{g_{\mu\nu} t^{\mu \nu\lambda  \rho}_{AVV}}{q^2-\lambda_\gamma^2}  &\supset &  \frac{ \tau^c_{ij} S^\rho_{\sigma \sigma^\prime}P^{abcd}}{48}\int \frac{\mathrm{d}^{d-1} \vec q}{\left( 2\pi \right)^{d-1}} \frac{4 Z_\pi F_\pi^2}{\sqrt{\vec q^2+m_\pi^2}}\left(\delta ^{Aa}\delta ^{Bb}-\delta^{ab}\delta ^{AB}+(a\leftrightarrow b)\right) \nonumber\\
&&\times \langle \pi^A(\vec q)\pi^B( -\vec q) N \left(k,\sigma',j \right)|  \overline{q}  \gamma^\lambda \gamma_5\tau^d q(0)| N \left(k,\sigma, i \right)\rangle \frac{i}{-2\sqrt{\vec q^2+m_\pi^2}+i\epsilon} \nonumber\\
&=& \frac{ \tau^c_{ij} S^\rho_{\sigma \sigma^\prime}P^{abcd}}{48} 2 Z_\pi F_\pi^2\left(\delta ^{Aa}\delta ^{Bb}-\delta^{ab}\delta ^{AB}+(a\leftrightarrow b)\right) \int \frac{i \mathrm{d}^d  q}{(2\pi)^d} \frac{-1}{ \left[q^2-m_\pi^2+i\epsilon\right]^2}\nonumber\\
&&\times 2q^0 \langle \pi^A(\vec q)\pi^B( -\vec q) N \left(k,\sigma',j \right)|  \overline{q}  \gamma^\lambda \gamma_5\tau^d q(0)| N \left(k,\sigma, i \right)\rangle \,,
\end{eqnarray}
where the factor $2q^0$ in the last line has been factored out in order to cancel the state-normalization factor $1/(2q^0)$. The low-energy representation of the last line can be deduced from chiral symmetry considerations and, as usual, is equivalent to using the HBChPT Lagrangian to deduce the $N\to N\pi\pi$ vertex. Exploiting this representation, the terms that contribute to Eq.~(\ref{eq:AVVZpi2}) read
\begin{align}
    & 2 q^0 \langle \pi^A(\vec q)\pi^B( -\vec q) N \left(k,\sigma',j \right)|  \overline{q}  \gamma^\lambda \gamma_5\tau^d q(0)| N \left(k,\sigma, i \right)\rangle  = - \frac{g^{(0)}_A \delta^{A B}}{2 F^2_\pi} \overline{N} S^\lambda \tau^d N \nonumber \\
    &+ \frac{\left(g^{(0)}_A \right)^3 \delta^{A B}}{F^2_\pi} \frac{ q^2 - \left(v \cdot q \right)^2} {\left(v \cdot q \right)^2} \overline{N} S^\lambda \tau^d N + 3 \frac{\left(g^{(0)}_A \right)^3 \delta^{A B}}{m_N F^2_\pi} q^2 \overline{N} S^\lambda \tau^d N i \pi \delta \left( v \cdot q\right) \nonumber \\
    &+ \frac{8 g^{(0)}_A \delta^{A B}}{F_\pi^2}   \overline{N}   S \cdot q \left[  \left( c_3 - \frac{1}{8 m_N} \right) q^\lambda  - \left(c_4 + \frac{1}{4 m_N} \right) \left[ S^\lambda,  S \cdot q \right] \right] \tau^d  N i \pi \delta \left( v \cdot q \right), 
\end{align}
(note that the second term is valid under the integration in Eq.~(\ref{eq:AVVZpi2})). This result reproduces the expected $Z_\pi$-enhanced $m_\pi$ dependence of the $AVV$ correlation function.

Similarly, for the two-pion state contribution to the $PVV$ correlation function $t^{\mu \nu \rho}_{P V V}$ we write down the Cottingham representation of $Z_\pi$ and obtain the integral of interest
\begin{eqnarray}\label{eq:VVPZpi2}
\int \frac{i\mathrm{d}^d q}{\left( 2\pi \right)^d}\frac{g_{\mu\nu} t^{\mu \nu \rho}_{PVV}}{q^2-\lambda_\gamma^2}  &\supset &   - \frac{ \tau^c_{ij} S^\rho_{\sigma \sigma^\prime}P^{abcd}}{24} 2 Z_\pi F_\pi^2\left(\delta ^{Aa}\delta ^{Bb}-\delta^{ab}\delta ^{AB}+(a\leftrightarrow b)\right) \nonumber\\
&& \int \frac{i \mathrm{d}^d  q}{(2\pi)^d} \frac{2 q^0 \langle \pi^A(\vec q)\pi^B( \vec{r}-\vec q) N \left(k,\sigma',j \right)|  \overline{q}  \gamma_5\tau^d q(0)| N \left(k,\sigma, i \right)\rangle}{ \left(q^2-m_\pi^2+i\epsilon\right)\left(\left(q - r \right)^2-m_\pi^2+i\epsilon\right)} \,.
\end{eqnarray}
As in the $AVV$ analysis, the low-energy representation of the matrix element of the pseudoscalar density consistent with chiral symmetry is most efficiently taken from the relevant chiral Lagrangian. The terms that contribute to Eq.~(\ref{eq:VVPZpi2}) at leading order in $r$ read
\begin{align}
    & 2 q^0 \langle \pi^A(\vec q)\pi^B( \vec{r}-\vec q) N \left(k,\sigma',j \right)|  \overline{q} \gamma_5\tau^d q(0)| N \left(k,\sigma, i \right)\rangle  = - \frac{g^{(0)}_A \delta^{A B}}{4 F_\pi^2 \overline{m}}  \overline{N} S \cdot r \tau^d N  \nonumber \\
    &+\frac{\left(g^{(0)}_A \right)^3 \delta^{A B}}{2 F^2_\pi \overline{m}}  \frac{ q^2 - \left(v \cdot q \right)^2} {\left(v \cdot q \right)^2}  \overline{N} S \cdot r \tau^d N + \frac{5}{2} \frac{\left(g^{(0)}_A \right)^3 \delta^{A B}}{m_N F^2_\pi \overline{m}} q^2 \overline{N} S \cdot r \tau^d N i \pi \delta \left( v \cdot q\right) \nonumber \\
    &+ \frac{f_1 + f_3}{6} \frac{g^{(0)}_A \delta^{A B}}{m_N \overline{m}} \frac{q^2}{Z_\pi \pi m_\pi}  \overline{N} S \cdot r \tau^d N i \pi \delta \left( v \cdot q\right) \nonumber \\
    &+ \frac{4 g^{(0)}_A \delta^{A B}}{F_\pi^2 \overline{m}}   \overline{N}   S \cdot q \left[  \left( c_3 - \frac{1}{8 m_N} \right) q \cdot r  - \left(c_4 + \frac{1}{4 m_N} \right) \left[ S \cdot r,  S \cdot q \right] \right] \tau^d  N i \pi \delta \left( v \cdot q \right),
\end{align}
where the $\left(g^{(0)}_A \right)^3/m_N$ term is valid only under the integration of Eq.~(\ref{eq:VVPZpi2}), and $t^{\mu \nu \rho}_{PVV,0}$ contributes only one term that can be obtain from $t^{\mu \nu \rho}_{PVV}$ by the replacement $f_1+f_3 \to 2 f_4$ in the contribution of $f_i$ coupling constants. Using $\delta^{A d} \tau^B + \delta^{B d} \tau^A - \delta^{A B} \tau^d \to 0$ after the projection with $P^{a b c d} \left(\delta ^{Aa}\delta ^{Bb}-\delta^{ab}\delta ^{AB}+(a\leftrightarrow b)\right)$, we reproduce the expected $Z_\pi$-enhanced $m_\pi$ dependence of the $PVV$ correlation function, which ensures the LECs do not depend on $m_\pi$.

We can also write the Cottingham representation for the contribution from $\tau^{\mu \nu}_{VV}$:
\begin{eqnarray}\label{eq:tauVVPZpi2}
\int \frac{i\mathrm{d}^d q}{\left( 2\pi \right)^d}\frac{g_{\mu\nu} \tau^{\mu \nu}_{VV}}{q^2}  &\supset &   - \frac{ \delta^{a b} \delta_{ij} \delta^{\sigma \sigma^\prime}}{12} 2 Z_\pi F_\pi^2\left(\delta ^{Aa}\delta ^{Bb}-\delta^{ab}\delta ^{AB}+(a\leftrightarrow b)\right) \nonumber\\
&& \int \frac{i \mathrm{d}^d  q}{(2\pi)^d} \frac{2 q^0 \langle \pi^A(\vec q)\pi^B(-\vec q) N \left(k,\sigma',j \right)| N \left(k,\sigma, i \right)\rangle}{ \left[ q^2-m_\pi^2+i\epsilon \right ]^2} \,.
\end{eqnarray}
The corresponding transition matrix element is given by
\begin{align}
    & 2 q^0 \langle \pi^A(\vec q)\pi^B( -\vec q) N \left(k,\sigma',j \right)| N \left(k,\sigma, i \right)\rangle  = -\frac{\left(g^{(0)}_A \right)^2}{8 F^2_\pi} \delta^{A B} q^2 \delta^{j i} \delta^{\sigma^\prime \sigma} i \pi \delta \left( v \cdot q\right).
\end{align}
Evaluating Eq.~(\ref{eq:tauVVPZpi2}) with this matrix element, we obtain the cancellation of $m_\pi$ dependence in the determination of $f_1+f_3$ by Eq.~(\ref{eq:f1f3_match_result}). The corresponding polology for the hadronic object $\tau^{\mu \nu}_{VV,0}$ is trivial, in agreement with vanishing IR contribution in Eq.~(\ref{eq:f4_match_result}).

\section{Ward identities}
\label{app:Ward}

The Ward identities can be obtained by using the fact that the Lagrangian is invariant under local transformations of the chiral symmetry group $G = SU(2)_L \times SU(2)_R \times U(1)_V$
\begin{align}
    q_L \to L (x)e^{i\alpha_V \left( x \right)}  q_L, \quad q_R \to R(x) e^{i\alpha_V \left( x \right)}  q_R\,,\quad     L \left( x \right) = e^{i\vec\alpha_L \left( x \right)\cdot \vec \tau}\,,\quad     R \left( x \right) = e^{i\vec\alpha_R \left( x \right)\cdot \vec\tau}\,,
\end{align}
(or the corresponding transformations of the nucleon and pion fields in HBChPT) if the external left, right, scalar, and pseudoscalar sources $l_\mu,~r_\mu,~s,$ and $p$, respectively, transform as
\begin{align}\label{eq:sourcetransf}
&l_\mu^a  \to l^a_\mu+\partial_\mu \alpha_L^a-2\varepsilon^{abc}\alpha_L^bl^c_\mu\,, &r_\mu^a &\to r^a_\mu+\partial_\mu \alpha_R^a-2\varepsilon^{abc}\alpha_R^br^c_\mu \,,\\
&l_\mu^0  \to l^0_\mu+\partial_\mu \alpha_V\,, &r_\mu^0 & \to r^0_\mu+\partial_\mu \alpha_V\,,\\
&s^a\to s^a+\varepsilon^{abc}s^b (\alpha_L^c+\alpha^c_R)-p^0 (\alpha^a_R-\alpha^a_L)\,, &p^a&\to p^a+\varepsilon^{abc}p^b (\alpha_L^c+\alpha^c_R)-s^0 (\alpha^a_L-\alpha^a_R)\,, \\
&s^0 \to s^0-\vec p\cdot(\vec \alpha_R-\vec\alpha_L) \,,\qquad &p^0& \to p^0-\vec s\cdot(\vec \alpha_L-\vec\alpha_R) \,,
\end{align}
where each external field is decomposed into isoscalar and isovector components as $j=j^0+j^a \tau^a$ for $j=\{ l_\mu, r_\mu, s, p\}$. The quark (or pion and nucleon) fields are integrated over in the generating functionals $W^{\rm LEFT}$ ($W^{{\rm HB}\chi {\rm PT}}$), respectively. Exploiting the invariance of the generating functionals under transformations of the external fields, we obtain
\begin{align} \label{eq:invariance_of_gen_functional}
0=\delta W[l_\mu,r_\mu,s,p]  = \sum_{j}\left[\alpha \frac{\delta j}{\delta \alpha}+\partial^\mu \alpha \frac{\delta j}{\delta (\partial^\mu \alpha)}\right]\frac{\delta }{\delta j}W[l_\mu,r_\mu,s,p] \,, 
\end{align}
where $\alpha$ stands for $\alpha^a_{L}$, $\alpha^a_R$, or $\alpha_V$. Using Eq.~\eqref{eq:sourcetransf} and rewriting the left- and right-handed external sources in terms of $v_\mu$ and $a_\mu$, we introduce the generator of infinitesimal transformations $\mathrm{\Gamma}$ and express Eq.~\eqref{eq:invariance_of_gen_functional} as
\begin{align}\label{eq:ward}
0&=\Gamma(x;\alpha_L,\alpha_R)W\equiv\Bigg\{\left(\alpha_L^a+\alpha_R^a\right)\left[\varepsilon^{abc}\left(v_\mu^c\frac{\delta}{\delta v_\mu^b}+a_\mu^c\frac{\delta}{\delta a_\mu^b}+s^c\frac{\delta}{\delta s^b}+p^c\frac{\delta}{\delta p^b}\right)+\overleftarrow{\partial}_\mu \frac{\delta}{\delta v^a_\mu}\right]\nonumber\\
&+\left(\alpha_L^a-\alpha_R^a\right)\left[\varepsilon^{abc}\left(v_\mu^c\frac{\delta}{\delta a_\mu^b}-a_\mu^c\frac{\delta}{\delta v_\mu^b}\right)+\overleftarrow{\partial}_\mu \frac{\delta}{\delta a^a_\mu}+p^0 \frac{\delta}{\delta s^a}-s^0 \frac{\delta}{\delta p^a}+\vec p^a \frac{\delta}{\delta s^0}-\vec s^a \frac{\delta}{\delta p^0}\right]\Bigg\}W\,,
\end{align}
where we suppressed the coordinate dependence of $\alpha$ and the sources. A similar relation can be derived for the isoscalar transformations, with substitutions $\alpha_L=\alpha_R=\alpha_V$ and $a=0$. $\Gamma$ can now be used to generate Ward identities by letting it act on the generating functional $W$.\footnote{Note that $W$ as used above is strictly speaking the generating functional for vacuum matrix elements, while the matching requires nucleon matrix elements. This discrepancy can be remedied by adding interpolating nucleon fields to $W$ and taking functional derivatives w.r.t.\ the corresponding sources. As the interpolating fields and the sources transform under chiral symmetry, one might worry that this would affect the Ward identity of Eq.\ \eqref{eq:ward}. However, as $S$ matrix elements are independent of the form of the interpolating field, the transformation of these terms drops out when evaluating on-shell matrix elements, ensuring that Eq.\ \eqref{eq:ward} is unaffected.} We provide several phenomenologically relevant examples below.

\subsection{Matrix elements of quark currents}
\label{app:Ward_one_point}

Exploiting $\left[\Gamma(x, e^{i q \cdot x}, \pm e^{i q \cdot x})W\right]_{j=\langle j\rangle }$ in LEFT, with $\langle j\rangle $ the physical values of the external sources,\footnote{The contributions from the electromagnetic and weak currents in $r_\mu$ and $l_\mu$ are suppressed as ${\cal O} (G_F e^2)$.} we obtain the well-known relations for the vector and axial-vector currents,
\begin{align}\label{eq:ward1pt}
e^{i q \cdot x}\left[i q^\mu\frac{\delta W}{\delta v_\mu^a }+\varepsilon^{ab3}s^3\frac{\delta W}{\delta s^b }\right]&= e^{i q \cdot x}\left[i q_\mu \overline{q}\gamma^\mu\frac{ \tau^a}{2} q(x)-\frac{m_u-m_d}{2}\varepsilon^{ab3}\overline{q} \tau^b q(x)\right]=0, \\
e^{i q \cdot x}\left[i q^\mu\frac{\delta W}{\delta a_\mu^a }-s^0\frac{\delta W}{\delta p^a }+s^a\frac{\delta W}{\delta p^0 }\right]&= e^{i q \cdot x}\left[-i q_\mu \overline{q}\gamma^\mu\gamma_5 \frac{\tau^a}{2} q(x)-\overline{m}~\overline{q} \tau^ai\gamma_5 q(x)+\frac{m_u-m_d}{2}\delta ^{a3}\overline{q} i\gamma_5 q(x)\right]=0\,.
\end{align}
These relations prove the conservation of the isoscalar vector current and prove that the conservation of the isovector components of the axial-vector (vector) current is broken by the quark mass (quark-mass differences), respectively.

\subsection{Two-point correlation functions}
\label{app:Ward_two_point}

Useful relations for two-point functions in LEFT can also be obtained by acting on $\Gamma(x;\alpha_L,\alpha_R) W$ with additional functional derivatives. For example, identities involving the  $t_{VA,0}^{\mu\nu\rho}$ and $t_{VA}^{\mu\nu\rho}$ correlators can be derived using an axial-vector transformation $\left[\frac{\delta}{\delta v_\nu ^b(y)}\Gamma(x; e^{i q \cdot \left( x - y \right)}, - e^{i q \cdot \left( x - y \right)})W\right]_{j=\langle j\rangle }$,
\begin{equation}\label{eq:ward2pt}
0 = e^{i q \cdot \left( x - y \right)}\left[ iq_\mu \frac{\delta^2 W}{\delta v_\nu ^b(y) \delta a_\mu ^a(x)} - \delta^d(x-y)\varepsilon^{abc}\frac{\delta W}{\delta a_\nu ^c(x)}-s^0 \frac{\delta^2 W}{\delta v_\nu ^b(y) \delta p ^a(x)}\right].
\end{equation}
After interchanging the coordinates, setting $y=0$, and changing the sign of the momentum $q^\mu$, we obtain the Ward identity
\begin{align}\label{eq:ward2pt1}
0&=\int \mathrm{d}^dx e^{i q \cdot  x}\left( i q_\mu T\left[\overline{q}\gamma^\nu\frac{\tau^b}{2} q(x) \,\overline{q}\gamma^\mu \gamma_5\frac{\tau^a}{2} q(0) \right] - \overline{m} T\left[\overline{q}\gamma^\nu\frac{\tau^b}{2} q(x) \,\overline{q} i \gamma_5\tau^a q(0) \right] \right)-\varepsilon^{abc} i \overline{q} \gamma^\nu\gamma_5\frac{\tau^c}{2} q(0)\,, 
\end{align}
which reproduces the left-hand side in Eqs.~\eqref{eq:Ward_like_t_VA}. Similarly, a vector transformation,  $\left[\frac{\delta}{\delta a_\nu ^b(y)}\Gamma(x; e^{i q \cdot \left( x - y \right)},  e^{i q \cdot \left( x - y \right)})W\right]_{j=\langle j\rangle }$, gives
\begin{equation}\label{eq:ward2pt2}
0 = e^{i q \cdot \left( x - y \right)}\left[ iq_\mu \frac{\delta^2 W}{\delta a_\nu ^b(y) \delta v_\mu ^a(x)}-\delta^d(x-y)\varepsilon^{abc}\frac{\delta W}{\delta a_\nu ^c(x)}+\varepsilon^{ajk}s^k \frac{\delta^2 W}{\delta a_\nu ^b(y) \delta s ^j(x)}\right].
\end{equation}
After interchanging the coordinates, setting $y=0$, and changing the sign of the momentum $q^\mu$, we obtain the Ward identity
\begin{align}\label{eq:ward2pt22}
0&= \int \mathrm{d}^dx e^{i q \cdot x}\left( iq_\mu T\left[\overline{q}\gamma^\nu\gamma_5\frac{\tau^b}{2} q(x) \,\overline{q}\gamma^\mu \frac{\tau^a}{2} q(0) \right] + \frac{m_u-m_d}{2}\varepsilon^{aj3} T\left[\overline{q}\gamma^\nu\gamma_5\frac{\tau^b}{2} q(x) \,\overline{q} \tau^j q(0) \right] \right) \nonumber \\
&- \varepsilon^{abc} i \overline{q} \gamma^\nu\gamma_5\frac{\tau^c}{2} q(0)\,,
\end{align}
which reproduces the expressions on the right-hand side of Eqs.~\eqref{eq:Ward_like_t_VA} in the isospin limit, i.e., $m_u=m_d$.

\subsection{Three-point correlation functions}
\label{app:Ward_three_point}

Finally, we can derive identities between the different representations of the three-point functions by using $\alpha_L=-\alpha_R$,
\begin{align}\label{eq:ward3pt}
0&=\left[\frac{\delta}{\delta v_\alpha^b(y)}\frac{\delta}{\delta v_\beta^c(z)}\Gamma(x; \alpha^a, -\alpha^a)W\right]_{j=\langle j\rangle}=\nonumber\\
&=2\alpha^a\Bigg[\Bigg(\overleftarrow{\partial_\mu}\frac{\delta}{\delta v_\alpha^b(y)}\frac{\delta}{\delta v_\beta^c(z)}\frac{\delta}{\delta a_\mu^a(x)}-\overline{m} \frac{\delta}{\delta v_\alpha^b(y)}\frac{\delta}{\delta v_\beta^c(z)}\frac{\delta}{\delta p^a(x)}\nonumber\\
&-\varepsilon^{acd} g^{\mu\beta}\delta^{d}(x-z)\frac{\delta}{\delta v_\alpha^b(y)}\frac{\delta}{\delta a_\mu^d(x)} -\varepsilon^{abd} g^{\mu\alpha}\delta^{d}(x-y)\frac{\delta}{\delta v_\beta^c(z)}\frac{\delta}{\delta a_\mu^d(x)} \Bigg)W\Bigg]_{j\to \langle j\rangle}\,.
\end{align}
This expression relates $AVV$ three-point functions (the first term) to $PVV$ three-point functions (the second term), as well as $VA$ two-point functions in the last line. After substituting $\alpha^a = e^{- i r\cdot \left( x - y \right)}$, integrating over $x$ and $y$, setting $z=0$, interchanging the coordinates and indices, and taking the derivative $\frac{\partial}{\partial r_\lambda}$, one obtains,
\begin{align}\label{eq:ward3pt2}
&\int \mathrm{d}^dx \mathrm{d}^d y \frac{\delta ^3 W}{\delta v^b_\alpha(x) \delta v^a_\beta(0) \delta a^d_\mu(y)} + r_\kappa \frac{\partial}{\partial r_\lambda} \left. \left[ \int \mathrm{d}^dx \mathrm{d}^d y  e^{i r\cdot \left( x - y \right)} \frac{\delta ^3 W}{\delta v^b_\alpha(x) \delta v^a_\beta(0) \delta a^d_\kappa(y)} \right] \right|_{r_\lambda = 0}\nonumber \\
&=i\frac{\partial}{\partial r_\lambda} \left. \left[ \overline{m} \int \mathrm{d}^dx \mathrm{d}^d y e^{i r\cdot \left( x - y \right)} \frac{ \delta ^3 W}{\delta v^b_\alpha(x) \delta v^a_\beta(0) \delta p^d(y)} \right] \right|_{r_\lambda = 0} + i\frac{\partial}{\partial r_\lambda} \left. \left[ \varepsilon^{acd} \int \mathrm{d}^dx e^{i r\cdot  x } \frac{ \delta ^2 W}{\delta v^b_\alpha(x) \delta a^c_\beta(0)}\right] \right|_{r_\lambda = 0}\,.
\end{align}
A similar relation holds for correlators involving isoscalar vector sources, $a=b=0$, in which case the second term on the right-hand side vanishes.

Multiplying Eq.~\eqref{eq:ward3pt2} by the projectors from Eq.~\eqref{eq:WA3PT_definition} leads to $W_A$ on the left-hand side. The right-hand side can be identified with a combination of the three-point function $W_P$ from Eq.~\eqref{eq:WP3PT_definition}, and two-point function $\Gamma_{VA}$ from Eq.~\eqref{eq:GammaVA}. All in all, we have
\begin{align}\label{eq:ward3pt3}
W_A + W_{A^\prime} \left( r \right) \Big |_{r_\lambda = 0} = W_P +\frac{1}{4}\Gamma_{VA}\,,
\end{align}
with the three-point function $W_{A^\prime} \left( r \right)$:
\begin{align}
W_{A^\prime} \left( r \right) &= - b_\rho  b_\lambda \frac{\tau^c_{ij} S^\rho_{\sigma \sigma^\prime}}{6} r_\kappa  \frac{\partial}{\partial r_\lambda} \int \mathrm{d}^d x \mathrm{d}^d y e^{i r \cdot \left( x - y \right)}\nonumber \\
& \langle N (k, \sigma^\prime, j)| P^{a b c d} \frac{\delta^3 W \left( \overline{a}, {\bf q}_V \right)}{ \delta {\overline{a}}_\kappa^d \left( y \right) \delta {\bf q}_V^{b} \left( x \right) \delta {\bf q}_V^{a} \left( 0 \right)}  +  \frac{\delta^3 W \left( \overline{a}, {\bf q}_V \right)}{\delta {\overline{a}}_\kappa^d \left( y \right) \delta {\bf q}_V^{0} \left( x \right)  \delta {\bf q}_V^{0} \left( 0 \right)} | N (k, \sigma, i) \rangle.
\end{align}
We verified the Ward identity~(\ref{eq:ward3pt3}) in HBChPT for tree-level and one-loop contributions at leading and next-to-leading orders in HBChPT. At this order, the function $W^{\rm HBChPT}_{A^\prime} \left( r \right)$ contributes to the Ward identity as
\begin{align}
     W^{\rm HBChPT}_{A^\prime} \left( r \right) \Big |_{r_\lambda = 0} &= \frac{e^2 g_A^{\left( 0 \right)}}{2 m_N} F^2_\pi \left[ f_1 + f_3 + 2 f_4 + 6 \left( g_A^{(0)}\right)^2 \frac{Z_\pi \pi m_\pi}{\left(  4\pi F_\pi\right)^2}    \right] + e^2 \mathcal{O} \left( \frac{m^2_\pi}{m_N^2}, \frac{m^2_\pi}{\left( 4 \pi F_\pi \right)^2} \right) \nonumber \\
     &= \frac{e^2 g_A^{\left( 0 \right)}}{2 m_N} \int \frac{ \mathrm{d}^4 q}{(2\pi)^4} \frac{ g_{\mu \nu} \left( \tau^{\mu \nu}_{V V} \left(q, v \right) + \tau^{\mu \nu}_{V V, 0} \left(q, v \right)\right)}{q^2} + e^2 \mathcal{O} \left( \frac{m^2_\pi}{m_N^2}, \frac{m^2_\pi}{\left( 4 \pi F_\pi \right)^2} \right).
\end{align}

After inserting the operator equations from Eq.~\eqref{eq:ward3pt2} between nucleon states in LEFT and multiplying by the projectors of Eqs.~\eqref{eq:tAVVdef} and~\eqref{eq:tAVV0def}, the first term in the left-hand side becomes proportional to the correlator $t_{AVV}^{\mu\nu\rho\lambda} \left( q, v \right)$,
\begin{align}
t_{AVV}^{\mu\nu\lambda\rho} \left( q, v \right) + \left. \left( r_\kappa \frac{\partial }{\partial r_\lambda } t_{AVV}^{\mu\nu \kappa \rho} \left(r, q, v \right) \right) \right|_{r_\lambda = 0} &= \overline m\, \frac{\partial }{\partial r_\lambda } \left. \left( t_{PVV}^{\mu\nu\rho} \left(r, q, v \right)+\frac{i}{2 \overline m} t_{VA}^{\mu\nu\rho} \left( q+r, v \right)\right) \right |_{r_\lambda = 0} \,, \label{eq:ward3pt3corr} \\
t_{AVV,0}^{\mu\nu\lambda\rho} \left( q, v \right) + \left. \left( r_\kappa \frac{\partial}{\partial r_\lambda }  t_{AVV,0}^{\mu\nu \kappa \rho} \left(r, q, v \right) \right) \right|_{r_\lambda = 0} &= \overline m\, \frac{\partial }{\partial r_\lambda } \left. \left( t_{PVV,0}^{\mu\nu\rho} \left(r, q, v \right) \right) \right|_{r_\lambda = 0}\,, \label{eq:ward3pt3corr0}
\end{align}
where we introduced the nonperturbative objects $t_{AVV}^{\mu\nu\lambda\rho} \left( r, q, v \right)$ and $t_{AVV, 0}^{\mu\nu\lambda\rho} \left( r, q, v \right)$:
\begin{align}
    t^{\mu \nu \lambda \rho}_{AVV} \left(r, q, v\right) &=  \frac{\tau^c_{i j} S^\rho_{\sigma \sigma^\prime}}{48} P^{abcd} \int \mathrm{d}^d x \mathrm{d}^d y e^{i r \cdot \left( x - y \right) + i q \cdot x} \nonumber \\
    &\langle N (k, \sigma^\prime, j) | T \left[ \overline{q} \gamma^\lambda \gamma^5 \tau^d q(y)\, \overline{q} \gamma^\mu \tau^b q (x) \, \overline{q} \gamma^\nu \tau^a q(0) \right] | N ( k, \sigma, i) \rangle, \\ 
    t^{\mu \nu \lambda \rho}_{AVV,\, 0} \left(r, q, v \right) &= \frac{\tau^c_{i j} S^\rho_{\sigma \sigma^\prime}}{48} \frac{1}{9} \int \mathrm{d}^d x \mathrm{d}^d y e^{i r \cdot \left( x - y \right) + i q \cdot x} \langle N (k, \sigma^\prime, j) | T \left[ \overline{q} \gamma^\lambda \gamma^5 \tau^c q(y)\, \overline{q} \gamma^\mu q (x) \, \overline{q} \gamma^\nu q(0) \right] | N ( k, \sigma, i) \rangle,
\end{align}
which are related to the objects in Eqs.~(\ref{eq:derivative_AVV}) and~(\ref{eq:derivative_AVV0}) as
\begin{align} 
    &r_\lambda t^{\mu \nu \lambda \rho}_{AVV} \left(r, q, v\right) \Big |_{r_\lambda = 0} + i t^{\mu \nu \rho}_{\partial AVV} \left(r, q, v\right) = \frac{i}{2}  \left( t^{\mu \nu \rho}_{VA} \left(q + r, v\right) + t^{\mu \nu \rho}_{VA} \left(-q, v\right)  \right), \label{eq:AVV_with_insertion1} \\
    &r_\lambda t^{\mu \nu \lambda \rho}_{AVV, 0} \left(r, q, v\right) \Big |_{r_\lambda = 0} + i t^{\mu \nu \rho}_{\partial AVV, 0} \left(r, q, v\right) = 0. \label{eq:AVV_with_insertion2}
\end{align}
Starting with identities in Eqs.~(\ref{eq:AVV_with_insertion1}) and~(\ref{eq:AVV_with_insertion2}), we derived the Eqs.~(\ref{eq:ward3pt3corr}) and~(\ref{eq:ward3pt3corr0}) and the relation between LEFT correlation functions in Eq.~(\ref{eq:correlation_functions_matching}). 

Multiplying the above expressions~(\ref{eq:ward3pt3corr}) and~(\ref{eq:ward3pt3corr0}) by the photon propagator in the Feynman-`t Hooft gauge and using the OPE expressions for the $VA$ and $AVV$ correlators in Eqs.~\eqref{eq:OPEtVA},~\eqref{eq:OPEtAVV}, and~\eqref{eq:OPEtAVV0} leads to
\begin{align}
& \int \frac{i \mathrm{d}^4 q}{(2\pi)^4} \frac{g_{\mu \nu} \overline{t}^{\mu \nu \lambda \rho}_{AVV} \left( q, v \right) }{q^2 - \lambda_\gamma^2} +  r_\kappa \frac{\partial }{\partial r_\lambda } \left. \left(\int \frac{i \mathrm{d}^4 q}{(2\pi)^4} \frac{g_{\mu \nu} t_{AVV}^{\mu\nu \kappa \rho} \left(r, q, v \right) }{q^2}   \right) \right|_{r_\lambda = 0} \nonumber \\
&=\overline{m} \frac{\partial}{\partial r_\lambda} \left. \left( \int \frac{i \mathrm{d}^4 q}{(2\pi)^4} \frac{g_{\mu \nu}  {t}^{\mu \nu \rho}_{PVV} \left( r, q, v \right)}{q^2 - \lambda_\gamma^2} \right)\right|_{r_\lambda = 0} +g_A^{(0)} \left( g^{\lambda\rho}-v^\lambda v^\rho \right) \left( \frac{1}{2} \frac{1}{(4\pi)^2}-\int \frac{i \mathrm{d}^4 q}{(2\pi)^4} \frac{{\overline t}_{VA} \left(q, v \right)}{ q^2} \right), \label{eq:relation_between_LEFT_correlators} \\
&\int \frac{i \mathrm{d}^4 q}{(2\pi)^4} \frac{g_{\mu \nu} \overline{t}^{\mu \nu \lambda \rho}_{AVV,0} \left( q, v \right) }{q^2}  + r_\kappa \frac{\partial }{\partial r_\lambda } \left. \left(  \int \frac{i \mathrm{d}^4 q}{(2\pi)^4} \frac{g_{\mu \nu} t_{AVV,0}^{\mu\nu \kappa \rho} \left(r, q, v \right) }{q^2}   \right) \right|_{r_\lambda = 0} \nonumber \\
&=\overline{m} \frac{\partial}{\partial r_\lambda} \left. \left( \int \frac{i \mathrm{d}^4 q}{(2\pi)^4} \frac{g_{\mu \nu} {t}^{\mu \nu \rho}_{PVV,0} \left( r, q, v \right)}{q^2} \right)\right|_{r_\lambda = 0}\,.\label{eq:relation_between_LEFT_correlators0}
\end{align}
Exploiting the derivation of Eq.~\eqref{eq:correlation_functions_matching}, we provide two new sum rules for hadronic objects $t_{AVV}^{\mu\nu\lambda\rho} \left( r, q, v \right)$ and $t_{AVV, 0}^{\mu\nu\lambda\rho} \left( r, q, v \right)$:
\begin{align}
    r_\kappa \frac{\partial }{\partial r_\lambda }  \left. \left(\int \frac{i \mathrm{d}^4 q}{(2\pi)^4} \frac{g_{\mu \nu} t_{AVV}^{\mu\nu \kappa \rho} \left(r, q, v \right) }{q^2}   \right) \right|_{r_\lambda = 0} &= \left( g^{\lambda\rho} - v^\lambda v^\rho  \right) \frac{g_A^{\left( 0 \right)}}{2 m_N} F^2_\pi \left[ f_1 + f_3 + 6 \left( g_A^{(0)}\right)^2 \frac{Z_\pi \pi m_\pi}{\left(  4\pi F_\pi \right)^2} \right] \nonumber \\
    &=  \left( g^{\lambda\rho} - v^\lambda v^\rho  \right) \frac{g_A^{\left( 0 \right)}}{2m_N} \int \frac{ \mathrm{d}^4 q}{(2\pi)^4} \frac{ g_{\mu \nu} \tau^{\mu \nu}_{V V} \left(q, v \right)}{q^2}, \\
    r_\kappa \frac{\partial }{\partial r_\lambda }  \left. \left(\int \frac{i \mathrm{d}^4 q}{(2\pi)^4} \frac{g_{\mu \nu} t_{AVV,0}^{\mu\nu \kappa \rho} \left(r, q, v \right) }{q^2}   \right) \right|_{r_\lambda = 0} &= \left( g^{\lambda\rho} - v^\lambda v^\rho  \right)  \frac{g_A^{\left( 0 \right)}}{m_N} F^2_\pi f_4  \nonumber \\
    &=  \left( g^{\lambda\rho} - v^\lambda v^\rho  \right) \frac{g_A^{\left( 0 \right)}}{2m_N} \int \frac{ \mathrm{d}^4 q}{(2\pi)^4} \frac{ g_{\mu \nu} \tau^{\mu \nu}_{V V,0} \left(q, v \right)}{q^2},
\end{align}
that are valid up to the order $\mathcal{O} \left( \frac{m^2_\pi}{m_N^2}, \frac{m^2_\pi}{\left( 4 \pi F_\pi \right)^2} \right)$.

\section{Low-energy  constants of ${\cal L}_{\pi N}$ at $\mathcal{O}\left(e^2 \right)$ and $\mathcal{O}\left(p^2 \right)$} \label{app:NLOChPT}

In this Appendix, we relate all next-to-leading order coupling constants in the one-nucleon sector of HBChPT to the single-nucleon matrix elements of time-ordered products of one and two quark bilinears (vector, axial-vector, scalar, and pseudoscalar). Exploiting these new relations, we are able to express the radiative corrections to the nucleon axial-vector coupling in terms of nucleon matrix elements of quark operators only, avoiding explicit dependence of the final result on the low-energy couplings of HBChPT. In addition, the matching relations we derive provide a new strategy for the direct lattice-QCD calculation of the LECs $c_{2,3,4}$, which play an important role for the determination of nuclear three-body forces and electroweak two-body currents~\cite{Hoferichter:2015hva}.

\subsection{Electromagnetic Lagrangian $\mathcal L^{e^2}_{\pi N}$}

In Section~\ref{sec:subsec315}, we derived the representations for the combination of LECs $f_{1}+f_3$ and $f_4$. To determine the remaining coupling constants in the $\mathcal L^{e^2}_{\pi N}$ Lagrangian, we define the following nucleon matrix elements of derivatives of the generating functional $W$ w.r.t. the vector and axial-vector charge spurions:
\begin{align}
    G_{VV} &= \frac{\delta^{a b} \delta^{i j}\delta^{\sigma \sigma^\prime}}{12} \int \mathrm{d}^d x \langle N(k, \sigma^\prime, j ) |  \frac{\delta^2 W  \left( {\bf{q}}_V, {\bf{q}}_A, {\bf{q}}_W \right)}{ \delta {\bf{q}}^b_{V} \left( x \right) \delta {\bf{q}}^a_{V} \left( 0 \right)}   \Bigg|_{{\bf{q}}=0} | N(k, \sigma, i) \rangle,  \label{eq:GVVapp} \\
    G_{AA} &= \frac{\delta^{a b} \delta^{i j}\delta^{\sigma \sigma^\prime}}{12} \int \mathrm{d}^d x \langle N(k, \sigma^\prime, j ) |  \frac{\delta^2 W  \left( {\bf{q}}_V, {\bf{q}}_A, {\bf{q}}_W \right)}{ \delta {\bf{q}}^b_{A} \left( x \right) \delta {\bf{q}}^a_{A} \left( 0 \right)}   \Bigg|_{{\bf{q}}=0} | N(k, \sigma, i) \rangle,  \label{eq:GAA}\\
    G_{V0V} &= \frac{\tau^{a b}_{i j}\delta^{\sigma \sigma^\prime}}{3} \int \mathrm{d}^d x \langle N(k, \sigma^\prime, j ) |  \frac{\delta^2 W  \left( {\bf{q}}_V, {\bf{q}}_A, {\bf{q}}_W \right)}{ \delta {\bf{q}}^0_{V} \left( x \right) \delta {\bf{q}}^a_{V} \left( 0 \right)}   \Bigg|_{{\bf{q}}=0} | N(k, \sigma, i) \rangle,  \label{eq:GV0V}\\
    G_{VV,0} &= \frac{\delta^{i j}\delta^{\sigma \sigma^\prime}}{4} \int \mathrm{d}^d x \langle N(k, \sigma^\prime, j ) |  \frac{\delta^2 W  \left( {\bf{q}}_V, {\bf{q}}_A, {\bf{q}}_W \right)}{ \delta {\bf{q}}^0_{V} \left( x \right) \delta {\bf{q}}^0_{V} \left( 0 \right)}   \Bigg|_{{\bf{q}}=0} | N(k, \sigma, i) \rangle, \label{eq:GVV0app} \\
    G_{VA,0} &=  \delta^{i j}\delta^{\sigma \sigma^\prime} \int \mathrm{d}^d x \langle N(k, \sigma^\prime, j ) |  \frac{\delta^2 W  \left( {\bf{q}}_V, {\bf{q}}_A, {\bf{q}}_W \right)}{ \delta {\bf{q}}^0_{V} \left( x \right) \delta {\bf{q}}^0_{A} \left( 0 \right)}   \Bigg|_{{\bf{q}}=0} | N(k, \sigma, i) \rangle. \label{eq:GVA0app}
\end{align}
Evaluating the tree-level and one-loop contributions in HBChPT at leading and next-to-leading orders, we present the correlation functions for the coupling constants in the Lagrangian $\mathcal L^{e^2}_{\pi N}$ as
\begin{align}
    \left. G_{VV} \right |^{\rm HBChPT} &= e^2 F^2_\pi \left[ f_1 + f_3  +  6 \left( g_A^{(0)}\right)^2 \frac{Z_\pi \pi m_\pi}{\left(  4\pi F_\pi\right)^2}   \right],  \label{eq:GVVHBChPT} \\
    \left. G_{AA} \right |^{\rm HBChPT} &= e^2 F^2_\pi \left[ - f_1 + f_3  - 6 \left( g_A^{(0)}\right)^2 \frac{Z_\pi \pi m_\pi}{\left(  4\pi F_\pi\right)^2}    + \frac{\xi }{2}  \left(g_A^{(0)}\right)^2 \frac{\pi m_\pi}{ \left( 4 \pi F_\pi \right)^2 }\right],  \label{eq:GAAHBChPT}\\
    \left. G_{V0V} \right |^{\rm HBChPT} &= 2 e^2 F^2_\pi  f_2,  \label{eq:GV0VHBChPT}\\
    \left. G_{VV,0} \right |^{\rm HBChPT} &= 2 e^2 F^2_\pi  f_4, \label{eq:GVV0HBChPT} \\
    \left. G_{VA,0} \right |^{\rm HBChPT} &= 2 e^2 F^2_\pi  f_5. \label{eq:GVA0HBChPT}
\end{align}
Exploiting the conservation of vector current and the Ward identities for the axial-vector current in LEFT, the correlation functions of interest can be expressed in terms of the ultraviolet-finite integrals from the two-current nucleon matrix elements as
\begin{align}
    \left. G_{VV} \right |^{\rm LEFT} &= e^2  \int \frac{ \mathrm{d}^4 q}{(2\pi)^4} \frac{ g_{\mu \nu} \tau^{\mu \nu}_{V V} \left(q, v \right)}{q^2},  \label{eq:GVVLEFT} \\
    \left. G_{AA} \right |^{\rm LEFT} &= e^2  \int \frac{ \mathrm{d}^4 q}{(2\pi)^4} \frac{ g_{\mu \nu} \tau^{\mu \nu}_{A A} \left(q, v \right)}{q^2} - e^2 \left( 1 - \xi \right) \int \frac{ \mathrm{d}^4 q}{(2\pi)^4} \frac{q_\mu q_\nu}{q^2} \frac{\tau^{\mu \nu}_{A A} \left(q, v \right)}{q^2},  \label{eq:GAALEFT}\\
    \left. G_{V0V} \right |^{\rm LEFT} &= e^2  \int \frac{ \mathrm{d}^4 q}{(2\pi)^4} \frac{ g_{\mu \nu} \tau^{\mu \nu}_{V0V} \left(q, v \right)}{q^2},  \label{eq:GV0VLEFT}\\
    \left. G_{VV,0} \right |^{\rm LEFT} &= e^2  \int \frac{ \mathrm{d}^4 q}{(2\pi)^4} \frac{ g_{\mu \nu} \tau^{\mu \nu}_{V V, 0} \left(q, v \right)}{q^2}, \label{eq:GVV0LEFT} \\
    \left. G_{VA,0} \right |^{\rm LEFT} &= - 4 e^2  \int \frac{ \mathrm{d}^4 q}{(2\pi)^4} \frac{ g_{\mu \nu} \tau^{\mu \nu}_{V A, 0} \left(q, v \right)}{q^2}.\label{eq:GVA0LEFT} 
\end{align}
The nonperturbative hadronic objects $\tau^{\mu \nu}_{VV}, \tau^{\mu \nu}_{AA}, \tau^{\mu \nu}_{V0V}, \tau^{\mu \nu}_{VV,0},$ and $\tau^{\mu \nu}_{VA,0}$ are evaluated at the momentum transfer $q$ and defined as
\begin{align}
    \tau^{\mu \nu}_{VV} \left(q, v \right) &= \frac{\delta^{a b} \delta^{i j} \delta^{\sigma \sigma^\prime}}{48} \int \mathrm{d}^d x e^{ i q \cdot x} \langle N (k, \sigma^\prime, j) | T \left[ \overline{q} \gamma^\mu \tau^b q (x) \, \overline{q} \gamma^\nu \tau^a q(0) \right] | N ( k, \sigma, i) \rangle \label{eq:tauVVapp}, \\
    \tau^{\mu \nu}_{AA} \left(q, v \right) &= \frac{\delta^{a b} \delta^{i j} \delta^{\sigma \sigma^\prime}}{48} \int \mathrm{d}^d x e^{ i q \cdot x} \langle N (k, \sigma^\prime, j) | T \left[ \overline{q} \gamma^\mu \gamma_5 \tau^b q (x) \, \overline{q} \gamma^\nu \gamma_5 \tau^a q(0) \right] | N ( k, \sigma, i) \rangle \label{eq:tauAA},  \\
    \tau^{\mu \nu}_{V0V} \left(q, v \right) &=  \frac{\tau^{a}_{i j}\delta^{\sigma \sigma^\prime}}{12} \frac{1}{3} \int \mathrm{d}^d x e^{ i q \cdot x} \langle N (k, \sigma^\prime, j) | T \left[ \overline{q} \gamma^\mu  q (x) \, \overline{q} \gamma^\nu \tau^a  q(0) \right] | N ( k, \sigma, i) \rangle \label{eq:tauV0V}, \\
    \tau^{\mu \nu}_{VV,0} \left(q, v \right) &=  \frac{\delta^{i j}\delta^{\sigma \sigma^\prime}}{16} \frac{1}{9} \int \mathrm{d}^d x e^{ i q \cdot x} \langle N (k, \sigma^\prime, j) | T \left[ \overline{q} \gamma^\mu  q (x) \, \overline{q} \gamma^\nu  q(0) \right] | N ( k, \sigma, i) \rangle \label{eq:tauVV0app}, \\
    \tau^{\mu \nu}_{VA,0} \left(q, v \right) &=  \frac{\delta^{i j}\delta^{\sigma \sigma^\prime}}{16}\frac{1}{9} \int \mathrm{d}^d x e^{ i q \cdot x} \langle N (k, \sigma^\prime, j) | T \left[ \overline{q} \gamma^\mu  q (x) \, \overline{q} \gamma^\nu \gamma_5  q(0) \right] | N ( k, \sigma, i) \rangle \label{eq:tauVA0}.
\end{align}
By equating LEFT and HBChPT expressions for the nonperturbative objects of Eqs.~(\ref{eq:GVVapp})-(\ref{eq:GVA0app}), we determine the scale- and scheme-independent $\mathcal O \left(e^2 \right)$ coupling constants as
\begin{align}
     f_1 &= \frac{1}{2 F^2_\pi} \int \frac{ \mathrm{d}^4 q}{(2\pi)^4} \frac{ g_{\mu \nu} \left( \tau^{\mu \nu}_{V V} \left(q, v \right) - \tau^{\mu \nu}_{A A} \left(q, v \right) \right)}{q^2} + \frac{1}{2 F^2_\pi}  \int \frac{ \mathrm{d}^4 q}{(2\pi)^4} \frac{q_\mu q_\nu}{q^2} \frac{\tau^{\mu \nu}_{A A} \left(q, v \right)}{q^2} - 6 \left( g_A^{(0)}\right)^2 \frac{Z_\pi \pi m_\pi}{\left(  4\pi F_\pi\right)^2} \nonumber \\
     &- \frac{\xi}{2 F^2_\pi}  \int \frac{ \mathrm{d}^4 q}{(2\pi)^4} \frac{q_\mu q_\nu}{q^2} \frac{\tau^{\mu \nu}_{A A} \left(q, v \right)}{q^2}  + \frac{\xi}{4}   \left(g_A^{(0)}\right)^2 \frac{\pi m_\pi}{ \left( 4 \pi F_\pi \right)^2 } ,\label{eq:f1} \\
     f_2 &= \frac{1}{2 F^2_\pi} \int \frac{ \mathrm{d}^4 q}{(2\pi)^4} \frac{ g_{\mu \nu}  \tau^{\mu \nu}_{V0V} \left(q, v \right)}{q^2},\label{eq:f2} \\
     f_3 &= \frac{1}{2 F^2_\pi} \int \frac{ \mathrm{d}^4 q}{(2\pi)^4} \frac{ g_{\mu \nu} \left( \tau^{\mu \nu}_{V V} \left(q, v \right) + \tau^{\mu \nu}_{A A} \left(q, v \right) \right)}{q^2} - \frac{1}{2 F^2_\pi}  \int \frac{ \mathrm{d}^4 q}{(2\pi)^4} \frac{q_\mu q_\nu}{q^2} \frac{\tau^{\mu \nu}_{A A} \left(q, v \right)}{q^2} \nonumber \\
     &+ \frac{\xi}{2 F^2_\pi}  \int \frac{ \mathrm{d}^4 q}{(2\pi)^4} \frac{q_\mu q_\nu}{q^2} \frac{\tau^{\mu \nu}_{A A} \left(q, v \right)}{q^2} -  \frac{\xi}{4} \left(g_A^{(0)}\right)^2 \frac{\pi m_\pi}{ \left( 4 \pi F_\pi \right)^2 },\label{eq:f3} \\
     f_4 &= \frac{1}{2 F^2_\pi} \int \frac{ \mathrm{d}^4 q}{(2\pi)^4} \frac{ g_{\mu \nu} \tau^{\mu \nu}_{VV,0} \left(q, v \right)}{q^2},\label{eq:f4} \\
     f_5&= - \frac{2}{F^2_\pi} \int \frac{ \mathrm{d}^4 q}{(2\pi)^4} \frac{ g_{\mu \nu}  \tau^{\mu \nu}_{VA,0} \left(q, v \right)}{q^2}\label{eq:f5}.
\end{align}

\subsection{Chiral Lagrangian $\mathcal L^{p^2}_{\pi N}$}

The Lagrangian $\mathcal L^{p^2}_{\pi N}$ depends on 7 LECs~\cite{Gasser:1987rb,Krause:1990xc,Ecker:1995rk,Bernard:1992qa,Meissner:1997ii}. $c_{1}$ and $c_{5}$ determine the coupling of the nucleon to isoscalar and isovector scalar sources as well as the quark-mass dependence of the nucleon mass and of the neutron-proton mass splitting. $\kappa_{0,1}$ represent the leading contributions to the nucleon anomalous magnetic moments. $c_{2,3,4}$ parameterize the coupling of two pions to two nucleons. These couplings are important for nuclear physics, as they determine pion-range contributions to the three-nucleon force. Chiral invariance relates them to the interaction of the axial-vector current with one pion and two nucleons, and thus to weak two-body currents, which are crucial for the understanding of nuclear $\beta$ decays~\cite{Pastore:2017uwc}. $c_{2,3,4}$ are currently extracted from pion-nucleon scattering data~\cite{Hoferichter:2015hva,Hoferichter:2016duk}. A first-principles determination from calculations of pion-nucleon scattering on the lattice has been attempted in Ref.~\cite{Bulava:2022vpq}. Here we propose an alternative strategy based on the calculation of the correlators of two axial-vector or two vector currents, which avoids the complications of multiparticle states on the lattice.

The coupling constants of $\mathcal L^{p^2}_{\pi N}$ can be determined by evaluating the following matrix elements of derivatives of the generating functional $W$ w.r.t. the external sources
\begin{align}
    G_{s,0} &= \frac{\delta^{i j} \delta^{\sigma \sigma^\prime}}{32 B_0}  \langle N(k, \sigma^\prime, j ) |  \frac{\delta W  \left( v, a, s, p \right)}{ \delta s^0 \left( 0 \right) }   \Bigg|_{v, a, s, p=0} | N(k, \sigma, i) \rangle \label{eq:Gs0}, \\
    G_{s} &= \frac{\tau^a_{i j} \delta^{\sigma \sigma^\prime}}{48 B_0} \langle N(k, \sigma^\prime, j ) |  \frac{\delta W  \left( v, a, s, p \right)}{\delta s^a \left( 0 \right)}   \Bigg|_{v, a, s, p=0} | N(k, \sigma, i) \rangle \label{eq:Gs},\\
    \Gamma^r_{v,0} &= \frac{\delta^{i j}  \left[ S_\nu,  S_\mu \right]^{\sigma \sigma'}}{2 \left( d - 2 \right) \left( d - 1 \right)}  \langle N(k + r, \sigma^\prime, j ) |  \frac{\partial}{\partial r^\mu} \left( \frac{\delta W  \left( v, a, s, p \right)}{\delta v^0_\nu \left( 0 \right)}   \Bigg|_{v, a, s, p=0} \right) | N(k, \sigma, i) \rangle \Bigg |_{r_\lambda = 0}\label{eq:Gammav0}, \\
    \Gamma^r_{v} &= \frac{\tau^a_{i j}  \left[ S_\nu,  S_\mu \right]^{\sigma \sigma'}}{6 \left( d - 2 \right) \left( d - 1 \right)}  \langle N(k + r, \sigma^\prime, j ) | \frac{\partial}{\partial r^\mu} \left( \frac{\delta W  \left( v, a, s, p \right)}{\delta v^a_\nu \left( 0 \right)}   \Bigg|_{v, a, s, p=0} \right) | N(k, \sigma, i) \rangle \Bigg |_{r_\lambda = 0} \label{eq:Gammav}, \\
    G^v_{aa} &= \frac{\delta^{a b} \delta^{i j} \delta^{\sigma \sigma^\prime}}{24} v_{\nu} v_{\mu} \int \mathrm{d}^d x \langle N(k, \sigma^\prime, j ) |  \frac{\delta^2 W  \left( v, a, s, p \right)}{ \delta a^b_\nu \left( x \right) \delta a^a_\mu \left( 0 \right)}   \Bigg|_{v, a, s, p=0} | N(k, \sigma, i) \rangle,  \label{eq:Gaav}\\
    G_{aa} &= \frac{\delta^{a b} \delta^{i j} \delta^{\sigma \sigma'}}{24} \frac{g_{\nu \mu} - v_\nu v_\mu}{d - 1} \int \mathrm{d}^d x \langle N(k, \sigma^\prime, j ) |  \frac{\delta^2 W  \left( v, a, s, p \right)}{ \delta a^b_\nu \left( x \right) \delta a^a_\mu \left( 0 \right)}   \Bigg|_{v, a, s, p=0} | N(k, \sigma, i) \rangle,  \label{eq:GaaS}\\
    \Gamma_{aa} &= \frac{\varepsilon^{a b c} \tau^c_{i j}  \left[ S_\mu,  S_\nu \right]^{\sigma \sigma'}}{12 \left( d - 2 \right) \left( d - 1 \right)}  i \int \mathrm{d}^d x \langle N(k, \sigma^\prime, j ) |  \frac{\delta^2 W  \left( v, a, s, p \right)}{ \delta a^b_\nu \left( x \right) \delta a^a_\mu \left( 0 \right)}   \Bigg|_{v, a, s, p=0} | N(k, \sigma, i), \label{eq:Gammaaa}  \\
    \Gamma_{vv} &= \frac{\varepsilon^{a b c} \tau^c_{i j}  \left[ S_\mu,  S_\nu \right]^{\sigma \sigma'}}{12 \left( d - 2 \right) \left( d - 1 \right)} i \int \mathrm{d}^d x \langle N(k, \sigma^\prime, j ) |  \frac{\delta^2 W  \left( v, a, s, p \right)}{ \delta v^b_\nu \left( x \right) \delta v^a_\mu \left( 0 \right)}   \Bigg|_{v, a, s, p=0} | N(k, \sigma, i) \label{eq:Gammavv}, 
\end{align}
where we imply the derivative of the expression without external spinors in Eqs.~(\ref{eq:Gammav0}) and~(\ref{eq:Gammav}). Eqs.~(\ref{eq:Gs0}) and~(\ref{eq:Gs}) contain the matrix elements of the isosinglet and isotriplet scalar charges, respectively, while Eqs.~(\ref{eq:Gammav0}) and~(\ref{eq:Gammav}) correspond to normalization of the nucleon isoscalar and isovector magnetic form factors, respectively. Eqs.~(\ref{eq:Gaav}) to~(\ref{eq:Gammavv}) contain the matrix elements of two axial-vector or two vector currents, at zero momentum transfer.

With the tree-level and one-loop contributions in HBChPT at leading and next-to-leading orders, the correlation functions for the coupling constants in the Lagrangian $\mathcal L^{p^2}_{\pi N}$ are given by
\begin{align}
    \left. G_{s,0} \right |^{\rm HBChPT} &= c_1 + \frac{9}{16} \left( g_A^{\left( 0 \right)} \right)^2 \frac{\pi m_\pi}{\left(4 \pi F_\pi \right)^2} \label{eq:Gs0HBChPT}, 
    \\
    \left. G_{s} \right |^{\rm HBChPT} &= c_5 \label{eq:GsHBChPT}, \\
    \left. \Gamma^r_{v,0} \right |^{\rm HBChPT} &=  - \frac{1 + \kappa_0}{4 m_N} \label{eq:Gammav0HBChPT}, \\
    \left. \Gamma^r_{v} \right |^{\rm HBChPT} &=  - \frac{1 + \kappa_1}{4 m_N} + \left( g_A^{\left( 0 \right)} \right)^2 \frac{\pi m_\pi}{\left(4 \pi F_\pi \right)^2} \label{eq:GammavHBChPT}, \\
    \left. G^{v}_{aa} \right |^{\rm HBChPT} &= c_2 - \frac{\left(g_A^{(0)}\right)^2}{8 m_N} + c_3 - \left[ 1 - \frac{3}{2} \left( g_A^{\left( 0 \right)} \right)^2 \right] \frac{\pi m_\pi}{\left( 4 \pi F_\pi\right)^2},  \label{eq:GaaSHBChPT}\\
    \left. G_{aa} \right |^{\rm HBChPT} &= c_3 + \left[ \frac{3}{2} + \left( g_A^{\left( 0 \right)} \right)^2 \right] \left( g_A^{\left( 0 \right)} \right)^2 \frac{\pi m_\pi}{\left( 4 \pi F_\pi\right)^2} - i\frac{\pi}{4} \left(g_A^{(0)}\right)^2 \delta \left( v \cdot r \right),  \label{eq:GaavHBChPT}\\
    \left. \Gamma_{aa} \right |^{\rm HBChPT} &= c_4 - \frac{\kappa_1}{4 m_N} - \left( g_A^{\left( 0 \right)} \right)^4 \frac{\pi m_\pi}{\left( 4 \pi F_\pi\right)^2} + i\frac{\pi}{2} \left(g_A^{(0)}\right)^2 \delta \left( v \cdot r \right), \label{eq:GammaaaHBChPT}  \\
    \left. \Gamma_{vv} \right |^{\rm HBChPT} &=  - \frac{1 + \kappa_1}{4 m_N} + \left( g_A^{\left( 0 \right)} \right)^2 \frac{\pi m_\pi}{\left(4 \pi F_\pi \right)^2} \label{eq:GammavvHBChPT}. 
\end{align}
Eqs.~(\ref{eq:Gammav0HBChPT}) and~(\ref{eq:GammavHBChPT}) agree with the NLO calculation of the nucleon magnetic moment in HBChPT, cf. Refs.~\cite{Bernard:1995dp,Bernard:1992qa}, while Eq.~(\ref{eq:Gs0HBChPT}) corresponds to the one-loop correction to the pion-nucleon sigma term, cf. Ref.~\cite{Hoferichter:2016duk}. The threshold behavior of nucleon vector form factors and relativistic calculations are described in Refs.~\cite{Bernard:1996cc,Becher:1999he,Gegelia:1999gf,Becher:2001hv,Fuchs:2003qc,Kaiser:2003qp}.

In LEFT, the correlation functions of interest can be expressed in terms of the nucleon matrix elements of one or two quark currents. We can write 
\begin{align}
    \left. G_{s,0} \right |^{\rm LEFT} &= - \tau_0, \label{eq:Gs0LEFT}  \\
    \left. G_{s} \right |^{\rm LEFT} &= - \tau, \label{eq:GsLEFT} \\
    \left. \Gamma^r_{v,0} \right |^{\rm LEFT} &=  \tilde{t}_{V,0}, \label{eq:Gammav0LEFT} \\
    \left. \Gamma^r_{v} \right |^{\rm LEFT} &=  \tilde{t}_{V}, \label{eq:GammavLEFT} \\
    \left. G^{v}_{aa} \right |^{\rm LEFT} &= \frac{i}{2} v_\nu v_\mu \tau^{\nu \mu}_{A A} \left( 0, v \right),  \label{eq:GaaSLEFT}\\
    \left. G_{aa} \right |^{\rm LEFT} &= \tilde{\tau}_{A A} \left( 0, v \right),  \label{eq:GaavLEFT}\\
    \left. \Gamma_{aa} \right |^{\rm LEFT} &= - \tilde{t}_{A A} \left( 0, v \right), \label{eq:GammaaaLEFT}  \\
    \left. \Gamma_{vv} \right |^{\rm LEFT} &=  - \tilde{t}_{V V} \left( 0, v \right), \label{eq:GammavvLEFT}
\end{align}
where the hadronic objects $\tau_0, \tau, \tilde{t}_{V,0}, \tilde{t}_{V}, \tilde{\tau}_{AA}, \tilde{t}_{AA},$ and $\tilde{t}_{VV}$ that are evaluated at one or a few points in momentum space
\begin{align}
    \tau_{0} &= \frac{\delta^{i j} \delta^{\sigma \sigma^\prime}}{32 B_0}  \langle {N} (k, \sigma^\prime, j )  | \bar{q} q | N (k, \sigma, i) \rangle, \label{eq:tau0} \\
    \tau &= \frac{\tau^a_{i j} \delta^{\sigma \sigma^\prime}}{48 B_0} \langle {N} (k, \sigma^\prime, j ) | \bar{q}  \tau^a q | N (k, \sigma, i) \rangle, \label{eq:tau} \\
    \tilde{t}_{V,0} &= \frac{\delta^{i j}  \left[ S_\nu,  S_\mu \right]^{\sigma \sigma'}}{4 \left( d - 2 \right) \left( d - 1 \right)}  \langle {N} (k +r , \sigma^\prime, j )  |\frac{\partial}{\partial r^\mu} \left[ \bar{q} \gamma^\nu q \left( 0 \right) \right] | N (k, \sigma, i) \rangle \Bigg |_{r_\lambda = 0}, \label{eq:tV0}\\
    \tilde{t}_{V} &= \frac{\tau^a_{i j}  \left[ S_\nu,  S_\mu \right]^{\sigma \sigma'}}{12 \left( d - 2 \right) \left( d - 1 \right)} \langle {N} (k + r, \sigma^\prime, j )  |\frac{\partial}{\partial r^\mu} \left[   \bar{q} \gamma^\nu \tau^a q \left( 0 \right) \right] | N (k, \sigma, i) \rangle \Bigg |_{r_\lambda = 0} , \label{eq:tV}  \\
    \tilde{\tau}_{AA} \left(q, v \right) &=  \frac{\delta^{a b} \delta^{i j} \delta^{\sigma \sigma'}}{96}  \frac{g_{\nu \mu} - v_\nu v_\mu}{d - 1} i \int \mathrm{d}^d x e^{ i q \cdot x} \langle {N} (k, \sigma^\prime, j )  | T \left[  \bar{q} \gamma^\nu \gamma_5 \tau^b q \left( x \right) \bar{q} \gamma^\mu \gamma_5 \tau^a q \left( 0 \right) \right] | N (k, \sigma, i) \rangle, \label{eq:tildetauAA} \\
    \tilde{t}_{AA} \left(q, v \right) &=  \frac{\varepsilon^{a b c} \tau^c_{i j}  \left[ S_\mu,  S_\nu \right]^{\sigma \sigma'}}{48 \left( d - 2 \right) \left( d - 1 \right)} \int \mathrm{d}^d x e^{ i q \cdot x} \langle {N} (k, \sigma^\prime, j )  | T \left[  \bar{q} \gamma^\nu \gamma_5 \tau^b q \left( x \right) \bar{q} \gamma^\mu \gamma_5 \tau^a q \left( 0 \right) \right] | N (k, \sigma, i) \rangle,  \\
    \tilde{t}_{VV} \left(q, v \right) &= \frac{\varepsilon^{a b c} \tau^c_{i j}  \left[ S_\mu,  S_\nu \right]^{\sigma \sigma'}}{48 \left( d - 2 \right) \left( d - 1 \right)} \int \mathrm{d}^d x e^{ i q \cdot x} \langle {N} (k, \sigma^\prime, j )  | T \left[  \bar{q} \gamma^\nu \tau^b q \left( x \right) \bar{q} \gamma^\mu \tau^a q \left( 0 \right) \right] | N (k, \sigma, i) \rangle. \label{eq:tVV}
\end{align}
We imply the derivative of the expression without external spinors in Eqs.~(\ref{eq:tV0}) and~(\ref{eq:tV}).

By equating LEFT and HBChPT expressions for the nonperturbative objects of Eqs.~(\ref{eq:Gs0})-(\ref{eq:Gammavv}), we determine the scale- and scheme-independent NLO coupling constants
\begin{align}
     c_1&= -\tau_0 - \frac{9}{16}\left( g_A^{\left( 0 \right)} \right)^2 \frac{\pi m_\pi}{\left(4 \pi F_\pi \right)^2} ,\label{eq:c1}\\
     c_2&= \frac{i}{2} v_\mu v_\nu \tau^{\mu \nu}_{A A} \left( 0, v \right)-  \tilde{\tau}_{AA}\left( r, v \right) + \frac{\left(g_A^{(0)}\right)^2}{8 m_N} + \frac{\pi m_\pi}{\left( 4 \pi F_\pi\right)^2} - i \frac{\pi}{4} \left(g_A^{(0)}\right)^2 \delta \left( v \cdot r \right),\label{eq:c2}\\
     c_3&= \tilde{\tau}_{AA} \left( r, v \right) - \left[ \frac{3}{2} +  \left( g_A^{\left( 0 \right)} \right)^2 \right] \left( g_A^{\left( 0 \right)} \right)^2 \frac{\pi m_\pi}{\left( 4 \pi F_\pi\right)^2} + i \frac{\pi}{4} \left(g_A^{(0)}\right)^2 \delta \left( v \cdot r \right),\label{eq:c3}\\
     c_4&=  \tilde{t}_{VV} \left( 0, v \right) - \tilde{t}_{AA} \left( r, v \right) - \frac{1}{4 m_N} + \left( g_A^{\left( 0 \right)} \right)^2 \frac{\pi m_\pi}{\left(4 \pi F_\pi \right)^2} + \left( g_A^{\left( 0 \right)} \right)^4  \frac{\pi m_\pi}{\left( 4 \pi F_\pi\right)^2} \nonumber \\
     &- i \frac{\pi}{2} \left(g_A^{(0)}\right)^2 \delta \left( v \cdot r \right),\label{eq:c4}\\
     c_5&= -\tau,\label{eq:c5}\\
     \frac{1+\kappa_0}{4 m_N}&= -\tilde{t}_{V,0}, \label{eq:kappa0} \\
     \frac{1+\kappa_1}{4 m_N}&= - \tilde{t}_{V} + \left( g_A^{\left( 0 \right)} \right)^2 \frac{\pi m_\pi}{\left(4 \pi F_\pi \right)^2} = \tilde{t}_{VV} \left( 0, v \right) + \left( g_A^{\left( 0 \right)} \right)^2 \frac{\pi m_\pi}{\left(4 \pi F_\pi \right)^2}. \label{eq:kappa} 
\end{align}
As in Section~\ref{sec:subsec35}, the calculation of the correlation functions can be performed by having a small amount of energy $v\cdot r$ flowing through the currents, and then by taking the limit $v \cdot r \rightarrow 0$ in the matching relations in Eqs.~\eqref{eq:c2}-\eqref{eq:c4}. After working out the polology for all correlation functions with nucleon, two-pion, and pion-nucleon intermediate states, we verified that all coupling constants $f_1$-$f_5,$ $c_1$-$c_5,$ and $\kappa_0,~\kappa_1$ in Eqs.~(\ref{eq:f1})-(\ref{eq:f5}) and Eqs.~(\ref{eq:c1})-(\ref{eq:kappa}) are free from the nucleon poles and do not depend on $m_\pi$.

\subsection{Determination of $\hat{C}_A$}

With the LEFT representation for the coupling constants $c_3$ and $c_4$ of Eqs.~(\ref{eq:c3}) and~(\ref{eq:c4}), respectively, we reexpress the $\mathcal{O}\left( \alpha \right)$ radiative correction to the nucleon axial-vector coupling. Considering the expression with the three-point function that involves the pseudoscalar density, cf. Eq.~(\ref{eq:matchingPS}), we obtain
\begin{align} \label{eq:CA_pseudoscalar2}
    &\hat{C}_{A} \left( a, \mu_\chi, \mu \right) =  \frac{Z_\pi}{2} \left( \left[ 1 + 3 \left( g_A^{(0)} \right)^2 \right] \left( 1 -  \ln \frac{\mu^2_\chi}{m_\pi^2}  \right) + 2 \left( g_A^{(0)} \right)^2 - 8 \pi m_\pi \left[ \tilde{t}_{VV} \left( 0, v \right) + \frac{1}{8 m_N} +  \frac{9}{16} \frac{\left( g_A^{(0)} \right)^2}{m_N}\right] \right) \nonumber \\
    &+ 4 Z_\pi \pi m_\pi \left( \tilde{\tau}_{AA} \left( r, v \right) + \tilde{t}_{AA} \left( r, v \right) +\frac{3}{4} i \pi \delta \left( v \cdot r \right) \left(g_A^{(0)}\right)^2\right)  \nonumber \\
    &+ \frac{1}{2}\left[ \frac{1}{2} \ln \frac{\mu^2}{\mu^2_0} - \frac{3}{2} \ln \frac{\mu^2_\chi}{\mu^2} - 4 B \left(a \right)- \frac{1}{4} \right] + \frac{16 \pi^2}{g^{(0)}_A} \int \frac{i \mathrm{d}^4 q}{\left( 2 \pi \right)^4}  \left( \frac{ \nu^2 - 2 Q^2}{3 Q^2} \frac{\overline{S}_1 (\nu, Q^2)}{Q^2} - \frac{\nu^2}{Q^2 } \frac{\overline{S}_2 (\nu, Q^2)}{m_N \nu} \right) \nonumber \\
    &+ \frac{8 \pi^2}{g_A^{(0)}} \left.\left[ \overline{m} b_\rho b_\lambda \frac{\partial}{\partial r_\lambda} \left( \int \frac{i \mathrm{d}^4 q}{(2\pi)^4} \frac{g_{\mu \nu} \left( {t}^{\mu \nu \rho}_{PVV} \left( r, q, v \right) + {t}^{\mu \nu \rho}_{PVV,0}\left( r, q, v \right) \right)}{q^2} \right) + 2 i \pi \delta(v \cdot r) \frac{g_A^{(0)}\Delta_{\rm em} m_N}{e^2} \right]\right|_{r_\lambda =0} \nonumber \\ 
    & + 16 \pi^2  \overline{m} \int \frac{i \mathrm{d}^4 q}{(2\pi)^4} \frac{{t}_{VP} \left(q, v \right)}{ q^2} -  \frac{4 \pi^2}{m_N} \int \frac{ \mathrm{d}^4 q}{(2\pi)^4} \frac{ g_{\mu \nu} \left( \tau^{\mu \nu}_{V V} \left(q, v \right) + \tau^{\mu \nu}_{V V, 0} \left(q, v \right) \right)}{q^2}.
\end{align}
The expression in terms of the three-point function that involves the axial-vector current, cf. Eq.~(\ref{eq:matchingAS}), results in
\begin{align} \label{eq:CA_axial_vector2}
    &\hat{C}_{A} \left( a, \mu_\chi, \mu \right) =  \frac{Z_\pi}{2} \left( \left[ 1 + 3 \left( g_A^{(0)} \right)^2 \right] \left( 1 -  \ln \frac{\mu^2_\chi}{m_\pi^2}  \right) + 2 \left( g_A^{(0)} \right)^2 - 8 \pi m_\pi \left[ \tilde{t}_{VV} \left( 0, v \right) + \frac{1}{8 m_N} + \frac{9}{16} \frac{\left( g_A^{(0)} \right)^2}{m_N}\right] \right) \nonumber \\
    &+ 4 Z_\pi \pi m_\pi \left( \tilde{\tau}_{AA} \left( r, v \right) + \tilde{t}_{AA} \left( r, v \right) +\frac{3}{4} i \pi \delta \left( v \cdot r \right) \left(g_A^{(0)}\right)^2\right)  \nonumber \\
    &+ \frac{1}{2}\left[\frac{1}{2} \ln \frac{\mu^2}{\mu^2_0} - \frac{3}{2} \ln \frac{\mu_\chi^2}{\mu^2}  - 4 B(a) + \frac{1}{4}  \right]  + \frac{16 \pi^2}{g^{(0)}_A} \int \frac{i \mathrm{d}^4 q}{\left( 2 \pi \right)^4}  \left( \frac{ \nu^2 - 2 Q^2}{3 Q^2} \frac{\overline{S}_1 (\nu, Q^2)}{Q^2} - \frac{\nu^2}{Q^2 } \frac{\overline{S}_2 (\nu, Q^2)}{m_N \nu} \right) \nonumber \\
    & + \frac{8 \pi^2}{g_A^{(0)}} \left. \left[  b_\rho b_\lambda \int \frac{i \mathrm{d}^4 q}{(2\pi)^4} \frac{g_{\mu \nu} \left( \overline{t}^{\mu \nu \lambda \rho}_{AVV} \left( q, v \right) + \overline{t}^{\mu \nu \lambda \rho}_{AVV,0}\left( q, v \right) \right)}{q^2} + 2 i \pi  \delta(v\cdot r) \, g_A^{(0)}\frac{\Delta_{\rm em} m_N}{e^2} \right] \right|_{r_\lambda =0} \nonumber \\ 
    & + 8 \pi^2  \int \frac{i \mathrm{d}^4 q}{(2\pi)^4} \frac{2 \overline{m}{t}_{VP} \left(q, v \right) - {\overline t}_{VA} \left(q, v \right)}{ q^2}.
\end{align}

\bibliography{beta_decay_EFT}{}

\end{document}